\documentclass[preprint2]{pp7}
\usepackage{graphicx,epsf,amsmath,amsfonts,amssymb,wrapfig,bm,natbib,color,url,hyperref}
\usepackage{enumitem}
\usepackage{comment}
\usepackage{ulem}
\usepackage[dvipsnames]{xcolor}
\usepackage{paralist}
\usepackage{upgreek}
\usepackage{threeparttable}

\definecolor{darkgreen}{rgb}{0.0,0.6,0.00}

\newcommand{\Macc}{\dot{M}_\mathrm{acc}}
\newcommand{\Mwind}{\dot{M}_\mathrm{wind}}
\newcommand{\OmegaK}{\Omega_\mathrm{K}}
\newcommand{\ass}{\alpha_\mathrm{S}}
\newcommand{\adust}{\alpha_\mathrm{D}}

\newcommand{\bp}{\beta_\mathrm{p}}
\newcommand{\vturb}{v_{\rm turb}/c_{\rm s}}
\newcommand{\au}{\mathrm{AU}}
\newcommand{\epsdust}{\epsilon_\mathrm{d}}
\newcommand{\epsgas}{\varepsilon}

\newcommand{\p}{\partial}
\newcommand{\rhog}{\rho_\mathrm{g}}
\newcommand{\Sigmag}{\Sigma_\mathrm{g}}
\newcommand{\vgas}{\bm{v}_\mathrm{g}}
\newcommand{\Hgas}{H}   %
\newcommand{\hgas}{\epsgas}%
\newcommand{\OmK}{\Omega_\mathrm{K}}
\newcommand{\cp}{c_P}
\newcommand{\Tref}{T_\mathrm{ref}}
\newcommand{\tcool}{\tau_\mathrm{cool}}
\newcommand{\bcool}{\beta_\mathrm{cool}}

\newcommand{\taus}{\tau_\mathrm{s}}
\newcommand{\Zc}{Z_\mathrm{c}}
\newcommand{\etaP}{\eta}

\usepackage{subfiles}
\graphicspath{{./figures/}{./Dust/figures/}{./HydroInstabilities/figures/}{./Introduction/figures/}{./MHD-wind/figures/}{./Observations/figures/}}

\input{pp7.h}

\begin{document}
\title{\textbf{\LARGE Hydro-, Magnetohydro-, and Dust-Gas Dynamics of Protoplanetary Disks }}

\author {\textbf{\large G. Lesur$^1$, B. Ercolano$^2$, M. Flock$^3$, M.-K. Lin$^{4,5}$, C.-C.~Yang$^6$,\\ J. A. Barranco$^7$, P. Benitez-Llambay$^8$, J. Goodman$^9$, A. Johansen$^{10,11}$, H. Klahr$^{3}$, G. Laibe$^{12,13}$, W. Lyra$^{14}$, P.  Marcus$^{15}$, R.P. Nelson$^{16}$, J. Squire$^{17}$, J. B. Simon$^{18}$, N. Turner$^{19}$, O.M. Umurhan$^{20,21}$, A.N. Youdin$^{22}$}}
\affil{\small\it $^1$ Univ. Grenoble Alpes, CNRS, IPAG, 38000 Grenoble, France \\
$^2$ Universit\"ats-Sternwarte, Fakult\"at f\"ur Physik,   Ludwig-Maximilians-Universit\"at M\"unchen, Scheinerstr.~1, 81679 M\"unchen, Germany\\
$^3$ Max-Planck-Institut f\"{u}r Astronomie, K\"{o}nigstuhl 17, 69117, Heidelberg, Germany\\
$^4$ Institute of Astronomy and Astrophysics, Academia Sinica, Taipei 10617, Taiwan\\
$^5$ Physics Division, National Center for Theoretical Sciences, Taipei 10617, Taiwan\\
$^6$ Department of Physics and Astronomy, University of Nevada, Las Vegas,
    4505 S.~Maryland Parkway, Box~454002,
    Las Vegas, NV~89154-4002, USA\\
$^7$ Dept. of Physics \& Astronomy, San Francisco State University, San Francisco, CA 94132, U.S.A.\\
$^8$ Niels Bohr International Academy, Niels Bohr Institute, Blegdamsvej 17, DK-2100 Copenhagen \O{}, Denmark \\
$^9$ Department of Astrophysical Sciences, Princeton University, Princeton NJ, U.S.A.\\
$^{10}$ Center for Star and Planet Formation, GLOBE Institute,
University of Copenhagen, \O ster Voldgade 5-7, 1350 Copenhagen, Denmark\\
$^{11}$ Lund Observatory, Department of Astronomy and Theoretical Physics, Lund University, Box 43, 221 00 Lund, Sweden \\
$^{12}$ Univ Lyon, Univ Lyon1, Ens de Lyon, CNRS, Centre de Recherche Astrophysique de Lyon UMR5574, F-69230, Saint-Genis,-Laval, France.\\
$^{13}$ Institut Universitaire de France\\
$^{14}$ Department of Astronomy, New Mexico State University, PO Box 30001, MSC 4500 Las Cruces, NM 88003, USA\\
$^{15}$ Dept. of Mechanical Engineering, University of California, Berkeley, CA, 94720, U.S.A.\\
$^{16}$ Department of Physics \& Astronomy, Queen Mary University of London, Mile End Road, London, E1 4NS, U.K.\\
$^{17}$ Physics Department, University of Otago, Dunedin 9010, New Zealand \\
$^{18}$ Department of Physics and Astronomy, Iowa State University, Ames, IA, 50010, USA \\
$^{19}$ Jet Propulsion Laboratory, California Institute of Technology, 4800 Oak Grove Drive, Pasadena, California 91109, U.S.A.\\
$^{20}$ SETI Institute, 339 Bernardo Ave., Suite 200, Mountain View, CA 94043, U.S.A.\\
$^{21}$ Space Sciences Division, NASA Ames Research Center, Moffett Field, CA 94035, U.S.A.\\
$^{22}$ University of Arizona, Departments of Astronomy and Planetary Sciences\\
}
\begin{abstract}
\baselineskip = 11pt
\leftskip = 1.5cm
\rightskip = 1.5cm
\parindent=1pc
{\small The building of planetary systems is controlled by the gas and dust dynamics of protoplanetary disks. While the gas  is simultaneously accreted onto the central star and dissipated away by winds, dust grains aggregate and collapse to form planetesimals and eventually planets. This dust and gas dynamics involves instabilities, turbulence and complex non-linear interactions which ultimately control the observational appearance and the secular evolution of these disks.

This chapter is dedicated to the most recent developments in our understanding of the dynamics of gaseous and dusty disks, covering hydrodynamic and magnetohydrodynamic turbulence, gas-dust instabilities, dust clumping and disk winds. We show how these physical processes have been tested from observations and highlight standing questions that should be addressed in the future.
 \\~\\~\\~}
\end{abstract}

\section{\textbf{INTRODUCTION}}
We focus in this chapter on the progress since PPVI in the fields of angular momentum transport, turbulence, dust growth, and winds from planet-forming disks. We begin by reviewing the basics of disk dynamics. The interested reader may consult the reviews and lecture notes provided in reference for more detailed discussion of these fundamental aspects.

\subsection{Framework and observational constraints}\label{framework_general}
The ongoing improvements in observational techniques, experiments, and simulations are opening a new window on our understanding of protoplanetary disks dynamics, as it becomes possible to test and validate scenarios that were, until recently, only theoretical concepts. In parallel, dynamical models are being refined, treating processes such as radiative heating and cooling, non-ideal magnetohydrodynamics (MHD), and outflows, to name a few. These efforts have led to significant revisions in our understanding of the mechanisms responsible for mass accretion. In addition, it has been realized that the dust/gas coupling is much richer than anticipated, leading to recognition of new multi-phase instabilities and dynamical phenomena that can enhance planet-formation rates. The goal of this chapter is to review these recent advances and to point out the open questions and some links between them.

In the following and unless stated otherwise, we consider an isolated protoplanetary disk, perturbed neither by a binary stellar companion nor by infalling material from a remnant envelope {   with mass $M_\mathrm{disk}$} orbiting a central young stellar object of mass $M_\mathrm{star}$ assumed to be near a solar mass. This restriction allows us to make general statements about the disk independent of environmental effects, which should be applicable to most class II disks. We note however that disks are not always isolated, as streamers are sometimes observed extending over thousands of astronomical units \cite[e.g.][]{pineda20}.
We assume the disk is low enough in mass to be non-self-gravitating, i.e. $M_\mathrm{disk}/M_\mathrm{star}<O(10^{-1})$ \citep{Armitage11}. More precise threshold disk masses can be found in, e.g. \citet{kratter16,haworth20}. The disk is accreting on the star, forming planets from dust grains, and ultimately dispersing on a timescale of a few million years. In this introduction, we define the main quantities and concepts used throughout the chapter relating to accretion, angular momentum transport (\ref{sec:intro:accretion}), and dust dynamics (\ref{sec:intro:dust}). In subsequent sections, we focus on instabilities of the gas phase leading to hydrodynamical turbulence (\S\ref{S:hydro}). We emphasize the conditions for existence of these instabilities which turn out to depend mainly on thermodynamics. We next move to the coupling with dust gains (\S\ref{section_dust}), describing instabilities and dust clumping arising because of dust-gas interactions. In \S\ref{mhd_section}, we discuss the roles of magnetic fields, magneto-rotational turbulence, and winds launched by gas pressure and magnetic forces. Finally, we link these theoretical processes to present and future observations in \S\ref{observations_section}. We conclude with a summary of the major achievements of the field since PPVI and highlight several key questions which should be addressed in the future.

\subsection{\textbf{Accretion, angular momentum transport, and turbulence}}\label{sec:intro:accretion}
Protoplanetary disks are accretion disks, delivering material onto their central stars at rates typically between $10^{-10}$ and $10^{-7}\,M_\odot/\mathrm{yr}$ \citep{Venuti14, Hartmann2016, Manara2016}. Their formation, evolution, and dispersal are closely linked to their angular momentum content and, since angular momentum is a conserved quantity, how the angular momentum is transported.

This question of angular momentum transport has been the subject of a great deal of work over the past 50 years  not only in protoplanetary disks, but also in the disks around stellar remnants and in active galactic nuclei. Essentially two transport mechanisms have been proposed: turbulence and magnetized disk winds (MDW). The former has attracted most attention because of the simplicity of treating it using the \cite{Shakura.Sunyaev73} pseudo-viscosity paradigm.  Winds, initially proposed by \cite{Blandford.Payne82}, have regained popularity over the last ten years as a way to transport angular momentum in otherwise "dead" discs.

The links between accretion flow, angular momentum transport, and disks' secular evolution can be seen in the equations of mass and angular momentum conservation, averaged through the disk thickness (e.g. \citealt{Lesur21}):
\begin{align}
\label{eq:acc_mass}
\frac{\partial \Sigma}{\partial t}+\frac{1}{2\pi R}\frac{\partial}{\partial R}\Macc&=-\zeta \Sigma \OmegaK ,\\
\nonumber
\frac{\Macc \OmegaK}{4\pi}&=\frac{1}{R}\frac{\partial }{\partial R}\left(R^2\ass P\right)\\
\label{eq:acc_ang}&\qquad+\zeta(\lambda-1) P \left(\frac{R}{H}\right)^2,
\end{align}
where we define the disk mass accretion rate $\Macc$, surface density $\Sigma$ and pressure $P$, Keplerian angular velocity $\OmegaK$ and geometrical thickness $H$. In these equations we include three unknown dimensionless coefficients: $\zeta$, the mass loss parameter due to a hypothetical outflow, $\ass$ which corresponds to \cite{Shakura.Sunyaev73} definition and describes radial angular momentum transport within the disk, such as by turbulence, and finally $\lambda$, the outflow lever arm \citep{Blandford.Payne82}, which quantifies the specific angular momentum extracted vertically by an outflow. All three of these coefficients may vary with radius and time.  Once the three coefficients are known, the secular evolution of the system can be entirely predicted and accretion theory is complete.

If the disk accretion is driven solely by turbulence of hydrodynamical or magnetohydrodynamical origin, then the accretion rate can be derived from (\ref{eq:acc_ang}) and is approximately \citep{Hartmann1998}
\begin{align}
 \nonumber   \Macc \sim \left(3\times 10^{-8}\,M_\odot/\mathrm{yr}\right)\left(\frac{\ass}{10^{-2}}\right) \left(\frac{\epsgas}{0.1}\right)^2 \\ \label{eq:mdot}\times\left(\frac{R}{10\,\au}\right)^{1/2} \left(\frac{M_\mathrm{star}}{M_\odot}\right)^{1/2}\left(\frac{\Sigma}{10\,\mathrm{g.cm}^{-2}}\right)
\end{align}
where we use the disk aspect ratio $\epsgas\equiv H/R$
and assume $\Sigma(R)$ follows a shallow power law (typically $\propto R^{-1/2}$). Hence typically $\ass\sim 10^{-2}$ is required to match the observed accretion rates by turbulence alone.

It is worth noting that gas turbulence need not always lead to outward angular momentum transport. One can obtain $\ass\sim 0$ from turbulence that is vigorous in the sense of large velocity fluctuations. This is particularly important in the context of dust transport and settling: some instabilities lead to efficient stirring of dust grains yet yield little angular momentum transport (e.g.\ \ref{vsi}).  Hence the $\ass$ parameter gives incomplete information on the presence and impact of turbulence.

\subsection{\textbf{Dust transport \& growth }}
\label{sec:intro:dust}

Dust grains have porous and fractal structures \citep{Dominik2007,Blum2008}. They are often modeled as spheres of radii $s \lesssim 10$ cm and densities up to $\rho \sim 1$ g.cm$^{-3}$ for compacted material \citep{Love1994}. Aggregates evolve as they undergo collisions. Sticking is favored for small ice-coated grains, while collisions between large aggregates involve redistribution of energy via elastic waves, which can trigger sliding or breaking at points of contact between substructures. Sticking, bouncing, compaction, abrasion or fragmentation, mass transfer, cratering, and erosion are all possible outcomes of collisions \citep{Blum2018}, as we discuss in \S~\ref{SS:barriers}.

 The drag stopping time  is $t_{\rm s} \sim (\rho_{\rm m} s) / (\rho_{\rm g} c_{\rm s})$, where $\rho_{\rm m}$ is the material density of the grain and $\rho_{\rm g}$ is the density of the surrounding gas (\citealt{Epstein1923}; ; \citealt{Baines1965,Clair1970}). Drag damps  eccentricity and makes them settle  and drift radially \citep{Adachi1976}. The competition between drag and the stellar gravity is  \citep{Safronov1969,Whipple1972,Whipple1973}
\begin{equation}
\mathrm{St} \equiv \tau_\mathrm{s} \equiv \OmegaK t_{\rm s}.
\label{eq:Stz}
\end{equation}
 for $\mathrm{St} \sim 1$. Solids  concentrate strongly as they  from the gas, making  for planet formation. In a vertically isothermal disk, Eq.~\eqref{eq:Stz}  $\mathrm{St}  = ((\rho_\mathrm{m} s) /  (\sqrt{2 \pi} \Sigma_{\mathrm{g}})) \exp(z^{2} / 2 H^{2})$,  that the  surface density $\Sigma_{\rm g}$ is key for setting values of $\mathrm{St} $ and thus, dust dynamics all through the disk. In the midplane of a disk  $\Sigma_{\rm g} \propto R^{-1}$ and $\Sigma_{\rm g} \sim 200$ g$.$cm$^{-2}$ at $R = 1$\,AU {   (which match the low end of the observed mass distribution, e.g.  \citealt{Andrews2009})} millimeter-sized grains have $\mathrm{St}_{0} \sim 0.04$ at $R \sim 50$\,AU. , most grains have $\mathrm{St} \gtrsim 1$. Since $\Sigma_g$ can not be inferred directly from observations, values of $\mathrm{St}$  major uncertainties .

In the vertical direction, and in the absence of turbulence, grains  damped harmonic oscillators of quality factors $\sim \mathrm{St}_{0} =  \mathrm{St}\left(z = 0 \right)$ driven by the mean flow of the gas at large scale. : the diffusivity $\adust$ (which is not necessarily equal to the angular momentum transport coefficient $\ass$; see Section~\ref{sec:obs:turbulence}) and the correlation time at eddy scales $t_{\rm e}$ \citep{YL07}. A classic choice is $\adust \sim \ass$ and  $t_{\rm e} \sim \OmegaK^{-1}$, the typical time for vortex stretching by differential rotation (see however \S~\ref{S:hydro} and \S~\ref{SSS:sisat}). In steady state, solids concentrate close to the midplane within a layer of height
\begin{equation} \label{E:h_ratio}
H_{\rm d} \simeq H \sqrt{\frac{\adust}{\adust + \mathrm{St}}}
\end{equation}%
\citep[see also][]{YMJ18}. As a result, larger grains .

In the radial direction,  between gas and dust.  for both phases. In an inviscid, non-magnetic, planet-free Keplerian disk, , given by
\begin{equation}
v_{\mathrm{d},R}=  \frac{ 1 }{\mathrm{St}+\mathrm{St}^{-1}\left(1+\epsdust\right)^2} \frac{1}{ \OmegaK\rho_{\mathrm g}}\frac{\partial P}{\partial R}  = - \epsdust^{-1} v_{\mathrm{g},R} ,
\label{eq:drift_vel}
\end{equation}
where $\epsdust = \rho_{\rm d} / \rho_{g}$ is the local dust-to-gas density ratio \citep{Nakagawa1986}.
 drifts towards pressure maxima -- either the inner  or local pressure bumps -- with optimal efficiency when $\mathrm{St}\sim 1$.
 grains travel through the disk in a typical time $t_{\rm drift} \sim \OmegaK^{-1} (H / R)^{-2} (1+\mathrm{St}^{2}) / \mathrm{St}$. %

\section{THERMO-HYDRODYNAMICAL INSTABILITIES}
\label{S:hydro}

\begin{center}
    \begin{table*}
    \caption{Candidates for hydrodynamic activity in PPDs: the Vertical Shear Instability (VSI), the Convective Overstability (COS), and the Zombie Vortex Instability (ZVI).
 Figures adapted from \cite{pfeil20}, \cite{lyra14}, and \cite{barranco2018}, respectively.
 The vertical component of the vorticity is shown for each, while the Reynolds stress is also shown for the VSI. Each instability requires a different gas cooling timescale $\tcool$, here scaled by the Keplerian orbital frequency $\OmK$. They also require different disk structures, here expressed for a vertically isothermal disk with a radial temperature profile $\propto R^{-q}$, midplane density profile $\propto R^{-p}$,
 adiabatic index $\gamma$, the height above the midplane $z$, and the gas pressure scale height $\Hgas$. $R$ is the cylindrical distance from the star. A Shakura-Sunyaev viscosity, $\alpha_\mathrm{SS}$, can be measured from simulations as a metric for radial angular momentum transport, but does not fully characterize the turbulence generated by these instabilities.
 }\label{hydro_summary}
    \begin{threeparttable}
     \begin{tabular}{|c c c|}
    \hline
  Vertical Shear Instability & Convective Overstability & Zombie Vortex Instability \\
 \hline\hline
   \includegraphics[width=0.31\linewidth,clip=true,trim= 0cm 0cm 0cm 0cm]{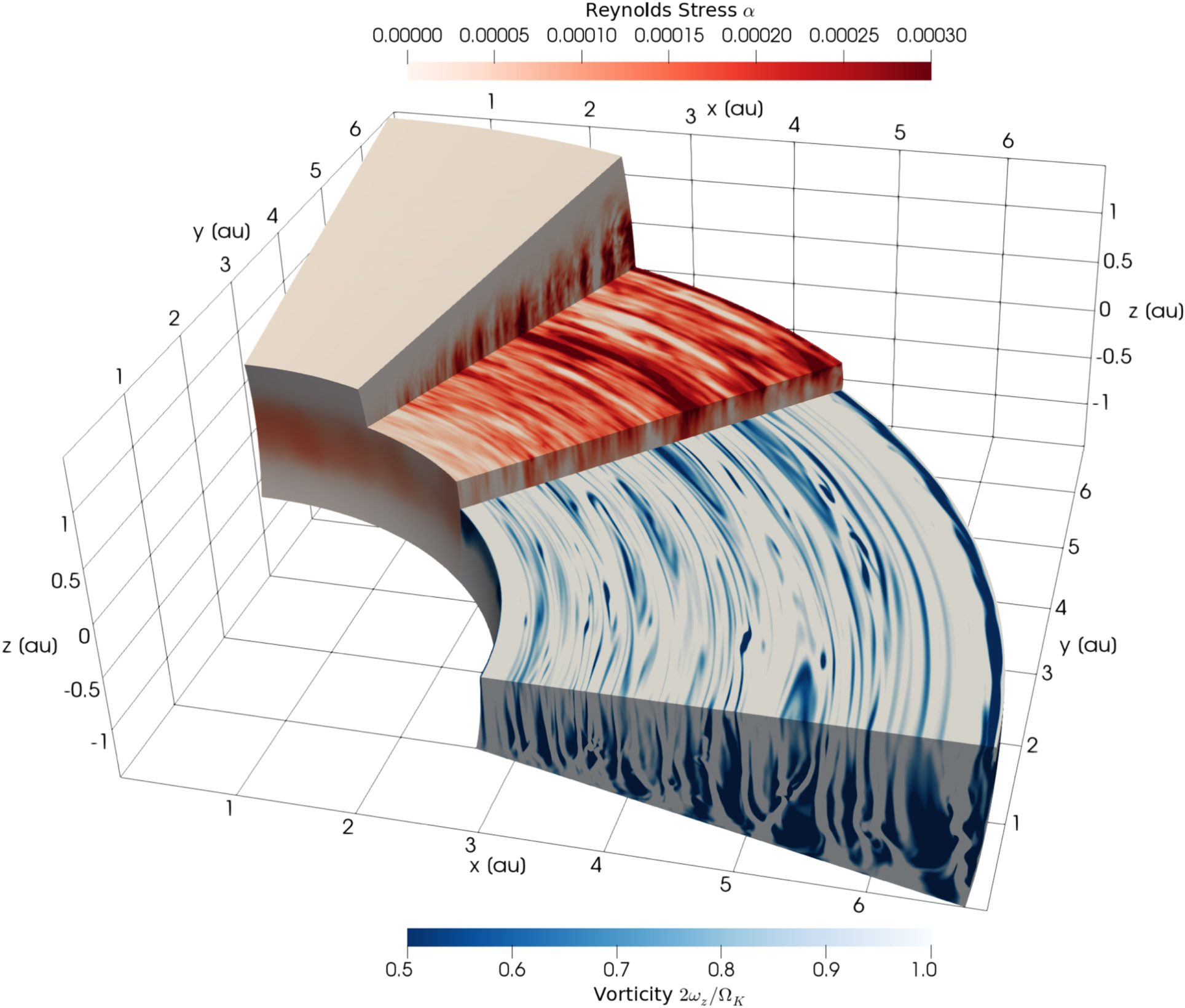}
   &
   \includegraphics[width=0.31\linewidth,clip=true,trim= 0cm 0cm 0cm 0cm]{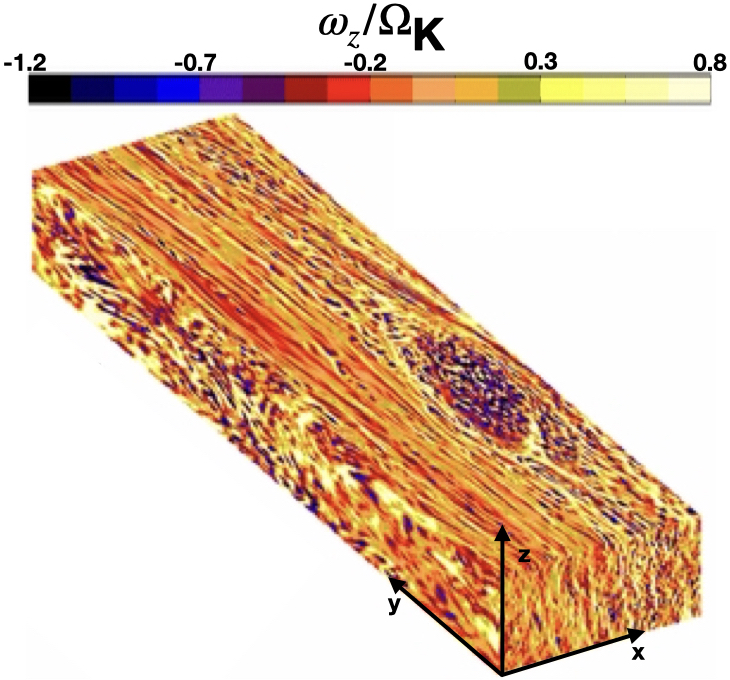}
   &
   \includegraphics[width=0.31\linewidth]{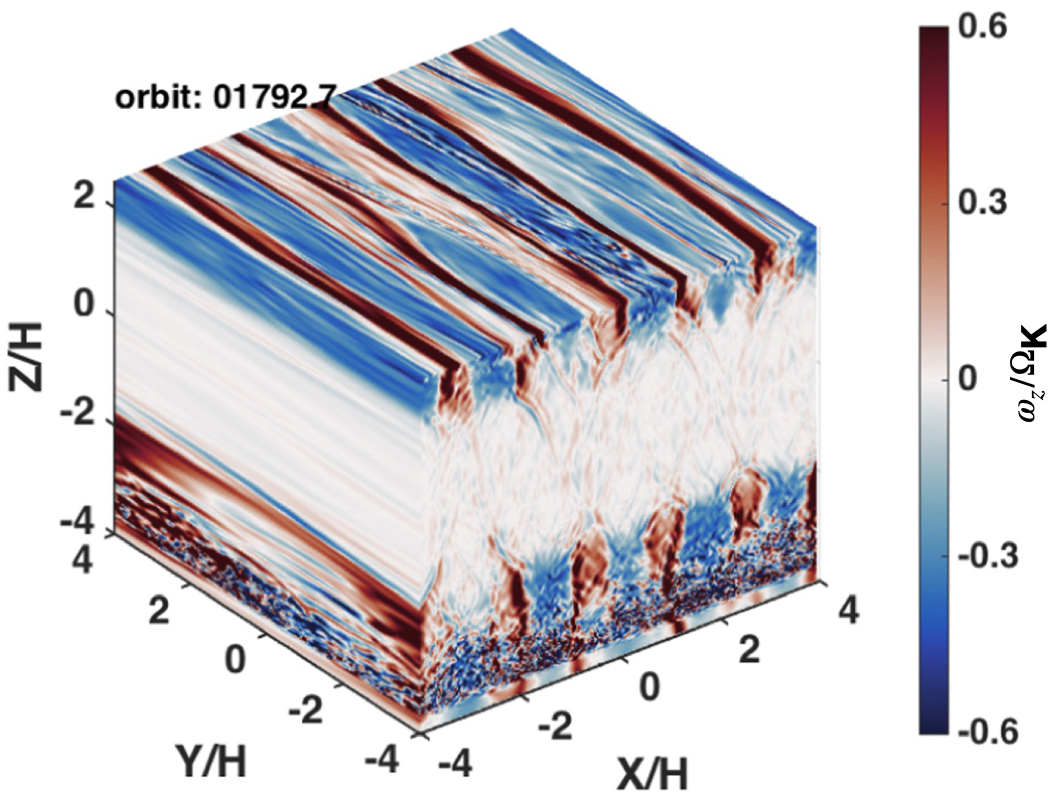} \\
   \hline
 $ \tcool\OmK \ll 1$ & $ \tcool\OmK \sim 1$ & $ \tcool\OmK \gg 1$ \\
\hline
$q\neq 0$ &
$-1 < p/q < 1/(\gamma-1)$ &
$|z| \gtrsim \sqrt{\gamma/(\gamma-1)}\Hgas $ \\
 \hline
  $\alpha_\mathrm{SS}\sim 10^{-4}$& $\alpha_\mathrm{SS}\sim 10^{-3}$ & $\alpha_\mathrm{SS}\sim  (10^{-5}$--$10^{-4})$\tnote{$\dagger$}\\
 \hline
 \hline
\multicolumn{3}{|c|}{Outcome: turbulence \& vortices}\\
 \hline
\end{tabular}
\begin{tablenotes}
\item[$\dagger$] Based on incompressible simulations \citep{barranco05}.
\end{tablenotes}
\end{threeparttable}
\end{table*}
\end{center}

PPDs are poorly ionized and weakly coupled to magnetic fields (see \S\ref{mhd_section} for details). The lack of magnetically-driven turbulence in some regions has led to a surge in the interest  in purely hydrodynamic origins of turbulence. To this end, three new hydrodynamic instabilities have been discovered:
\begin{inparaenum}[1)]
\item the Vertical Shear Instability (VSI);
\item the Convective Overstability (COS); and
\item the Zombie Vortex Instability (ZVI).
\end{inparaenum}
Key features of these instabilities are summarized in Table \ref{hydro_summary}. Which of these operate depends on the disk's structure and the gas thermal timescales. However, taken together they likely pertain to large swaths of PPDs (see \S\ref{hydro_application}) and are expected to produce weak to moderate turbulence, transport, and large-scale structures such as vortices, which are all relevant to interpreting disk observations as well as planetesimal formation  (\S\ref{section_dust}).

To anchor the following discussion, we consider a three-dimensional (3D) PPD orbiting a central star of mass $M_*$. Cylindrical $(R,\phi,z)$ coordinates are centered on the star with the disk plane located at $z=0$. The equilibrium disk is steady and axisymmetric with a purely azimuthal flow $\vgas=R\Omega(R,z)\hat{\bm{\phi}}$, where $\Omega$ is the gas' angular velocity, which is set by centrifugal balance and is close to Keplerian rotation but not exactly.  Note that we do not consider warped or eccentric disks.
The gas density $\rhog$ is set by vertical hydrostatic balance and we take a power-law midplane density profile $\rho_\mathrm{g0}(R) \propto R^{-p}$.
For an ideal gas its temperature $T$ is defined through $P = \mathcal{R}\rho T/\mu$, where $\mathcal{R}$ is the gas constant and $\mu$ is the mean molecular weight. The gas' specific entropy is $S\equiv \cp\ln\left(P^{1/\gamma}/\rhog\right)$, where $\cp$ is the specific heat capacity at constant pressure and $\gamma$ is the constant adiabatic index. %

We consider vertically isothermal disks with {  power-law} equilibrium {  temperature profiles} $T\propto R^{-q}$, which are often used to model the {  disk bulk or interior in the outer regions} of irradiated PPDs \citep{hubeny90, chiang97}. The isothermal sound-speed is $c_s\equiv \sqrt{\mathcal{R}T/\mu}$ and $\Hgas = c_s / \OmK$. Then the gas surface  density $\Sigmag \propto R^{-\sigma}$ with $\sigma = p + q/2 -3/2$. PPDs are thin with aspect-ratios $\hgas\equiv \Hgas/R\ll 1$. {  Finally, we assume gravitationally stable disks with Toomre parameters $Q_g= c_s\Omega/\pi G \Sigma_g \gg 1$, see \citet{kratter16} for a discussion of gravitational instabilities and the Chapters by Bae et al. and Pinte et al. for their potential to explain spiral structures observed in PPDs.}

Prior to {  PPVI}, PPDs were mostly considered hydrodynamically inactive, at least to small-scale, local, infinitesimal perturbations. %
This notion stems from the classic Solberg-H{\o}iland (SH) stability criteria \citep[e.g.][]{tassoul78}, which can be written as $\kappa^2 + N_R^2 + N_z^2 > 0$ and $-\p_zP\left(\kappa^2\p_zS - R\p_z\Omega^2\p_RS\right)> 0 $. Here, $\kappa ^2 = R^{-3}\p_R\left(R^4\Omega^2\right)$
is the square of the radial epicycle frequency; and  $N_{R,z}^2=-\left(c_P\rhog\right)^{-1}\p_{R,z}P\p_{R,z}S$ are the squares of the radial and vertical
contributions to the buoyancy {  (or Brunt-V\"{a}is\"{a}l\"{a})} frequency, respectively. For radially smooth, vertically thin disks, the SH stability criteria are generally satisfied \citep{lin15} because usually $N_z^2\geq 0$ (so the disk is stable against vertical convection as well) and $|N_R^2|\ll \kappa^2$, while the second criterion reduces to $\gamma > 1 + O(\hgas^2)$, which is satisfied for typical values of  $\gamma = 5/3$ or $7/5$.

However, the SH stability criteria assume the gas is adiabatic, i.e. it does not experience heat gains or losses, and disturbances are axisymmetric and infinitesimal. The key to the new class of thermo-hydrodynamic instabilities in Table~\ref{hydro_summary} is the violation of these assumptions (see discussion in \citealt{Fromang2019}). Both the VSI and COS require finite thermal losses. The ZVI applies to adiabatic gas, but requires finite amplitude, non-axisymmetric perturbations. These conditions can be expected in PPDs and thus the SH criteria are not applicable. Furthermore, parts of PPDs may have a non-standard radial structure such that $N_R^2<0$, which is necessary for the COS.

The revived interest in hydrodynamic instabilities in PPDs thus resulted from more careful treatments of the disk's thermal structure and evolution. To this end, we write the gas energy equation as
$\p_t S + \vgas\cdot\nabla S = -\Lambda.$
 Here, $\Lambda$ represents all heating and cooling processes, but for simplicity we will refer to it as `cooling'. Physically, cooling is mediated by radiation and its form depends on the gas and dust properties \citep[e.g.][see also \S\ref{hydro_application}]{malygin17,barranco2018,pfeil19}. However, for discussion purposes it is sufficient to consider the Newtonian cooling prescription,
 \begin{align}\label{beta_cooling}
     \Lambda = \frac{c_P}{\gamma T} \frac{\left(T-\Tref\right)}{\tcool},
 \end{align}
 where the reference temperature $\Tref$ is usually taken to be the initial temperature field; and $\tcool$ is the {\it cooling timescale} over which the local temperature relaxes back to $\Tref$. We define the dimensionless cooling time $\beta_\mathrm{cool}\equiv \tcool\OmK$.

The beta cooling prescription encapsulates all possible gas thermodynamic responses. This enables the three hydrodynamic instabilities to be considered by varying $\bcool$. The limit $\bcool\to\infty$ corresponds to adiabatic evolution, where the specific entropy is materially conserved, which enables the ZVI provided $|N_z|$ is sufficiently large. In the isothermal limit, $\bcool \to 0$, or instant cooling, $T\to\Tref$ and the gas temperature does not evolve, which enables the VSI provided there is a radial temperature gradient to produce vertical shear (see below). In the intermediate case, $\bcool \sim 1$, the disk cools on the orbital  timescale, which enables the COS if $N_R^2<0$ also.

\label{sec:hydro:turbulence_parameters}
The onset of these hydrodynamic instabilities thus has both structural and thermodynamical requirements (Table~\ref{hydro_summary}). If these are met, they can lead to hydrodynamic turbulence and vortex formation. In simulations, it is straightforward to measure the conventional Shakura-Sunyaev $\ass$ to describe radial angular momentum transport mediated by said turbulence. However, it is important to note that $\ass$ alone is insufficient to fully characterize the ensuing turbulence. {  The fundamental measure of turbulence is the} root-mean-squared velocities or Mach numbers $\mathcal{M}$ at various scales. For planetesimal formation, one is often interested in the {  resulting} turbulent diffusion of particles, $\adust$ (see \S \ref{sec:intro:dust} and \S\ref{section_dust}). One often assumes $\mathcal{M}^2$, $\ass$, and $\adust$ are equal and isotropic, but this is not necessarily the case, especially for the anisotropic and inhomogeneous turbulence associated with the three instabilities.
Observationally, this means that the turbulence parameters associated with mass accretion, dust distributions, and line broadening can all be different, even when they have the same physical origin.

\subsection{Vertical shear instability }\label{vsi}
The VSI \citep{arlt04,nelson13} is a linear, axisymmetric instability that is a disk-variation of the Goldreich-Schubert-Fricke instability in  differentially rotating stars \citep[GSFI,][]{goldreich67,fricke68}. The VSI  can be considered as the $\mathrm{Pr}\rightarrow 0$ limit of the GSFI
\citep{Knobloch_Spruit_1982}, where the Prandtl number $\mathrm{Pr}$ is the ratio of viscosity to thermal diffusivity.

The VSI is powered by the disk's {\it vertical} differential rotation, i.e. $\p_z\Omega\neq 0$, which is exhibited by any baroclinic equilibrium wherein surfaces of constant density and pressure are misaligned ($\nabla \rhog\times\nabla P \neq \bm{0}$). This includes vertically isothermal disks with a radial temperature gradient ($q\neq 0$), for which one finds
 \begin{align}
     R\frac{\p\Omega^2}{\p z} \simeq q\hgas \frac{z}{\Hgas} \OmK^2.
 \end{align}
Although vertical shear is weak ($\left|R\p_z\Omega\right|\sim \hgas \OmK\ll \OmK$), the associated free energy can be released through radially-short, vertically-elongated disturbances \citep{umurhan13} provided that the stabilizing effect of vertical buoyancy is either absent ($N_z^2=0$) or eliminated by rapid cooling when $N_z^2>0$ \citep{nelson13}, as is also the case
for the GSFI when Pr$\rightarrow 0$ \citep{Knobloch_Spruit_1982}. In the case  $N_z^2>0$, \cite{lin15} find the criterion
\begin{align}
    \tcool \lesssim \frac{|q|\hgas}{\gamma-1}\OmK^{-1},
\end{align}
or $\bcool \lesssim \hgas\ll 1$ for typical disk parameters. That is, cooling timescales must be much shorter than the dynamical timescale.

 Linear VSI theory was initially developed in the local approximation for axisymmetric disturbances \citep{urpin98,urpin03}. More recently,   non-axisymmetric \citep{volponi14,latter18} and radially-local, vertically global models have been developed \citep{nelson13,barker15,mcnally15,lin15,umurhan16c}. The linear instability mechanism can be explained in terms of angular momentum, vorticity, and energetic considerations \citep{yellin21}. Weakly non-linear theories have been developed by \cite{latter18} and \cite{shtemler19,shtemler20}.

 Semi-global linear analyses find:
\begin{inparaenum}[1)]
\item destabilized inertial waves or `body modes' that permeate the disk column, and
\item `surface modes' concentrated near the vertical boundaries. %
\end{inparaenum}
Body modes tend to dominate nonlinear simulations, which are typically characterized by narrow bands of large-scale vertical motions \citep{nelson13}, similar to that exhibited by low-order body modes with widths $\sim \hgas \Hgas$.
Interestingly, local, linear VSI modes are also nonlinear solutions, which partly explains why they can persist with large amplitudes \citep{latter18}.

Nevertheless, linear VSI modes are subject to secondary, Kelvin-Helmholtz-type instabilities, which are often associated with vortex formation \citep{nelson13,latter18}. Indeed, vortices -- non-axisymmetric flow features -- are readily observed in full 3D simulations \citep{richard16,manger18}, although there it has been attributed to the Rossby Wave Instability \citep[RWI][]{Lovelace1999,li00,li01} of narrow vorticity rings directly formed by the VSI or pressure bumps that arise from radial variations in the rate of angular momentum transport by the VSI. These vortices may be subsequently amplified by the Sub-critical Baroclinic Instability \citep[][see also \S\ref{cov}]{klahr03,peterson07a,peterson07b,lesur10} for steep radial temperature profiles and moderate cooling timescales;
or attacked by 3D Elliptic Instabilities \citep[][]{kerswell04,lesur09b}, especially if their aspect-ratios $\chi$ are small ($\lesssim 4$).
Recent simulations with beta and radiative cooling find both compact, short-lived (a few orbits) vortices and large-scale, elongated ($\chi\gtrsim 8$) vortices that survive for hundreds of orbits \citep{manger20,flock20}. The latter can arise from pressure bumps at the boundary of `VSI dead zones' \citep{flock20}, which develops from the strong dependence of the VSI on the cooling time that leads to a situation where the VSI does not operate at small radii ($\lesssim 10$~au) while being active at larger radii. High resolution simulations also find meridional vortices \citep{flores20}.

VSI-active disks are turbulent with moderate outwards angular momentum transport: the radial stress-to-pressure ratio, or an alpha parameter, $\ass$ of order $ 10^{-4}$ is found for $\hgas=0.05$ and is an increasing function of $\hgas$ \citep{manger20}. However, VSI-turbulence is {\it strongly anisotropic} because typically $|v_z/v_R|\gg 1$. This results in a
{\it vertical} stress-to-pressure ratio of $\alpha_z \equiv \langle \rhog v_z\delta v_\phi \rangle/P \sim 10^{-2}$ (here $\delta$ denotes the deviation from equilibrium), i.e. two orders of magnitude larger than radial transport \citep{stoll17}.
Furthermore, both $\ass$ and $\alpha_z$ are non-uniform. These are important distinctions from standard, constant-alpha disks as the anisotropy of the VSI implies that the gas' and dust's radial and vertical evolution can differ significantly.

The basic properties of the VSI can be captured by locally isothermal disks. A first approximation to modeling the VSI in more realistic PPDs is to use the beta cooling prescription with an estimate of physical cooling timescales (see \S\ref{hydro_application}) in linear theory \citep{lin15,fukuhara21} or numerical simulations \citep{pfeil20}. Beyond this, one can explicitly account for radiative losses and stellar irradiation by solving the radiative hydrodynamic equations \citep{stoll14,flock17,flock20}. Such models show that the VSI can indeed be expected in the outer regions of PPDs with an $\ass\sim 10^{-4}$ and flow structures largely consistent with locally isothermal models. Turbulent velocities on the order of $10^{-2}{c_s}$ are found in the midplane regions{ ;} whereas in the coronal regions {  ($|z|\gtrsim 0.2R$, or roughly beyond two pressure scale heights)} these increase to $\sim 0.06c_s$ and can {  reach $O(10^{-1}c_s)$} at some points \citep{flock17}.  These are consistent with the upper limits obtained from ALMA observations of turbulent line widths  (\S\ref{sec:obs:turbulence}). However, identifying VSI turbulence this way appears difficult owing to the VSI's strong anisotropy; instead, its vertical motions can be detected by spatially resolved observations of the CO line emission \citep{barraza21}.

Recent studies have also begun to incorporate the VSI in models of planetesimal formation and planet evolution. Axisymmetric vorticity perturbations induced by the VSI, and discrete vortices in the disk plane, have been shown to act as effective traps of mm dust \citep{stoll16,flock17,flock20}. The former gives rise to prominent ring-like structures, while the latter produces localized enhancements of dust. Large-scale vertical gas motions induced by the VSI can effectively lift up passive dust grains to the disk atmosphere  \citep{stoll16,flock17,flock20}. When dust-on-gas feedback is accounted for, solids can settle against VSI-stirring if the overall dust abundance is sufficiently super-solar \citep{lin19}. The dense midplane dust layer can further undergo the Streaming Instability \citep[][see also \S\ref{section_dust}]{Youdin2005}, in which case the VSI can enhance radial dust concentrations \citep{schafer20}. Following planet formation, disk-planet interaction and pebble accretion with VSI turbulence have been investigated by \cite{stoll17b} and \cite{picogna18}, respectively. For this particular problem they find results  similar to viscous disk models with an equivalent $\ass$ as measured in VSI runs, but in the case of pebble accretion one also needs to include stochastic kicks onto the solids.

Finally, while the VSI is fundamentally a hydrodynamic instability, PPDs are magnetized. In ideal MHD, a strong field with a plasma beta parameter $\gtrsim 1/\hgas$ can stabilize the VSI \citep{latter18,cui21}, which may apply to the well-ionized disk atmospheres. On the other hand, in the disk bulk where non-ideal effects dominate, the VSI persists and can co-exist with magnetized disk winds \citep{cui20}. The dynamics of dust grains under such conditions remains to be investigated.

\subsection{Convective overstability and baroclinic vortex amplification}\label{cov}

The COS is a linear, axisymmetric instability that can be interpreted as radial convection mediated by cooling \citep{klahr14, lyra14}. Unstratified models show that the COS requires
\begin{inparaenum}[1)]
\item a radial buoyancy such that $N_R^2<0$, and
\item cooling on dynamical timescales, $\bcool\sim 1$. %
\end{inparaenum}
For power-law disks the first criterion becomes
\begin{align}
    -1 < \frac{p}{q} < \frac{1}{\gamma-1} \quad \text{for $N_R^2 < 0$}.\label{cov_nr_cond}
\end{align}
This analytical criterion applies to the disk midplane where $N_z^2=0$. For $q>0$, Eq. \ref{cov_nr_cond} translates to
$-(q+3)/2 < \sigma < [q(\gamma+1)/(\gamma-1)-3]/2$,
implying a shallow or even rising surface density profile.
For example, in irradiated disks with $q=0.5$, one requires $-1.75 <\sigma < 0$ for $\gamma = 7/5$. Such atypical profiles may be realized at special locations such as gap edges {  or the outer edge of dead zones \citep[e.g.][]{delage22}.}

However, even if $N_R^2 > 0 $ at $z=0$, in typical disk models the fact that $\hgas$ increases with $R$
can drive $N_R^2 < 0$ away from the midplane, where $N_z^2>0$. Unlike the VSI, though, vertical buoyancy is not expected to suppress COS modes, which have little vertical motion  \citep{lyra14}. A similar situation applies to the symmetric instability in the Earth's atmosphere \citep{schultz99}.
Nevertheless, the existence and saturation of the COS in vertically stratified disks off the midplane still need to be demonstrated explicitly, which is not trivial as potentially also VSI can be operative in this region.

The COS corresponds to growing epicyclic oscillations. In a thin disk, rotation is strongly stabilizing. This means that, even if a gas parcel is radially buoyant ($N_R^2<0$), it will still perform radial oscillations on the epicyclic frequency. Indeed, in the absence of cooling the SH conditions imply stability as $\kappa ^2\gg |N_R^2|$ since typically $|N_R|\sim  \hgas\OmK$.
However, if cooling occurs on the timescale on the order of the orbital period, the gas parcel can exchange heat with the surrounding such that it experiences a buoyant acceleration, thereby increasing its oscillation amplitude
\citep{latter16}. On the other hand, if cooling is too rapid then the parcel experiences no buoyancy and epicycles are again stable. Ultimately, the COS is mainly powered by the unstable entropy gradient (cf. vertical shear for the VSI). Also unlike the VSI, the COS prefers vertically short, radially extended length scales.

The linear analysis for the COS was possible as there are, just like for the VSI, axisymmetric modes which are unstable  \citep{klahr14,lyra14,latter16,volponi16}. Weakly nonlinear theory \citep{latter16} suggested that COS modes might be limited to small amplitudes due to secondary parasitic instabilities. {  However,} recent {  axisymmetric} numerical simulations \citep{TeedLatter21} showed a progression in the nonlinear regime with increased Reynolds number as the system transitions from a weakly nonlinear state with relatively ordered waves, to wave turbulence, and finally to the formation of intermittent and then persistent zonal flows. {  This process is} likely what {  spawned} vortices in 3D, non-axisymmetric models of the COS \citep{lyra14}. The axisymmetric simulations of \cite{TeedLatter21} also showed the existence of "elevator modes", in which a mode of uniform vertical velocity dominates the vertical extent of the (unstratified) box. Whether or not this leads to a large-scale meridional circulation in a stratified, global disk needs to be clarified.

Yet, well before the discovery of the axisymmetric, linear COS, the amplification of vortices had already been observed in simulations of radially stratified disks with cooling  \citep{klahr03,peterson07a,peterson07b, lesur10,lyraklahr2011}. These authors find that such amplification of the vortices also requires $N_R^2<0$ and cooling that is neither too fast nor slow. \citeauthor{lesur10} described this effect as a Sub-critical Baroclinic Instability (SBI), where `sub-critical' refers to the fact that a vortex is not an infinitesimal perturbation in the flow, but needs a finite initial size to have the amplification be stronger than viscosity and shear. That is, SBI is a non-linear phenomenon. In addition, the SBI operates in 2D, thin disks, unlike the linear, axisymmetric COS that requires the vertical dimension.

Ultimately, however, the criteria for COS and SBI are qualitatively the same, suggesting that both instabilities share the same physical driving, i.e. radial buoyancy and cooling rates on their individual radial oscillation frequencies. Whereas for the linear COS, this frequency is the epicyclic frequency ($\kappa$); vortices typically rotate much slower than the disk rotation, with a turnover frequency of order $\OmK/\chi$ (recall $\chi$ is the vortex aspect ratio). Thus the larger the vortex aspect ratio, %
the longer the rotation time, which gives a wide range of cooling times that will lead to the amplification of a range of vortex shapes \citep{raettig13}.

Global numerical simulations of the SBI in 2D and 3D disks readily show the development of large-scale vortices that can persist for hundreds of orbits \citep{klahr03,peterson07a,peterson07b,lyraklahr2011,Barge2016}. Simulations also find that COS/SBI leads to outwards angular momentum transport at a typical level of $\alpha_\mathrm{SS}\sim 10^{-3}$, but can vary by an order of magnitude depending on buoyancy and cooling parameters \citep{raettig13}. Local 2D models suggest that vortices can be amplified even when radial buoyancy is only weakly unstable \citep[][]{raettig13}. High resolution, local 3D simulations
show that SBI growth is balanced by secondary, 3D Elliptic Instabilities \citep{lesur09b}, resulting in a vortex with a turbulent core \citep{lesur10}. %

Nevertheless, the ability for COS-induced or SBI-amplified vortices to act as efficient dust traps have been demonstrated by \cite{raettig15} in 2D, and
\cite{lyra18} and \cite{raettig21} in 3D.
On the other hand, \cite{gomes15} find the interaction between massive planets and radially buoyantly unstable, cooling disks renders $N_R^2>0$, in which case planet-induced gaps become smoother and leads to weaker vortices that are associated with the RWI of the gap edge, rather than COS or SBI. However, models of planetesimal formation and disk-planet interaction that explicitly incorporate COS/SBI-induced turbulence and vortices are still lacking (cf. as being developed for the VSI), which should be pursued in the future, especially given the relevance of vortices for planet formation, and the predicted occurrence of the COS in the 1-10 AU region. %

\subsection{Zombie vortex instability }\label{zvi}

Unlike the VSI and the COV, the ZVI is a non-axisymmetric, non-linear instability, the latter meaning that it requires finite-amplitude perturbations. The ZVI can develop if
there is sufficient vertical buoyancy. In practice, this translates to \begin{inparaenum}[1)]
\item $N_z^2\gtrsim \OmK^2$, which is generally the case for $|z|\gtrsim 1.5\Hgas$; and \item $\bcool
\gg 1$, otherwise buoyancy is diminished by cooling. \end{inparaenum}

\citet{barranco05} first observed coherent, long-lived vortices naturally form in the strongly stratified regions above and below the PPD midplane in their local, shearing box simulations with a spectral method.  At first, they hypothesized that the vortices were created by breaking internal gravity waves, but \citet{MPJH13}, \citet{MPJBHL15}, and \citet{MPJB16} demonstrated the existence of a new type of purely hydrodynamic instability associated with ``baroclinic critical layers": narrow structures in a stratified shear flow where a wave's phase speed matches the shear velocity plus/minus the {  vertical buoyancy}
frequency divided by the wavenumber.  They named the instability the ``Zombie Vortex Instability'' (ZVI) not only because it may occur in dead zones, but also because of the way one zombie vortex ``infects'' neighboring baroclinic critical layers, spawning new zombie vortices, which ``infect'' farther critical layers, and so on, filling a region of the disk with vigorous turbulence.  ZVI is not an artifact of the numerical method as it was computed with spectral codes and finite-volume codes, with Boussinesq, anelastic, and fully compressible treatments of the continuity condition, and with and without the shearing box.  One may ask, if it is so robust, how was it missed in previous numerical calculations?  Prior work often lacked one or more of the crucial ingredients: ZVI requires strong vertical stratification, high resolution to resolve the narrow critical layers {  ($\gtrsim 256$ cells or spectral modes per $H$)}, a broad spectrum of perturbations (i.e., Kolmogorov, but not Gaussian-peaked, so that the vorticity peaks on the small scales), and enough simulation time to allow the critical layers to amplify perturbations.

\citet{lesurlatter2016} confirmed the existence of the instability with their own spectral simulations, but also pointed out the sensitivity of the ZVI to dissipation processes and small scale physics: fast cooling and/or sufficient viscosity suppresses the ZVI in numerical simulations (with Reynolds number $\mathrm{Re}\lesssim 10^6$). However, the relevant Reynolds number in PPDs is $\mathrm{Re}\sim 10^{14}$, so viscosity may not inhibit the ZVI in PPDs, but does make it difficult to produce ZVI in a laboratory \citep{wangmarcus2016}. \citet{barranco2018} investigated ZVI under a wider range of realistic scenarios, including with nonuniform vertical stratification and radiative cooling, and demonstrated that the off-midplane regions throughout the planet-forming regions of protoplanetary disks may be susceptible to ZVI so long as the cooling time is longer than a few orbital periods, which may be likely with modest grain settling and/or growth.  \citet{umurhanshariff2016} and \citet{wangbalmforth2020} have undertaken theoretical investigations into the earliest phases of the instability considering both linear growth and the nonlinear dynamics within the baroclinic critical layers.  Most recently, \citet{Wang_Balmforth_2021} developed a reduced model for the nonlinear dynamics within forced baroclinic critical layers and have demonstrated that perturbations grow secularly, generating jet-like defects within the shear which are then later susceptible to secondary instabilities that yield coherent vortical structures.

Turbulence from ZVI has some unique characteristics that may be highly relevant to dust transport.  At late times, ZVI turbulence results in significant vertical mixing that tends to homogenize the background stratification resulting in a nearly uniform {  $N_z$}.  Similar behavior is seen in atmospheric and oceanic flows in which the breaking of internal gravity waves creates step-like or staircase patterns in stratification \citep{orlanski1969,phillips1972,pelegri1998}. While the region in the immediate vicinity of the midplane lacks the requisite stratification for the excitation of baroclinic critical layers, \citet{barranco2018} observed that zombie turbulence from the ZVI susceptible regions can penetrate into the midplane, albeit with a smaller magnitude.  At late times, ZVI turbulence resulted in the creation of azimuthal quasi-steady-state zonal flows.  The zonal flows consisted of 5-6  pairs of dipolar vortex layers within a radial extent of $8H$ (see the third figure in Table \ref{hydro_summary}).  The radial thickness of the cyclonic layers is approximately one-third the thickness of the anticyclonic layers, but this is compensated by the fact that the cyclonic layers are roughly three times more intense than the anticyclonic layers.  Fully-developed zombie turbulence shows intermittency where the flow cycles through near-laminar phases of zonal flow punctuated by chaotic bursts of new zombie vortices.  In some simulations, the bursting is quasi-periodic in time (with periods between 100-150 orbits), whereas in other cases, it appeared more stochastically.

The ultimate impact of ZVI on PPD dynamics is still to be demonstrated.  Future topics include: (1) how efficient is ZVI at transporting angular momentum in compressible simulations ($\ass$);
(2) the properties of ZVI turbulence as a function of the Reynolds number, especially beyond $O(10^7)$;
(3) how does ZVI turbulence transport and mix dust and how does dust affect the ZVI, as well its mutual interaction with planetesimals and planets; (4) what does the ZVI look like in global simulations; and (5) what are the observational signatures of ZVI turbulence, e.g., broadening of molecular lines, and can that be distinguished from turbulence from other purely hydrodynamic instabilities.

\subsection{Occurrence in physical disk models }\label{hydro_application}

As Table \ref{hydro_summary} summarizes, the onset of the above hydrodynamic instabilities depends on both the disk structure and thermal relaxation or cooling times ($\tau_{{\rm cool}}$), which are intimately interdependent on one another. As discussed in \S\ref{vsi} -- \ref{zvi}, the structural requirements for the VSI, $\p_z\Omega\neq 0$, can be met by a non-zero radial temperature gradient ($q\neq 0$), that for the COV where the radial buoyancy $N_R^2 < 0$, and that for the ZVI where the vertical buoyancy $|N_z| > \OmK$.
As for their thermodynamic requirements, with a broad brush stroke, we can say that VSI requires cooling times significantly shorter than an orbital period, COS needs cooling times of order the orbital period, whereas ZVI is operable when the cooling time is longer than a few orbital periods. {  In this section, we describe the ingredients for estimating cooling times and discuss recent efforts to do so in order to assess the occurrence of the above instabilities. We will see that at the current stage, results are still model-dependent as they are subject to a number of uncertainties.}

Establishing the relevant cooling times for the mechanism in question depends upon the spatial lengthscale of
the fastest growing disturbance $\ell_m$
and how $\ell_m$ compares to the photon mean-free path $\ell_{{\rm ph}} = 1\big/(\kappa_\mathrm{op}\rho)$, where $\kappa_\mathrm{op}$ is
the material's frequency integrated opacity. This places the dynamics in either the optically thin ($\ell_{{\rm m}} \ll \ell_{{\rm ph}}$) or thick
($\ell_{{\rm m}} \gg \ell_{{\rm ph}}$)
regime \citep[e.g., see the discussion in ][]{barranco2018}.
Since H$_2$ and other molecules are inefficient
radiators compared to dust particles \citep[however molecular line cooling may play a role in the disk's coldest regions, e.g.,][]{Woitke_etal_2016},
the primary cooling pathway %
involves the collisional transfer
of thermal energy from gas to particles, the latter of which efficiently
radiates
\citep[e.g., see discussions in][]{barranco2018,pfeil19,lyra19}.

In  practice, $\tau_{{\rm cool}}$ can be thought of as the longer timescale of the particle-gas collisional energy exchange timescale
$\tau_{{\rm gas}}^{{\rm col}}$
and the radiative timescale $\tau_{_{\rm cool}}^{{\rm rad}}$, i.e., $\tau_{{\rm cool}} = {{\rm max}}\left(\tau_{{\rm cool}}^{{\rm rad}},\tau_{{\rm gas}}^{{\rm col}}\right)$. Furthermore,
$\tau_{{\rm cool}}^{{\rm rad}}$
 depends upon whether the
lengthscale of interest is in the optically thick
or thin regimes embodied in
the expression \citep{Spiegel_1957}
\begin{equation}
\frac{1}{\tau_{{\rm cool}}^{{\rm rad}}}
=
\frac{\mu}{\tau_{{\rm cool}}^{{\rm thin}}}, \qquad
\mu \equiv
\left[1-\frac{\tan^{-1}\left(2\pi\ell_\mathrm{ph}/\ell_m\right)}
{2\pi\ell_\mathrm{ph}/\ell_m}\right];
\end{equation}
where the frequency integrated optically thin cooling timescale is given by
$\tau_{{\rm cool}}^{{\rm thin}}
\equiv {\ell_{{\rm ph}} \rho_g C_P}\big/{16\sigma_\mathrm{B} T^3},$
and $\sigma_\mathrm{B}$ is the Stefan-Boltzmann constant.
A further complication is how to estimate the frequency-integrated opacity $\kappa_\mathrm{op}$ as a function of disk location and epoch that, in turn, depends on the disk's turbulence-mitigated spatiotemporal particle distribution $n(a)$, where $a$ is particle size  \citep[e.g.][]{Birnstiel2012,Estrada_etal_2016}.
This reveals a causality dilemma as the key determinant driving disk hydrodynamic instabilities, i.e. $\tau_{{\rm cool}}\left(\kappa_{{\rm op}},n(a),\cdots\right)$ and associated  disk structure, depends on the turbulence
generated by the very same instabilities.  These, in turn, are sensitive to the assumed disk mass and maximum particle size as a function of disk epoch -- two quantities that are currently only weakly constrained by observations.

The currently adopted strategy in confronting the problem of determining $\kappa_{{\rm op}}$, and subsequently
$\tau_{{\rm cool}}$, is to assume
an alpha-type viscous disk model in which vertical particle settling is balanced
by upward turbulent transport \citep{Dubrulle1995}.  Together with evolutionary models of dust mass and size distributions -- and
their resulting frequency-integrated Rosseland or Planck mean opacities \citep[c.f.,][]{Semenov_etal_2003,Woitke_etal_2016,Cuzzi_etal_2014ApJS..210...21C}  -- several recent studies have made
preliminary forays into mapping out disk hydrodynamic instabilities using the
derived thermal relaxation times based on the aforementioned opacity models
\citep{malygin17,barranco2018,pfeil19,fukuhara21}.
Each of these studies extracts various dynamical and thermodynamical
profiles resulting from and/or are inputs to various 1+1 global disk evolutionary models
-- e.g., quantities like density, pressure, total disk mass ($M_{{\rm disk}}$), particle and gas surface densities
($\Sigma_g, \Sigma_d$ respectively), particle size distribution, the turbulent $\ass$ {  for gas  accretion, the turbulent $\adust$ for dust diffusion,} etc.,  all as functions
of position and disk age. One can then assess whether the structure and thermodynamic criteria for the onset of the instabilities discussed in \S\ref{vsi}--\ref{zvi} are met, at least marginally.

\begin{figure}
\centering
\includegraphics[width=1.0\linewidth,clip=true,trim=3cm 0cm 3cm 0cm]{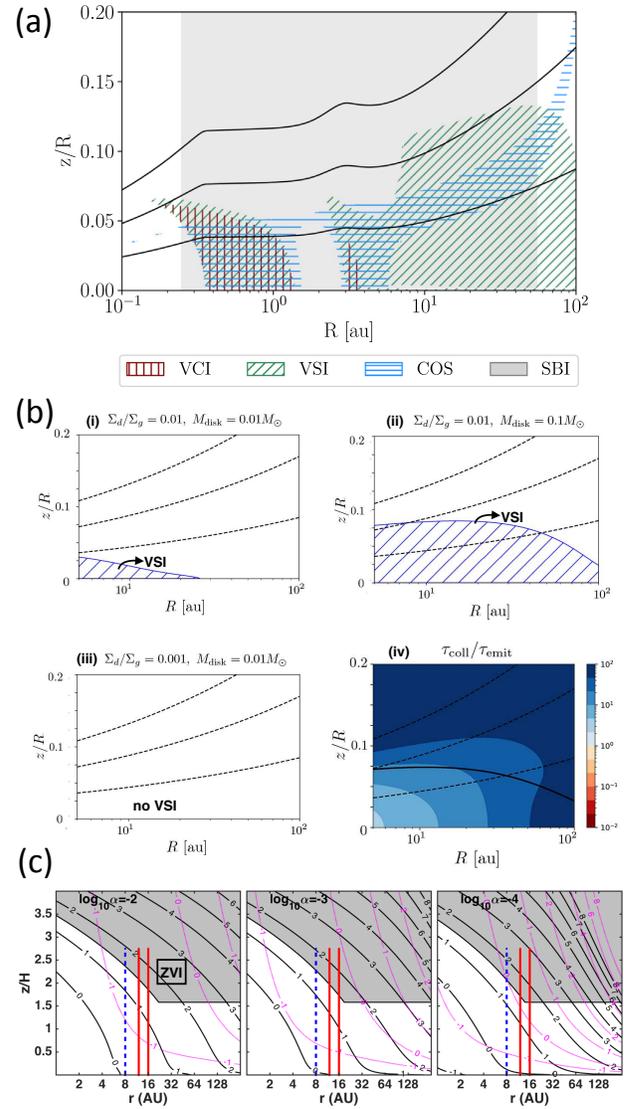}
\caption{Occurrence of hydrodynamic instabilities in {  different PPD} models. (a) {  From \citet{pfeil19}.} This monodisperse ($a\approx 1\mu$m) grain population model considers $M_{{\rm disk}} = 0.1 M_\odot$ and $\ass =  10^{-3}$. Black lines delineate successive pressure scale heights. {  In their models, a Vertical Convective Instability (VCI) can develop because $N_z^2<0$}, but this is not considered in the main text. (b) {  From \citet{fukuhara21}. Panels (i)--(iii) show VSI activity
for $M_\mathrm{disk}/M_\sun=10^{-2},\, 10^{-3}$, $\Sigma_d/\Sigma_g = 10^{-3}, 10^{-2}$, and polydisperse grains with $a_{{\rm max}} = 1$mm, $s=3.5$, $\adust = 10^{-4}$.
Panel (iv) shows $\tau_{_{\rm cool}}^{{\rm col}}/\tau_{_{\rm cool}}^{{\rm rad}}$ for $a_{{\rm max}} = 10\mu$m, $M_{{\rm disk}} = 0.01 M_\odot$,  and $\alpha_\mathrm{D} = 10^{-4}$}. Dashed black lines delineate successive pressure scale heights. (c) {  From \citet{barranco2018}. ZVI-active regions are shaded. Black contours are $\log(\Omega \tau_{{\rm cool}})$ while dotted magenta contours signify $\log (\Omega\tau_{_{\rm cool}}^{{\rm rad}})$. Here $M_{{\rm disk}} \approx 0.04 M_\odot$, $\Sigma_d/\Sigma_g = 0.01$, and grains have $a_{{\rm max}} = 10$~cm and $s=3.25$. Panels correspond to different $\adust$ (labeled as $\alpha$)}. {  Simulations with vertical extent and at the radii of the red lines develop ZVI turbulence, while those of the blue dashed lines do not.} {  Note that in (a) $\ass$ parameterizes mass transport, while in (b) and (c)  $\adust$ parameterizes the dust settling.}
\label{occurence_figure}}
\end{figure}

\citet{pfeil19}, following the framework described in \citet{malygin17},  assume a submicron-sized monodisperse population of grains and primarily analyze disks with mass $M_\mathrm{disk} =0.1 M_\odot$.  Fig \ref{occurence_figure}a is a corresponding map where
their results predict that the VSI is widespread across the vast majority of the disk
(1AU$<R<$100AU), with some coincidental potential for active COS for 1AU$<R<$10AU in regions
containing the midplane and for regions extending away from the midplane for  $R>$10AU.  This study implies that $\tau_{{\rm cool}}$ is limited by $\tau_{{\rm cool}}^{{\rm rad}}$  as the grain number density and disk mass is sufficiently high
that the collision time is considerably short, i.e.,
$\tau_{{\rm cool}}^{{\rm rad}} \gg\tau_{{\rm gas}}^{{\rm col}}$.
\par

\citet{fukuhara21} treat disk models with a polydisperse population of grains with a number density size distribution $n(a)\propto a^{-s}$ with a constant power-law index $s=3.5$ between a minimum and maximum grain sizes $a_{{\rm min}}$ and $a_{{\rm max}}$, respectively. From this they construct an effective mean molecule travel length $\ell_{gd}$, from which they estimate
the gas-grain collisional energy exchange timescale in terms of a simpler collisional timescale, with $\tau_{{\rm gas}}^{{\rm col}} \equiv \ell_{gd}/v_{{\rm th}}$, where  $v_{{\rm th}}$ is a typical thermal velocity
($\propto T^{1/2}$).  Their analysis reveals the useful approximate form
$\ell_{gd} \propto a_{{\rm max}}^{1/2}\Sigma_g^{-1/2}
(\Sigma_d/\Sigma_g)^{-1}
\propto
M_{{\rm disk}}^{-1/2}(\Sigma_d/\Sigma_g)^{-1}$.  As noted earlier in \citet{barranco2018}, as grains get incorporated
into larger particles, the total effective collisional cross-section of the grain population decreases.  Thus, a molecule must travel further before the next collision, which leads to
increasingly inefficient cooling.  However, this reduction in the local medium's
effective collisional cross-section may be halted if particles grow as porous fractal aggregates instead of as zero-porosity spheres \citep{Okuzumi2012}.

 For a range of $M_{{\rm disk}}$ values,
\citet{fukuhara21} examine the onset of the VSI, finding that it is absent in
low mass disks  ($M_{{\rm disk}} = 0.001 M_\odot$), while for
relatively high disk masses  ($M_{{\rm disk}} = 0.1 M_\odot$) it can potentially take root in fairly narrow zones
sandwiching the midplane and out to the disk's outer edge ($|z|/R \lessapprox 0.05-0.1$, see also Fig \ref{occurence_figure}b).
They show how the suppression of the VSI, especially in the low disk mass models, is due to the inability of gas molecules to efficiently communicate with
the dust particles causing $\bcool$ to greatly exceed the critical value of 0.1.  However their results would suggest that the increased cooling times could make their disks susceptible to the COS or ZVI instead.
\par

In a prior study, \citet{barranco2018} take a more sophisticated approach to estimating cooling times by explicitly considering the finite time for energy to be exchanged between gas molecules and dust grains via collisions \citep[e.g.,][]{Hollenbach_McKee_1979,Burke_Hollenbach_1983,Glassgold_etal_2004}. They allow for a variable power-law index ($3<s<4$) in the particle size distribution and find
$\tau_{{\rm gas}}^{{\rm col}} \propto a_{{\rm S}}\rho_g^{-1} T^{-1/2}
(\Sigma_d/\Sigma_g)^{-1}$, where $a_{{\rm S}}$ is the Sauter-mean radius of grains (i.e.\ the radius of a monodisperse population of grains that would have the same total surface area and total volume of a given polydisperse population of grains).  Their study makes predictions for a Minimum Mass Solar Nebula disk model
with $M_{{\rm disk}} \approx 0.042 M_\odot$
\citep{Cuzzi1993} for a variety of $\alpha_\mathrm{SS}$ and $a_{{\rm max}}$ values, finding under these
parameter conditions that
 $\tau_{{\rm cool}}$  is generally controlled by $\tau_{{\rm gas}}^{{\rm col}} $ .
 Fig \ref{occurence_figure}c
exhibits a set of results for large values of $a_{{\rm max}}$ ($\sim10$~cm) in which
the VSI is entirely ruled out as $\bcool \ge 0.1$ everywhere in the model,
while allowing for the feasibility of the ZVI generally away from the midplane where $\bcool \gtrapprox 20$, and possibly the COS in midplane layers where $\bcool\sim 1$.  For smaller $a_{{\rm max}}$ ($\sim 1$~cm) their results indicate that the VSI is feasible closer
to the midplane and closer to the star ($<$ 10 AU) where $\bcool \le 0.1$.

\par
The general trends reported
in  \citet{fukuhara21} -- especially the limiting behavior posed
by collisional exchange timescales (e.g., see bottom right panel of
Fig. \ref{occurence_figure}b) -- are largely consistent with the findings of
\citet{barranco2018}; while
their mutual differences are likely due to the use of different disk model parameters as well as in the approaches applied in estimating $\tau_{{\rm gas}}^{{\rm col}}$. These are matters that need future resolution.
\par
In summary: a systematic examination is required to determine what type of
instability is operative when and where in a PPD.  The strategies
used in the aforementioned studies ought to be applied toward more realistic global evolution models that improve upon simplistic power-law formalisms. The maps in Fig. \ref{occurence_figure} should not be taken as definitive but are meant to illustrate the diverse outcomes depending on model assumptions. However, some trends may be inferred from such recent studies:  the picture of the disk during its earliest stages -- i.e., when the $a_{{\rm max}}$ are still small and $M_{{\rm disk}}$  still relatively high -- may be more like Fig. \ref{occurence_figure}a, while as the disk evolves with $M_{{\rm disk}}$ decreasing and $a_{{\rm max}}$ inexorably growing the turbulent state of the disk may be driven by processes designated like those shown in Figs. \ref{occurence_figure}b-c.  In any case, given the possible range of
$\beta_{{\rm cool}}$ values likely relevant across the disk bulk,
some type of hydrodynamical instability that leads to turbulence  -- whether it be ZVI and/or COS, and/or VSI (or perhaps something yet to be discovered) -- is likely active in PPDs.

As a final remark, we emphasize that alpha-type viscous disk models are generally no replacement for direct numerical simulations for modeling the turbulence generated by these hydrodynamic instabilities. It should be kept in mind that PPDs are essentially inviscid but computational limitations prohibit one to
probe realistic Reynolds numbers. An important future direction is to develop detailed models of the hydrodynamic turbulence generated by these instabilities that are suitable for use in large-scale, long-term global simulations of PPDs. Until then, alpha-type viscous disk models -- while convenient -- should be interpreted with care. %

\section{MULTI-PHASE INSTABILITIES WITH DUST-GAS INTERACTIONS}\label{section_dust}

\subsection{Barriers to planetesimal formation in turbulent disks} \label{SS:barriers}

Bridging the gap between dust grains and solid planetary cores in the core accretion paradigm involves intermediate gravitationally-bound bodies of typical sizes $\sim$0.1--100\,km called planetesimals . Conversion of pebbles into planetesimals requires for solids to overcome several physical barriers as they grow, and is constrained by spatially resolved observations of young disks.
Strong concentrations of solids must be driven and retained in the disk.
Privileged places for large dust enrichments are long-lived pressure maxima, where dust grains tend to drift into.
Moreover, dust grains settle towards the mid-plane of the disk.
Turbulent  of order $\adust\sim\ass \gtrsim 10^{-5}$--$10^{-4}$  solid-to-gas density ratio $\epsdust \lesssim 1$ in the mid-plane may prevent the gravitational collapse of the dust layer .
This turbulence can be hydrodynamical, MHD, or even of dust origin since the dust layer may itself be Kelvin-Helmholtz unstable .

\subsubsection{\textbf{The radial-drift barrier}}
\label{sec:rd_barrier}

Disks need to somehow retain their solids as the solids grow in size and reach , at which their radial drift onto the star becomes  (Eq.~\eqref{eq:drift_vel}).
Pure growth could in principle assist grains for decoupling from the gas before they get accreted, but bouncing or fragmentation probably limit this possibility in practice (\citealt{Brauer2008,Birnstiel2009,Zsom2010}; see also \S\ref{SSS:gb}).
Local dust traps offer powerful alternatives to concentrate grains (see chapter by Bae et al.\ in the same book). Some traps are axisymmetric.
Among suggestions: disk boundaries, dead zones, snow-lines, zonal flows, and edges of planetary gaps.
 the dust layer, , modifies the transport of gas according to
\begin{align*}
\frac{\partial \Sigma}{\partial t}  - \frac{1}{R} \frac{\partial}{\partial R} &\left[ a Z\frac{\partial \left(c_\mathrm{s}^{2} \Sigma \right)}{\partial R} \right]\\
&+ \frac{1}{R} \frac{\partial }{\partial R} \left( \frac{ 1 + Z b }{1+Z} R \, v_{\rm visc} \right) = 0 ,
\end{align*}
where $a =1 / \left[ \mathrm{St} + (1 + Z)^2 \mathrm{St}^{-1}\right]$, $b = \mathrm{St}^{2} / \left[\left(1 + Z\right)^2 + \mathrm{St}^{2}\right]$, and vertical integration being restrained to the dust layer \citep{Gonzalez2017}.
Relative intensities of  are of order $\sim a Z / \alpha $ in smooth disks, meaning that for $10^{-2} \lesssim\mathrm{St}\lesssim 10^{2}$ and a large $Z \gtrsim 0.1$, gas dynamics can be dominated by  viscosity, favoring the formation of self-induced dust traps \citep{Gonzalez2017}.

 spirals, vortices and shocks induced by a companion (see chapter by Bae et al.\ in the same book).
Studying the resilience of these traps  is key for understanding planetesimal formation.
The existence and the morphology of these traps combined with the location of the dust outer radius, and/or the remaining dust flux outside the observed traps provide indirect but degenerated constraints on the parameters $\adust$, $\rm{St}$ and $\epsdust$ \citep{Birnstiel2014,Rosotti2019}.
Models of dust traps should be compatible with the constraints provided by structural data.
Observation of silicate  reveals crystallinity in the grains in the cold outer regions , although it requires high temperatures to develop.
Explanations invoke outward drift mechanisms, redistribution of material by , without any major consensus so far .
Isotopic analysis reveals that the early Solar System has likely been split into two parts in less than $\sim$3 Myrs \citep{Kruijer2017}. This puts a drastic constraint on the existence of a dust trap in its early life, with subsequent consequences for the formation of terrestrial and giant planets \citep{Desch2018,Kruijer2020}.

\subsubsection{\textbf{The growth barriers}} \label{SSS:gb}

Lab experiments suggest that  make dust aggregates grow up to a few centimeters in size by hit-and-stick collisions.  occur without restructuring  at first, but .
Above a few centimeters in size, collisions are found to be on average either non-adhesive,  or destructive \citep{Blum2018}, which is referred as the bouncing, erosion or fragmentation barriers.
Fragmentation occurs for collision velocities of order $\sim$1--10\,m.s$^{-1}$, when monomer bonds are broken.
The fragmentation thresholds appear to  for icy particles, but this is not in general true and depends on temperature and surface coating \citep{Homma2019,Musiolik2019,Steinpilz2019,Bischoff2020,Arakawa2021}
The lab experiments appear to be in agreement with a maximum size of order $\sim$1\,mm for the size distribution of dust grains analyzed on the Comet 67P/Churyumov-Gerasimenko \citep{Blum2017}, and with a maximum size of $\sim$1\,cm for meteoritic inclusions.

Besides the dust-gas instabilities discussed in the following sections, there may be other possible pathways to overcome these growth barriers.
One involves rare low-velocity events that seed the growth of large bodies \citep{Windmark2012,Garaud2013,Booth2018}.
However, the growth timescales may be too long for solids to resist both the radial-drift barrier and erosion \citep{Schrapler2018}.
Porous aggregates made of small ultra-sticky icy monomers have also been invoked \citep{Potapov2020}. The robustness of this scenario remains to be validated experimentally with matrix monomers of size 1--10\,$\mu$m.
Electrostatic charges such as resulting from photoelectric and plasma charging were thought to inhibit growth efficiency \citep{Okuzumi2009,Akimkin2020}. Triboelectric charging may relieve this constraint, given collision velocities $\lesssim$0.1\,m$.$s$^{-1}$ \citep{Steinpilz2020a,Steinpilz2020b,Teiser2021}.

In any case, the lithospheric pressure of comets is only compatible with material of low tensile strength. Experiments reveal that such tensile strength can not originate from collisional growth through mass transfer, but fits with a gentle collection of solids via a local gravitational collapse \citep{Blum2017}. So far, observations have probed continuum emission of solids up to a few centimetres \citep{Lommen2009,Casassus2019}. Although the forthcoming Square Kilometre Array will allow to probe larger wavelengths, it is not clear whether continuum emission of decimeter-sized grains will be lost into free-free electron emission or not \citep{Dewdney2009}.

Another possibility to overcome the growth barriers is to concentrate grains aerodynamically by means of dust-gas instabilities into the form of dusty clouds, up to the stage where the local mass of solids is such that it collapses gravitationally \citep{war00,Shariff2011,Youdin2011,shi13,Latter2017}.
We hereby dedicate the rest of the section to this pathway.

\subsection{Streaming instability \& resonant drag instabilities }\label{sec:dust:streaming}

\subsubsection{\textbf{Linear phase}}\label{SSS:silin}
A linear analysis of the SI proceeds from the equations of motion for gas and dust under the local-shearing-box approximation, treating dust as a pressureless fluid that drifts inwards
\citep[see, e.g.,][Equations~(1)--(4)]{Youdin2007}.
Considering a small, homogenous patch at $z=0$ within the midplane dust layer, one linearizes the equations to obtain algebraic equations for each %
 Faster-growing, larger-scale modes are considered likely to have a stronger nonlinear influence, although the link between linear-SI growth rates and planetesimal formation remains highly uncertain at this time.

\begin{figure}[!th]
\begin{center}
\includegraphics[width=1.0\columnwidth]{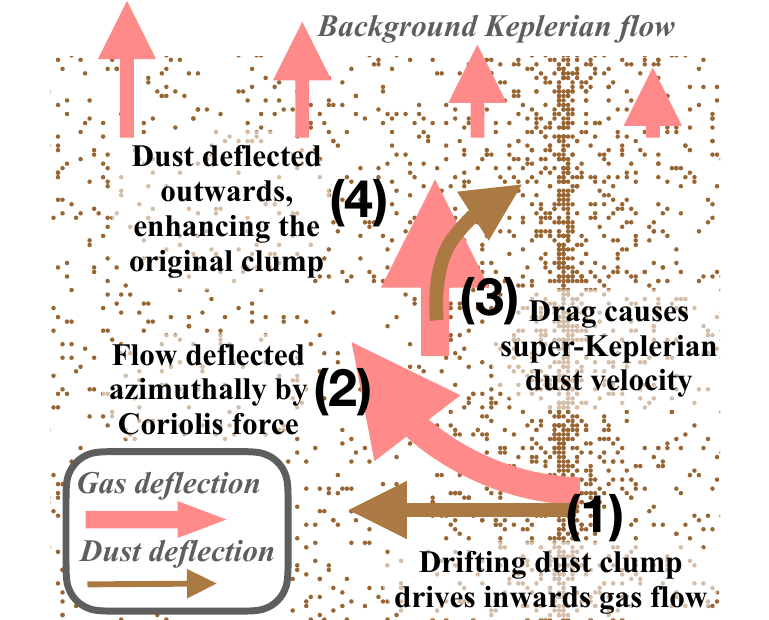}
\caption{The basic ingredients involved in the mechanism of the streaming instability. The feedback cycle (steps (1)--(4)) shows
how an asymmetric clump of dust can dynamically drive gas and dust flows that subsequently enhance the density of the clump.
We show the radial-azimuthal plane with dust density shown using brown dots and the motions of gas and dust relative to the background drifts shown with pink and brown arrows, respectively. During this process, the dust continues its bulk inwards drift; the pictured deflections are on top of this drift.  For this feedback to operate,
the perturbation must have a vertical structure, otherwise radial gas motions in step (1) are halted by the gas pressure because they are compressive.
While the key physical principles shown here are valid in both the low- and high-$\epsdust$ regimes, it is necessarily overly simplified; more detailed versions are given in \citet{Squire2020a}.}
\label{fig: SI mechanism}
\end{center}
\end{figure}

\paragraph{Ideal streaming instability}Motivated by the study of \citet{Goodman2000}, the analysis described above was first carried out by \citet{Youdin2005}, who discovered the SI and suggested its importance to planetesimal formation.
They showed that the key parameters governing the instability are the dimensionless stopping time $\taus$ and the solid-to-gas density ratio $\epsdust$,%
\footnote{The pressure support parameter $\etaP$ \citep[defined by][]{Nakagawa1986} sets the velocity scale via the differential speed $\eta v_\mathrm{K}$ between gas and dust, but otherwise does not change the properties of the instability.}
demonstrated that the motions induced by the SI act to clump the dust, and argued that in most relevant regimes the growth rates are large enough for the instability to reach its nonlinear stages well before the radial dust drift significantly modifies the equilibrium.
A key feature uncovered in their analysis was the sudden increase in the SI's growth rate for $\epsdust \gtrsim 1$.

Most of the analysis of \citet{Youdin2005} solved for the eigenvalues numerically, which can be inconvenient for making simple estimates across a wide parameter range.
We see that the SI
grows faster -- although at smaller scales -- for near vertical modes ($k_{z} \gg k_{r}$, $\theta_{k}\approx0$), and grows more slowly for smaller grains (smaller $\taus$).
At $\epsdust \gtrsim 1$, likely the more relevant regime for planetesimal formation, \citet{Squire2018a} showed that the SI changes its character and mechanism, .
For $\epsdust\gg1$ and $\taus\ll1$, its  maximum growth rate is ${\rm Im}(\omega)/\Omega\approx \sqrt{\epsdust-1}$ at radial wavenumber $k_{r} \etaP r \approx0.4\epsdust^{19/8}\taus^{-5/4}$.
This growth is very rapid (${\rm Im}(\omega)>\Omega$), with a rate that increases monotonically with $\epsdust$, although it operates at very small scales for small $\taus$.

Even in its most basic form, the mechanism for the streaming instability is more complex
than most astrophysical fluid instabilities. Various works have presented simplified analyses,
exploring reduced equations and the behavior of the roots of the dispersion relation to understand how the instability operates
\citep[see, e.g.,][]{Jacquet2011,Squire2018a,Zhuravlev2019,Jaupart2020,Pan2020a,Squire2020a}. These have
shown that  rotation, dust drift, and  two-dimensional structure are key to the SI, although the background shear flow is not. The basic mechanism of its operation is sketched in Fig.~\ref{fig: SI mechanism}, which shows how a radial clump of dust can drive a gas flow that -- due to the Coriolis force -- drives a  radial  drift of dust outwards (relative to the background dust drift), thus enhancing the original clump.
In order for the associated gas flow to be incompressible, and thus not strongly resisted by gas pressure, the flows must
have vertical structure. .
Note also that the gas pressure perturbations, which are driven in step (2) , are not in  phase with the  dust-density perturbations in most regimes, complicating the picture
beyond dust trapping in pressure bumps.
This is consistent with \cite{Lin2017}'s finding that an isothermal gas embedded with small grains is equivalent to a single fluid with a modified cooling, in which case a phase lag between pressure and dust-density perturbations is a necessary feature of growing oscillations, such as the SI.
Interestingly, this feature seems to carry over to the nonlinear regime, where an anti-correlation between dust and gas density has been observed (see figure 7 of \citealt{YJ14} as well as \citealt{LYS18}).
Thus, the conventional notion of dust-trapping at pressure maxima does not necessarily apply under dynamical (i.e., non-steady) conditions such as instabilities.

\paragraph{Extensions: polydisperse grains, turbulence, and other drag instabilities}
Any real disk will involve a wide spectrum of grain sizes ($\taus$) at a given radial location, which is not taken into account in the above analysis.
\citet{Krapp2019} first explored this ``polydisperse'' streaming instability, uncovering the unexpected result that in regimes with $\epsdust\lesssim 1$ the maximal growth rate of the SI decreases monotonically as the number of discrete grain species is increased.
Very low growth rates, or only upper bounds in some cases, are found with a true continuous distribution.
Their results, which have been further explored in \citet{Zhu2021} and \cite{Paardekooper2020,Paardekooper2021}, suggest that the polydisperse SI is primarily controlled by $\epsdust$ and the maximum $\taus$ of the distribution ($\tau_{{\rm s, max}}$).
Specifically, \cite{Zhu2021} found that the instability is robust (converged growth rates $\gtrsim 0.1\Omega$) only above a sharp, distinct boundary in $\epsdust$-$\tau_{{\rm s, max}}$ space, with $\epsdust\gtrsim 1$ or $\tau_{{\rm s, max}}\gtrsim1$ (see also \S\ref{SSS:sisat}).
The results also suggest that calculations involving polydisperse grains must carefully check convergence in the number of grain species.

The presence of background turbulence, even at very low levels, may also adversely affect the SI by diffusing small-scale
motions and dust density variations.
Simple estimates can be obtained by modeling the effect of turbulence as causing a gas viscosity, dust pressure, and dust density diffusion, each parameterized by the disk $\ass$.  The approach, explored by \cite{Umurhan2020} and \cite{chen20}, suggests that, because turbulence decreases its growth rate and increases its spatial scale, the  SI can operate only relatively far out in the disk beyond $\simeq \!10{\rm AU}$, and only for very low (though realistic) levels of turbulence $\ass\lesssim 10^{-4}$ and relatively large grains \citep[see also][for further mathematical details]{Squire2018a,Jaupart2020,Zhuravlev2020}.
On the other hand, turbulence can also act to clump grains itself, an effect that cannot be captured by diffusive treatments \citep[e.g.,][]{CH01,PP11,HCW17,HC20}.

Finally, it is worth noting several extensions to the standard SI analysis, which have revealed related instabilities. \citet{Pan2020}
considered non-axisymmetric motions in the radial-azimuthal plane (perturbations without vertical structure across the dust layer). They found a similar instability, as expected from numerical simulations \citep{Schreiber2018}, although there are some unique features and different scalings from the axisymmetric SI. \citet{Auffinger2018} found an interesting related
instability when the background profile is modified by a pressure bump, although its details and relationship to the standard SI remain unclear. The vertically global SI study of
\citet{Lin2021} revealed a faster-growing instability driven by the %
vertical gradient of the dusty gas' rotation velocity
that is not captured by standard analyses \citep[see also the unpublished study of ][]{ishitsu09}.
This may dominate the SI and control the dust-layer thickness, showing interesting similarities to results of nonlinear simulations; more investigation is needed.
Finally, the RDI method of \citet{Squire2018a} also uncovered a related  instability, termed the ``settling instability,''
which occurs as grains settle towards the midplane of the disk \cite[see also][]{Lambrechts2016}.
Unlike standard SI, the instability grows rapidly even for the smallest grains at low $\epsdust$, suggesting the possibility of early seeding of grain growth through clumping before grains reach the midplane.
It is also more robust than SI to polydisperse grains, although the nonlinear clumping it causes may be rather weak \citep{Krapp2020}.

\subsubsection{\textbf{Nonlinear saturation and dust concentration}\label{SSS:sisat}}

The discussion in \S\ref{SSS:silin} considers how a system of coupled gas and dust in a Keplerian disk responds to small perturbations.
As the instability drives these perturbations to grow exponentially with time, they should ultimately become nonlinear, leading to a saturated, potentially turbulent, state.

\cite{JY07} first studied the nonlinear saturation of the streaming instability and the resulting turbulence with an unstratified local shearing box and a single dust species.
Depending on the dust size, the system could reach two distinctly different saturation states, which continues to be phenomenologically accurate for more complex systems.
When the dimensionless stopping time $\taus \gtrsim 1$, the dust undergoes traffic jams and organizes into axisymmetric dust filaments.
When $\taus \lesssim 0.1$, by contrast, the system generates numerous dust-gas vortices vertically, collecting dust in between.
The former could lead to a $\sim$10$^3$ enhancement in dust concentration, while the latter a much weaker $O(10)$ enhancement.
The gas remains incompressible to a high degree ($\lesssim$0.1\%).
These saturated states are qualitatively consistent across various simulations with different numerical methods, although some issues in numerical convergence at high-end tail of the particle density distribution might exist \citep{Bai2010,YJ16,BKP19}.

As mentioned in \S\ref{SSS:silin}, two distinct regimes occur in the linear growth of the streaming instability with polydisperse dust grains -- fast and slow growth --, and it appears that this dichotomy also carries over to their nonlinear saturation in unstratified disks \citep{YZ21}.
When $\epsdust \gtrsim 1$ or $\tau_\mathrm{s,max} \gtrsim 1$, the dust-gas dynamics at the saturation state is similar to that driven by monodisperse dust grains with $\taus \sim \tau_\mathrm{s,max}$ discussed above \citep[see also][]{SJL21}.
On the contrary, when $\epsdust \lesssim 1$ and $\tau_\mathrm{s,max} \lesssim 1$, the system at the saturation state appears to be close to laminar.
Interestingly, \cite{YZ21} found that when the saturation state is turbulent (i.e., $\epsdust \gtrsim 1$ or $\tau_\mathrm{s,max} \gtrsim 1$), significant dust segregation by size occurs and the mean radial drift of dust grains of different size is noticeably altered from the drag-force equilibrium of \cite{Nakagawa1986}.

When vertical gravity from the central star is considered, the saturated turbulence driven by the streaming instability sustains a vertically stratified layer of dust with a finite scale height.
The same dust-gas vortices could be seen near the mid-plane in the stratified simulations of \cite{YJC17} with single dust species.
With multiple species, \cite{BS10c} and \cite{SYJ18} demonstrated clear scale separation between different species \citep[see also][]{YZ21}.
If the initial solid loading is not sufficiently high, this equilibrated stratified dust layer can be maintained for more than thousands of orbital periods, when simulations ended.

If the solid loading, quantified by the dust-to-gas \emph{column} density ratio $Z \equiv \Sigma_\mathrm{d} / \Sigma_\mathrm{g}$, is above some critical value $\Zc$, strong concentration of solids occurs \citep{JYM09}.
This leads to nearly \emph{axisymmetric}, dense, narrow filamentary structures of solids (\citealt{YJ14,LYS18}; see also \citealt{FM21}), which may have observational consequences \citep{SBC21}.
Without external turbulence, typical radial separation between adjacent filaments is about 20\% of the gas scale height $ H_\mathrm{g}$.
With external turbulence such as non-ideal MHD, the separation can be on the order of about $H_\mathrm{g}$ \citep{YMJ18}.
In any case, the separation appears to decrease with increasing $Z$ \citep{YJC17,YMJ18}.
As mentioned in \S\ref{SSS:silin}, these dense filaments are \emph{not} correlated with local pressure maxima \citep{YJ14,LYS18} and hence the latter may not be responsible for driving the concentration of solids in this scenario.
Furthermore, the boundary between strong clumping of solids when $Z \gtrsim \Zc$ and no strong clumping when $Z \lesssim \Zc$ appears to be sharp (see also below).
However, the relationship of this boundary to the linear streaming instability discussed in \S\ref{SSS:silin} (e.g., its change at $\epsdust \simeq 1$) is tenuous \citep[see][\S3.5]{LY21}.
It remains to be understood whether this dichotomy arises because other linear instabilities are important (e.g., that of \citealt{Lin2021} with vertical stratification), or whether clumping to Roche densities is inherently nonlinear.
Further study of this important, but subtle, issue is needed.

\begin{figure}[h!]
\centering
\includegraphics[width=\columnwidth]{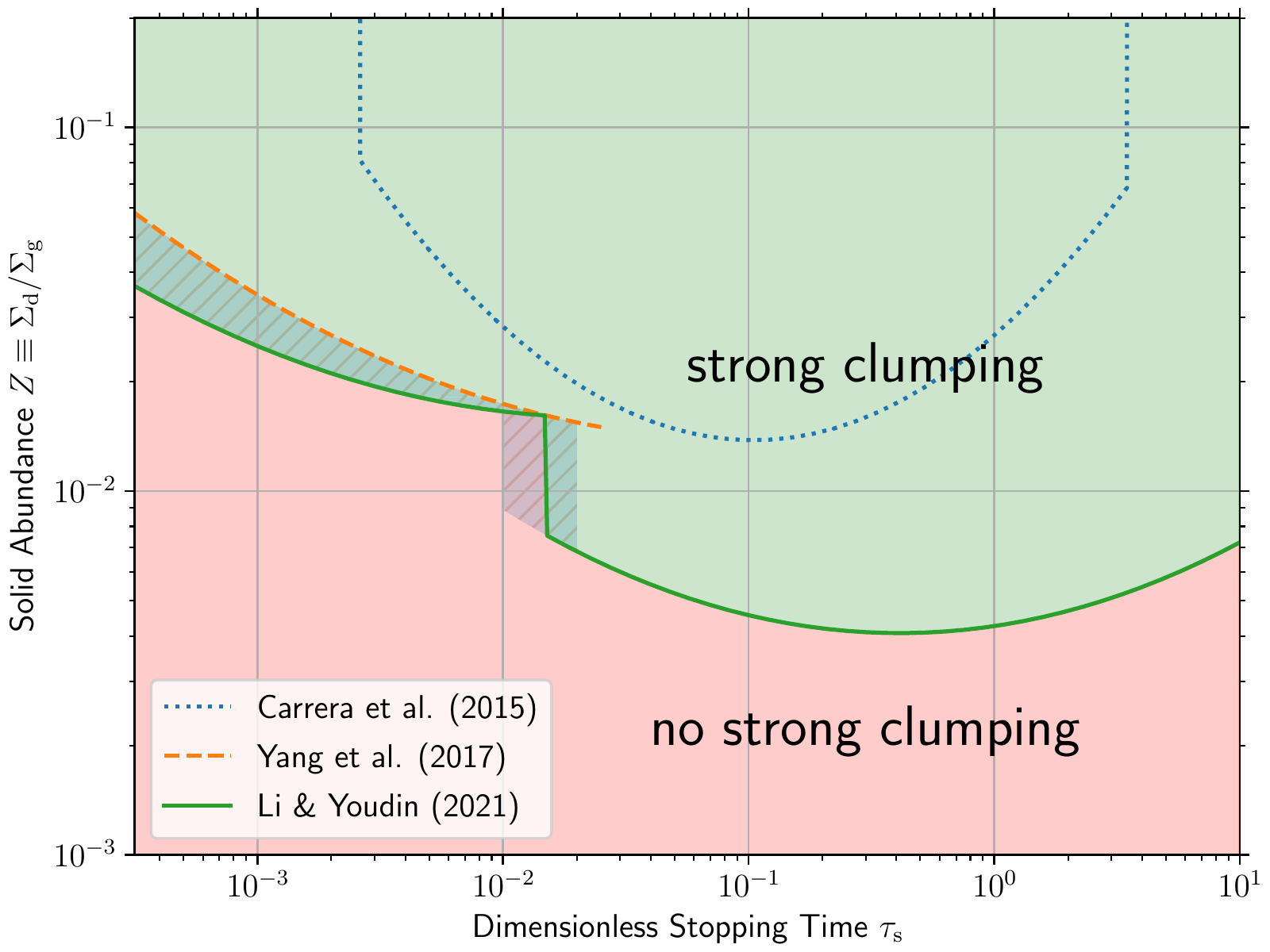}
\caption{Critical solid abundance $\Zc$ as a function of dimensionless stopping time $\taus$.
    Each line is the best fit to the numerical simulations conducted in different work, and above the line may strong clumping of solids occur and trigger the formation of planetesimals.
    All works compiled here assumed a dimensionless radial pressure gradient of $\Pi = 0.05$.\label{F:zcrit}}
\end{figure}

The value of $\Zc$ should depend on several parameters, including the dimensionless stopping time $\taus$ (single or multiple species), the radial pressure gradient, as well as the dynamics of external disturbances.
The radial pressure gradient is often quantified by the dimensionless number
$\Pi \equiv \Delta u_\phi / c_\mathrm{s}$ \citep{BS10c}
or
$\eta \equiv \Delta u_\phi / v_\mathrm{K}$ \citep{Nakagawa1986}, where $\Delta u_\phi$ is the reduction in the equilibrium azimuthal velocity of the gas , and $c_\mathrm{s}$ and $v_\mathrm{K}$ are the local speed of sound and Keplerian speed, respectively.
At a given and representative radial pressure gradient $\Pi = 0.05$, without external disturbances, and with single species, \cite{CJD15} first measured $\Zc$ as a function of $\taus$ by gradually removing the gas over the course of the simulations.
By fixing $Z$ instead in each simulation, the measurement for $\taus < 0.1$ was later modified by \cite{YJC17}, and \cite{LY21} further found significantly lower threshold for $\taus \gtrsim 0.015$.
The comparison between the three works is shown in Fig.~\ref{F:zcrit}.
To trigger strong solid concentration for small particles with $\taus \lesssim 10^{-2}$, a value of $Z$ greater than a few percent is in general required.
The timescale for forming dense solid filaments when $Z > \Zc$ appears reciprocal to $\taus$, i.e., proportional to the timescales of radial drift and vertical sedimentation, and in fact a smaller dust species may lead to a much stronger concentration than a larger one \citep{YJC17}.
When varying radial pressure gradient, it appears that the stronger the gradient, the more difficult to drive strong concentration of solids \citep{BS10b,AS19}, and $\Zc$ may simply scale linearly with $\Pi$ \citep{SO18}.
Finally, for a system with a dust-size distribution, it remains unclear how to best quantify the critical condition \citep{BS10c,SJL21}.

The evaluation of this critical solid abundance $\Zc$ becomes complicated when external disturbances to dust-gas dynamics driven by (magneto-)hydrodynamical instabilities are present (see \S\ref{S:hydro} and \S\ref{mhd_section}).
For large particles with $\taus \sim 1$, it was found that the MHD turbulence driven by ideal MRI in general assists local concentration of solids and hence triggers their strong clumping with a lower $\Zc \simeq 1\%$ (as compared to $\Zc \simeq 2$--3\% with a similar numerical setup but without MHD turbulence) \citep{JO07,JKH11}.
However, it appears that under such an ideal MHD turbulence, strong clumping of particles with $\taus \sim 0.1$ is significantly more difficult to trigger, potentially making $\Zc \gtrsim 8\%$ \citep{YMJ18}.

 the \emph{anisotropic} turbulence driven by the  non-ideal MHD or the VSI might alleviate this difficulty for planetesimal formation. \cite{YMJ18} found that in a layered accretion disk with an Ohmic dead zone, $\Zc \simeq 2\%$ for particles of $\taus \sim 0.1$.
Similar $\Zc$ was also found in disks dominated by ambipolar diffusion \citep{XB21}.
\cite{schafer20} found that in some cases, the interaction between the VSI and the streaming instability makes the clumping of solids even stronger as compared to the latter alone.
In both scenarios, the turbulence near the mid-plane is highly anisotropic, with the vertical stirring of particles much stronger than the radial one \cite[see also][]{ZSB15,stoll16,RL18}.
This also raises the question of how sensitively the triggering of strong clumping depends on vertical sedimentation of solids and hence the local dust-to-gas density ratio $\epsdust$ near the mid-plane.

\subsection{Secular gravitational instabilities}\label{sec:dust:sgi}
The idea that gravitational instabilities (GI) in the dust layer cause solids to collapse into planetesimals has a long history (\citealp{Safronov1969,gw73,ys02}; reviewed in more detail by \citealp{cy10}).  Several Solar System observations support the hypothesis of planetesimal formation by gravitational collapse \citep{morbidelli09, NYR10, NL19, blum17, mckinnon20} .  %

 The gaseous component of protoplanetary disks is at best marginally gravitationally unstable, and the dust component has less self-gravitating mass at early times.  If the gas could be ignored, then collisional damping of particle motions ensures gravitational fragmentation \citep{tanga04, michikoshi07}.  Gas cannot be ignored in the early stages of planet formation, which introduces both stabilizing and destabilizing effects on a self-gravitating dust layer.

Gas drag can also lead to strong particle clumping by the SI or in pressure bumps (\S\ref{SS:barriers} and \S\ref{SSS:sisat}; see also \citealp{pinilla17}).  Strong particle clumping clearly facilitates GI.  The conditions for gravitational collapse in a dynamically clumping medium is addressed  below.  Dust gravitational instabilities  in an otherwise smooth disk of gas and dust  strong clumping, e.g. outside pressure bumps and below SI clumping thresholds \citep[Fig.~\ref{F:zcrit};][]{LY21}.  %

Dust that is perfectly coupled to the gas gives a  limiting case of GI.  %
\citet{sek83} analyzed a midplane layer in this limit and found that  GI requires a total density, $\rho_0 >  0.17 \rho_{\rm R}$, where $\rho_{\rm R} = 3.53 M_\ast/r^3$ is the Roche limit below which a hydrostatic fluid body is tidally disrupted (see \citealp{cy10} for discussion).  If the midplane layer were dust-free, then applying this \citet{sek83} condition to the midplane of a gas disk (that is vertically isothermal with $\gamma = 7/5$) translates to $Q_{\rm g} = c_s\Omega/(\pi G \Sigma_{\rm g}) \lesssim 0.25$, stricter than usual due to finite thickness, the assumed incompressibility and other details of the sublayer analysis.  With dust (more relevantly), instability occurs for midplane dust-gas ratios $\epsdust \gtrsim 4 Q_{\rm g} -1$, which is large for gravitationally stable gas disks.

For larger grains that slip relative to gas, a new mode of secular gravitational instability (SGI) exists \citep{war76,war00,y05a, michikoshi12}.  Due to the transfer of angular momentum from dust to gas,  SGI modes can have wavelengths  longer than the standard Toomre (dust) upper limit of  $\lambda_{\rm T,d} = 4\pi^2 G \Sigma_{\rm d}/\Omega^2$.  The possibility of SGI forming wide dust rings has been proposed as an explanation for observed structures in protoplanetary disks \citep{Takahashi2016} and --- after the rings fragment into many planetesimals --- for the radial zonation of asteroid spectral classes \citep{Youdin2011}.

However, the viability of the SGI in real protoplanetary disks is unclear.  Radial turbulent diffusion was shown to have a strong stabilizing effect on SGI, especially for smaller solids \citep{Youdin2011, Shariff2011}.   \citet{takahashi14} analyzed ``two fluid" models that include drag feedback (similar to the linear SI, but with self-gravity and without the vertical motions) and showed that sufficiently long wavelength modes are stabilized.  Their stability criterion, valid for $\adust \ll \tau_{\rm s} \ll 1$, i.e.\ weak turbulent diffusion and small solids, is
\begin{align}\label{eq:SGI}
\epsdust(1+\epsdust) &> Q_{\rm g}^2
\end{align}
where we use $Q_{\rm g} = Q_{\rm g, sub} H_{\rm p} /H_{\rm g} = Q_{\rm g, sub} \sqrt{\adust/\tau_{\rm s}}$ to replace the Toomre $Q_{\rm g, sub}$ of the gas subdisk that appears in the original works (see also \citealp{Latter2017, tominaga20}).  If we require the gas disk to be gravitationally stable, $Q_{\rm g} \gtrsim 2$, then the instability condition becomes $\epsdust \gtrsim Q_{\rm g}$, i.e.\ similar to the Roche-like criteria of \citet{sek83}.

In general, an effective viscosity that transports momentum is known to facilitate GI \citep{lin16}.  \citet{tominaga19} discovered a viscous GI in a gas-dust sublayer with an approximate stability criterion
\begin{align}
\frac{\epsdust}{\sqrt{1+\epsdust}}  &> Q_{\rm g} \sqrt{\frac{3 \alpha}{\tau_{\rm s} }} \, ,
\end{align}
where $\alpha \sim \adust \sim \ass \ll \taus$ is assumed.
Since $\alpha \ll \tau_{\rm s}$ is again required, this criterion is easier to satisfy than Equation~\eqref{eq:SGI}.  Simulations with turbulence that could act as an effective viscosity are thus strongly motivated.

In general, investigations that go beyond height-integrated models of the isolated dust sublayer  are needed to better determine the relevance of the SGI and related viscous GI for observed disk structures and planetesimal formation.

\subsection{Formation of planetesimals and their expected properties}\label{sec:dust:planets}

Since PPVI, it has become feasible to include the self-gravity of pebbles in the simulations of the streaming instability to trigger gravitational collapse of local pebble concentrations into a large number of bound objects and hence planetesimals, leading to a statistically meaningful mass function of newborn planetesimals.
At similar conditions, the result appears insensitive to numerical methods, or the use of sink particles \citep{JM15} or clump finding \citep{SA16} to identify the bound objects, although some variability may exist between different realizations of the same system \citep{RW21}.

Excluding the high-mass cutoff, it was found that the differential mass function could be represented by a power law $\mathrm{d}N / \mathrm{d}M \propto M^{-q}$ with $q \simeq 1.6$, where $M$ is the mass of a planetesimal.
Therefore, small planetesimals dominate in number, while a few largest planetesimals dominate the total mass.
This function seems not particularly sensitive to the pebble size ($\taus$) or the solid abundance ($Z$) \citep{SA17}.
On the other hand, the \emph{efficiency} of planetesimal formation does appear to depend on the radial pressure gradient \citep{AS19} and the strength of turbulence (\citealt{GS20}; see also the linear analyses by \citealt{chen20} and \citealt{Umurhan2020}; \S\ref{SSS:silin}).
The higher the pressure gradient or the stronger the turbulence, the less efficient the formation of planetesimals is.
In fact, the former could be understood by the more turbulent saturation state of the streaming instability \citep{AS19}, and hence indicating that the region near a pressure bump may be a favorable location for planetesimal formation \citep{CS21}.
Moreover, there may exist a maximum turbulent forcing at $\ass \sim 10^{-3}$ above which planetesimal formation is inhibited \citep{GS20}.

In contrast to the differential mass function of planetesimals, the cumulative mass function showed that the largest few planetesimals are situated with a much steeper slope, indicating a high-mass cutoff.
It was convenient to express this cutoff as an exponential tapering \citep{JM15}, while it is perhaps more accurate to describe the whole mass function by a broken power law \citep{LYS19}.
In any case, the ``knee'' of the mass function defines the characteristic mass $M_\mathrm{c}$ of newborn planetesimals.
\cite{SYJ17} found that $M_\mathrm{c}$ correlates with the separation between axisymmetric dense filaments of pebbles driven by the streaming instability, and hence that $M_\mathrm{c}$ is directly proportional to the total mass reservoir inside each filament.
Moreover, $M_\mathrm{c}$ could also depend on the balance between gravitational contraction and turbulent diffusion of a collapsing pebble cloud \citep{Schreiber2018,GM20,KS20,KS21}.
In an attempt to consolidate the simulations in the literature, \cite{LL20} fit and gave an expression for $M_\mathrm{c}$ as a function of $Z$, $\Pi$, disk aspect ratio, as well as strength of the disk gravity.

Even though the mass function of planetesimals could be produced and analyzed with hydrodynamical simulations, these simulations could not resolve the process of gravitational collapse of the concentrated pebble clouds.
To study the internal structure of a planetesimal resulting from such a process, these pre-collapse pebble clouds are taken as initial conditions and evolved using statistical or $N$-body methods.
Factors under consideration included planetesimal mass, initial pebble size or size distribution, pebble porosity, pebble collision, and gas drag.
In general, the timescale for the collapse decreases with increasing planetesimal mass and is limited by the free-fall time when the planetesimal radius $R \gtrsim 100$\,km \citep{WJ14}.
At such a large size $R \gtrsim 100$\,km, though, the collisions between pebbles become fragmenting and much less primordial pebbles are preserved \citep{WJBB17}.
When $R \lesssim 10$\,km, by contrast, most of the collisions may be bouncing and the resulting planetesimal could be a rubble pile.
Nevertheless, the pebbles are compressed at bouncing collisions.
Icy pebbles may be compacted up to a volume filling factor of $\sim$20\%, while silicate pebbles $\sim$40\% \citep{LG16}.
The porosity of a planetesimal then depends on the final size distribution of pebbles; the more small dust, especially resulting from fragmenting collisions, the less porous the planetesimal is.
Finally, gas drag causes size sorting of pebbles in a collapsing pebble cloud \citep{WJ17,VDD21}.
While small pebbles have slow terminal velocities, large pebbles have long deceleration times.
The combination of both effects along with the nebula conditions determine the pebble size distribution as a function of radial distance to the planetesimal center.

Another essential factor is the initial angular momentum of a collapsing pebble cloud.
It appears that almost all pebble clouds in nonlinear simulations of the streaming instability with self-gravity contained too much angular momentum and should fragment into planetesimal binaries \citep{NL19}.
The distribution of the angular momentum vector showed that about 80\% of the binaries were prograde.
By simulating the gravitational collapse of these pebble clouds using $N$-body methods, these clouds tend to form binaries with similar-sized objects and appreciable eccentricity $e \lesssim 0.6$ \citep{NYR10,RF20,NL21}.
Clouds with sufficiently low angular momentum may result in low contact speeds between the two components of the binaries and hence form a pathway to contact binaries (\citealt{RF20,NL21}; see also \citealt{LYJ21}).
On the other hand, clouds with high mass or angular momentum may lead to a hierarchical system or multiple binaries.
The orbital inclination of the binary should be similar to that of the initial pebble cloud, which is limited by the vertical scale height of the pebbles and hence should be small.

\section{\textbf{MAGNETOHYDRODYNAMICAL PROCESSES \& WINDS }}\label{mhd_section}
\subsection{Framework}\label{sec:mhd:framework}
\subsubsection{\textbf{Ionization}}
Protoplanetary disks are observed to have both inflows (i.e., accretion) and outflows (winds).  While purely hydrodynamic and thermal processes might be responsible for both phenomena, various arguments suggest that magnetohydrodynamic (MHD) processes are important.  For example, known hydrodynamic instabilities that might give rise to a turbulent viscosity are weak (\S\ref{S:hydro}), except gravitational instabilities of disks that are quite massive and therefore probably quite young (Class 0).  Also, accelerating molecular outflows to speeds of several $\mathrm{km\,s^{-1}}$ is difficult to do purely thermally without heating the gas to such a temperature that the molecules dissociate, whereas magnetocentrifugal winds can in principle expel gas without heating it at all, at least in ideal MHD. MHD, however, requires the gas to support electrical currents, which in turn requires
mobile charge carriers: free electrons, atomic and molecular ions, and even small charged grains or polycyclic aromatic hydrocarbons (PAH). In a typical low-mass Class II system, temperatures are high enough for (partial) thermal ionization only at small radii ($\lesssim 0.1\textsc{au}$), where only a small fraction of the mass of the disk resides.
The thermally ionized region becomes larger both for systems with high luminosity (e.g. for a Herbig Star with 55 solar luminosity this region can extend to 1 au \citep{2016ApJ...827..144F}), and for systems with high accretion, as here the MRI heating can heat up and so ionize the disk further outward \citep{2018ApJ...861..144M}.
Elsewhere in the disk, as in much of the interstellar medium (ISM), ionization relies on nonthermal sources.  These include X-ray \citep{Glassgold+2017} and far ultraviolet \citep{Perez-Becker+Chiang2011} photons emitted by the protostar's corona
and accretion layers, cosmic rays \citep{Rab+2017,Seifert+2021}, radionuclides such as $\mathrm{^{26}Al}$ \citep{Umebayashi+Nakano2009}, and perhaps strong electric fields associated with MRI \citep{Inutsuka+Sano2005,Okuzumi+2019} or even lightning \citep{Whipple1966,Desch+Cuzzi2000}.
\cite{Guedel2015} gives a pedagogical review of many of these processes.

The degree of ionization of protoplanetary disks is uncertain because of the uncertain strengths of the various ionizing processes listed above, but also because of the uncertain abundance and size distribution of grains, which tend to trap free charges and promote recombination \citep[and references therein]{Ivlev+2016,Thi+2019}.
Close to the star where midplane temperatures are 500-1500~K, however, thermionic emission from grains may actually increase the abundance of free electrons \citep{Desch+Turner2015,2022MNRAS.509.5974J}; at still hotter temperatures where the grains sublime, the main source is thermal ionization of alkali metals (K, Na, $\ldots$), which have low first-ionization potentials.

\subsubsection{\textbf{Plasma conductivity}}
At magnetic field strengths and ionization levels sufficient to drive accretion at observed rates, the electrical conductivity behaves as a tensor rather than a scalar \citep{Wardle+Ng1999}.  That is to say, the electric field $\bm{E'}$ in the local rest frame of the gas and the current density $\bm{J}$ are generally not parallel, because the mobile charges are deflected by the magnetic field to varying degrees depending on their charge-to-mass ratios and collision rates with the neutral gas.  When the mass density in the charged species is negligible compared to that of the neutrals, the tensorial Ohm's Law $\bm{E'}=\bm{\upsigma^{-1}\cdot J}$ reduces to three terms \citep{Balbus+Terquem2001, Bai2014_Hall}
\begin{equation}\label{eq:etaOHA}
\bm{E'}=\eta_{\mathrm{O}}\bm{J} + \eta_{\mathrm{H}}\bm{J\times\hat{B}} + \eta_{A}\underbrace{\bm{\hat{B}\times}(\bm{J\times\hat{B}})}_{\bm{J}_\perp},
\end{equation}
in which $\bm{\hat{B}}=\bm{B}/|\bm{B}|$ is a unit vector parallel to the magnetic field.

The Ohmic diffusivity $\eta_{\mathrm{O}}$ is independent of $|\bm{B}|$ and therefore dominates when the field is weak compared to the pressure of the gas and the ionization fraction is very low; these conditions often prevail at the disk midplane.  $\eta_{\mathrm{O}}$ is also the only one of the three diffusivities that contributes to $\bm{E'\cdot B}$, and therefore the only one that can alter magnetic flux; the others preserve flux but allow it to drift relative to the neutral gas.  Insofar as the charged particles collide with the neutrals more often than one another, $\eta_{\mathrm{O}}$ is inversely proportional to the fractional ionization, which typically needs to be $n_e/n_{\textsc{H}}\gtrsim 10^{-13}$ to support MRI in protoplanetary disks.

The ambipolar diffusivity $\eta_{A}\propto|\bm{B}|^2$ dominates the slippage between the charged and neutral components when the field is strong and the ionization fraction is well above $10^{-13}$ (but still $\ll 1$), as tends to be the case at high altitudes in the disk and in their MHD winds.  Although it preserves magnetic flux, ambipolar diffusion contributes to $\bm{E'\cdot J}$, and therefore dissipates magnetic energy (not flux), which heats the gas.

Finally, the Hall diffusivity $\eta_{\mathrm{H}}$ scales linearly with field strength.  It can be important when the dominant charge carriers of one sign are magnetized---meaning that their collision rate with the neutrals is less than their rate of gyration around field lines---while those of the other sign are not.  At inferred field strengths, this tends to be the case at intermediate altitudes in the disk: i.e. above the midplane, but below the FUV dissociation front where disk winds may be launched \citep{Wardle2007}, and $\eta_{\mathrm{H}}$ may be important even where subdominant \citep{Wurster2021}.
Unlike the other two diffusivities, the Hall term contributes neither to $\bm{E'\cdot B}$ nor to $\bm{E'\cdot J}$ and is therefore not directly dissipative. But it does contribute to flux migration, yielding an $\bm{E'\times B}$ drift along $-\bm{J}_\perp$.

\subsubsection{\textbf{Magnetic field}}
Once the gas is coupled to magnetic fields, the magnetic field strength and topology become a key control parameters to the dynamics of the system. It is usual to quantify the field strength with the dimensionless plasma $\beta\equiv 8\pi P/B^2$ parameter. {  While it is in principle possible to generate poloidal field loops {\it in situ}  by dynamo action, such a dynamo remains difficult to sustain in the regions $R>1\,\mathrm{AU}$ because of the strong magnetic diffusion \citep{Simon.bai.ea13b,Riols2021}. Hence, the} large-scale poloidal field threading the disk $B_\mathrm{p}$ (that is to say averaging out the turbulent part) is conserved: i.e it can only be accreted or diffused away from the central star during the system lifetime. As a result, it is often the key control parameter of MHD processes in disks, quantified in terms of $\bp\equiv 8\pi P_\mathrm{mid}/B_p^2$, where we have used the mid-plane pressure $P_\mathrm{mid}$. While it is a useful parameter from a theoretical standpoint, it is also useful to connect it to physical values. Using a ``typical'' T-tauri disk with $\Sigma=200\,\mathrm{g.cm}^{-2}R_\mathrm{AU}^{-1}$, one gets $B_p\simeq10\,R_\mathrm{AU}^{-11/8}\bp^{-1/2}\,\mathrm{G}$ \citep[eq. 2.4]{Lesur21} . Hence, the typical value expected in disks with $\bp\sim 10^4$ is $B_p\sim 5\,\mathrm{mG}$ at 10$\,\au$. If the large-scale field dominates angular momentum transport, and hence a MHD wind is launched (\S\ref{sec:mhd:mhdw}), it can be shown that the total field $B$ is tightly linked to the accretion rate. It is then possible to deduce a minimum field strength for a given accretion rate: $B\gtrsim 0.31\,R_\mathrm{AU}^{-5/4} (\Macc/10^{-7}M_\odot.\mathrm{yr}^{-1})^{1/2}\,\mathrm{G}$ \citep{Bai2009}. Note that this lower limit should be taken with care given that in most MHD winds, $B$ is dominated by its toroidal component, which changes sign on both sides of the disk, and hence is relatively difficult to measure directly from optically thin tracers.

\subsection{MRI-driven turbulence }\label{sec:mhd:mri}

The magneto-rotational instability \citep{1991ApJ...376..214B} plays an important role for the accretion processes and overall for the protoplanetary disk evolution. Since the last 30 years, many works have investigated the MRI in the context of protoplanetary disks, using theoretical models \citep{Turner.Fromang.ea14} and lab experiments \citep{2018PhR...741....1R}. From our current understanding, we believe that in the inner (T $>$ 1000 K) and outer protoplanetary disk regions ($\Sigma \leq 15 \rm \, g cm^{-2}$), the conditions for the MRI are met, which are: a weak magnetic field $\bp \ge 1$ \citep[eq. 6.64]{Lesur21}, shear and high enough ionization (\S\ref{sec:mhd:framework}).

 In this section we highlight the most important achievements since the last PPVI, more specifically, the convergence of the MRI under a net-vertical field, the MRI with realistic thermodynamics, and the MRI in the regime from ideal to non-ideal MHD using local and global simulations.

\subsubsection{\textbf{Quasi-Ideal regime}}
In the quasi-ideal MHD regime (that is to say when diffusive length-scales are much smaller than $H$) and with a net-flux vertical magnetic field present, the saturation of the MRI is reached by the super-exponential growth of parasitic instabilities, halting the further growth of the MRI \citep{2015ApJ...802..139M,2018ApJ...853..174H}. Such conditions of saturated MRI activity are present in the inner and outer regions of protoplanetary disks, and many global simulations confirmed the efficiency of the MRI using a net-vertical magnetic field \citep{2014MNRAS.441.2078P,2014ApJ...784..121S,2017ApJ...835..230F,Zhu.Stone18}. Fully saturated MRI turbulence shows an angular momentum transport coefficient between $\ass =0.01$ and $1.0$ depending on the strength and topology of the initial magnetic field. For instance, \cite{Salvesen16} found $\ass\simeq 11\bp^{-0.53}$ for $10<\bp<10^5$ in stratified shearing box simulations threaded by a vertical field, indicating that $\ass$ is roughly proportional to $B_p$. While this result holds in "ideal" MHD, in protoplanetary disks we expect the magnetic Prandtl number $\mathrm{Pm}$, the ratio between the viscous and the magnetic diffusion rate, to be very small, even in the regions where ideal MHD applies. This small $\mathrm{Pm}$ regime was investigated using non-stratified high-resolution simulations, reaching for the first time convergence \citep{2015A&A...579A.117M} with a value of $\ass = 3.2 \times 10^{-2}$ and a net horizontal magnetic flux with $\beta=10^3$. In contrast, zero-net flux MRI simulations in the low $\mathrm{Pm}$ limit show turbulent decay \citep{2020ApJ...904...47M} and no sustained turbulence. Further studies are needed to understand the characteristics of the MRI at different levels of $Pm$ and different magnetic configurations.

\subsubsection{\textbf{Non-ideal regime}}
Moving radially outward in the protoplanetary disk, the non-ideal terms of Ohmic, Hall and Ambipolar diffusion become more important (\S\ref{sec:mhd:framework}), efficiently reducing the level of MRI turbulence \citep{2015ApJ...798...84B,2018ApJ...865...10S} down to levels of several $\sim 10^{-4}$ for $\ass$. For these non-ideal terms, many disk parameters play now an important role, since magnetic diffusivities (\ref{eq:etaOHA}) depend on the magnetic field strength, local density of gas and dust and the local high-energetic radiation field of X-Ray,Cosmic ray and even FUV \citep{2018ApJ...869...84G}. These non-ideal MHD regions are prone to magnetic flux concentrations, triggering zonal flows which appear as axisymmetric long-lived perturbation and which appear in the Ohmic \citep{2014ApJ...796...31B}, Hall \citep{2013MNRAS.434.2295K} and Ambipolar regimes \citep{2014ApJ...784...15S}. At the transition between dead and active zones, these perturbations and flux concentrations are enhanced \citep{2015A&A...574A..68F} halting the radial drift of pebbles \citep{2016A&A...590A..17R}. The study of the non-ideal MHD evolution of the disk has become more important in the recent years, especially due to the ability to cause variations in the surface density. Future studies should extend the parameter regime of the non-ideal terms and investigate their importance for the magnetic evolution of the disk.

\subsubsection{\textbf{Thermodynamics}}
In recent years, local and global simulations investigated the effect of combined radiative transfer and magnetohydrodynamics, capturing the important heating and cooling processes together with the MRI using radiation MHD simulations. In optical thin regimes of protoplanetary disks, the temperature is mostly defined by the irradiation heating, and the MRI strength adapts to the local profile of temperature and pressure, as shown by \citep{2013A&A...560A..43F} using global radiation magnetohydrodynamical simulations of protoplanetary disks including irradiation. In the optically thick regime \citep{2014ApJ...787....1H,2018A&A...620A..49S} the heating by the MRI leads to an increase of the disk temperature and so to the disk scale height which again enhances the turbulence and leads to convection. Most of the MHD heating is taking place in the magnetic reconnection layers \citep{2018MNRAS.477.3329R} for which in general high resolution is required \citep{2014ApJ...791...62M} to resolve the heating rate. In protoplanetary disks the inner gas disk is believed to remain mostly optically thin for its own thermal radiation \citep{2017ApJ...835..230F} however a detailed set of gas opacities remain under investigation. Especially the inner gas disk, between the silicate sublimation line and the co-rotation radius, depends heavily on the heating and cooling processes and so the gas opacity includes various molecular lines \citep{2014A&A...568A..91M}. At the same time, MRI activity in this region controls the last stage of the accretion process, transporting matter onto the star while at the same time feeding the magnetically driven winds in this inner region. Future studies of the thermodynamics and the MRI should improve the gas opacity to show the importance of the heating and cooling processes, especially for the evolution of the inner, MRI-active, gas disk.

\subsubsection{\textbf{Outlook}}
The saturation level, the strength and the appearance of MRI turbulence dependents on many disk parameters. Most important here are the local degree of ionization and the large-scale magnetic field $B_p$ threading the disk, as they determine the appearance and the strength of MRI activity, both in the ideal and non-ideal regimes. Future studies should focus on the non-ideal terms, the long-term magnetic flux evolution and the proper thermodynamics to study the importance of the MRI for the disk evolution. Solving for detailed thermodynamics is crucial not only to obtain realistic levels of turbulence but also to enable synthetic observations. Heating by the non-ideal MHD terms \citep{2019ApJ...872...98M} is significant and should be captured in future studies.

\subsection{Outflows}

Outflows have been proposed since the 1990s to explain the atomic jets observed in several YSOs \citep{Frank.Ray.ea14}. The distinction between jets and winds is relatively unclear in the literature, but it is usually implicitly linked to the outflow collimation and velocity: while jets are assumed to be fast (100-1000km/s) and well collimated, and therefore appear narrow at large distances, winds are usually seen as slow (1-30km/s) conical outflows. These slow outflows are sometimes called ``molecular`` since they are detected in molecular lines. {  Here, we make the distinction between two kinds of outflows: thermally driven outflows (photoevaporation), for which the energy source is primarily an external radiation field (which could be due to the central star or the immediate environement), and magnetised outflows, which are driven by accretion energy.

\subsubsection{\textbf{Thermal outflows}}\label{sec:mhd:thermal}

Thermal (photoevaporative) winds are covered in detail in the Chapter by Pascucci et al., where current efforts to detect these winds in gas line observations are also reviewed. While significant progress has been done on theoretical modeling of these winds since PPVI , the broad-brush picture of how thermal winds work and their effect on the surface density evolution of disks has been largely confirmed. Recent calculations, however, have highlighted some important discrepancies in the prediction of wind properties, which will need to be resolved before the models can be used to quantitatively interpret observations, as detailed in the Chapter by Pascucci et al.

In the simple picture gas can only be thermally unbound, if it is heated to temperatures that approach the local escape temperature, $T_{\rm esc}=GM_*\bar m/r k_{\textsc{b}}$, where $\bar m$ is the mean mass per gas particle; this is $\approx 10^4$~K for $M_*=M_\odot$, $r=1\textsc{au}$, and $\bar m=0.68 m_p$ (ionized hydrogen at solar abundance).  Depending on the irradiating spectrum, gas is heated to different temperatures. As an example, Extreme-Ultraviolet (EUV) radiation ($13.6eV < E < 100 eV$) heats the gas mainly by photoionization of hydrogen and helium, generally yielding an almost isothermal gas with temperature around $10^4\,\mathrm{K}$. For geometrically thin, non self-gravitating disks around a 1$M_{\odot}$ star, the radius at which the escape temperature of the gas equals $10^4\,\mathrm{K}$ is approximately $1\,\au$. The mass loss profile, in this case, is peaked around the so-called gravitational radius of the system, which is the cylindrical radius at which the sound speed of the heated gas equals the Keplerian orbital speed \citep{Alexander2014}.  and has a total rate of about 10$^{-10}M_{\odot}/yr$ assuming an EUV flux of 10$^{41} phot/sec$ \citep{Alexander2006a, Alexander2006b}, as demonstrated by isothermal hydrodynamical simulations. Soft X-ray radiation ($100 eV < E < 1k eV$) is not efficient at ionizing hydrogen directly, but it ejects inner-shell electrons from abundant metals (e.g. Carbon and Oxygen); the ejected photoelectrons have suprathermal energies and result in secondary ionizations. This yields a more weakly ionized gas of a range of temperatures from a few thousand to $10^4\,\mathrm{K}$. Two-dimensional radiation-hydrodynamic models yield mass loss profiles that are more extended with respect to the EUV case and total mass loss rates of order 10$^{-8}M_{\odot}/yr$, assuming a soft X-ray flux of 10$^{30} erg/sec$ \citep{Owen2010, Picogna2019}, or even higher for Carbon-depleted disks \citep{Woelfer2019}. The dominant gas heating mechanism for FUV photons ($E < 13.6 eV)$) is photoelectric heating from dust grains, which is proportional to the grain surface. Thus in the presence of abundant small grains or PAHs in the disk atmosphere, FUVs can efficiently heat the gas and may also contribute to drive a thermal wind, as suggested by hydrostatic 1+1D models \citep[e.g.]{Gorti2009}. However, uncertainties in the atmospheric abundance of PAHs, which are rarely observed in T-Tauri disks \citep{Seok2017}, and the variable nature of the FUV flux, which is linked to the accretion rate onto the central object, make it difficult to assess the role of FUV-driven winds on the final disk dispersal.

 Recent calculations of thermal winds have combined two-dimensional (axisymmetric) hydrodynamics with explicit, though simplified, treatments of the chemical and thermal state of the gas under the influence of these radiative processes \citep{Wang.Goodman17,Nakatani+2018a,Nakatani+2018b}.  These works find that X-rays alone are less effective at driving photoevaporation than EUV \citep{Wang.Goodman17} or FUV \citep{Nakatani+2018b}, except at very low gas metallicities or if some cooling processes are suppressed, contrary to \cite{Owen2010}.  However, the former assumed harder X-ray spectra than the latter: hard ($E\ge 1$~keV) X-rays penetrate more deeply than softer ones, thereby warming a larger volume of gas to a lower temperature, possibly $<T_{\rm esc}$, for the same luminosity.  \cite{Wang.Goodman17} and \cite{Nakatani+2018b} also differ among themselves as to the relative importance of EUV and FUV.  Clearly, more work is needed, but it might be helpful to propose a fiducial set of radiation fields and disk structures against which different groups could test their codes.}

\subsubsection{\textbf{MHD outflows}} \label{sec:mhd:mhdw}

The lack of any systematic process to drive turbulence in protoplanetary disks has revived the interest in the magnetised outflow (MO) scenario, especially in the weakly ionized regions, following an idea initially proposed by \cite{Wardle.Koenigl93} and subsequently developed by \cite{Konigl2010} and \cite{Salmeron2011}. These MOs launched from weakly ionized disks received more support from numerical simulation, first in shearing box models. \cite{Bai.Stone13} first found that a magnetically dead disk dominated by Ohmic and ambipolar diffusion could still emit a MO thanks to the ionized layer localized at the disk surface. This result was quickly generalized to the outer, more turbulent, regions ($R>30$ AU, \citealt{Simon.Bai.ea13a}) and to models including the Hall effect \citep{Lesur.Kunz.ea14}. Despite their importance, these early shearing box models suffer from an oversimplification of the gravitational potential at large $z$ \citep{Turner.Fromang.ea14} which makes them unreliable to study outflows. Hence, since PPVI, a lot of effort has been devoted to confirming these results in global models, which do not suffer from these drawbacks.

The first generation of global simulations including non-ideal MHD effects \citep{Gressel.Turner.ea15,Bethune.Lesur.ea17,Bai17} demonstrated that these outflows were not a mere artifact of the shearing box model, and could indeed explain accretion rates of the order of $10^{-8}\,M_\odot/\mathrm{yrs}$, assuming a disk surface density $\Sigma\simeq 10\,\mathrm{g.cm}^{-2}$ at $R=10\au$ with a disk magnetization $\bp\simeq 10^5$. It was also found that the mass loss rate in the wind $\Mwind=2\pi \int \,R\mathrm{d}R \zeta \Sigma \Omega_K$ (see eq.~\ref{eq:acc_mass}) was of the order of the mass accretion rate $\Macc$. A way to interpret these high mass ejection rates is to use the locally defined ejection efficiency
\begin{align*}
\xi=\frac{1}{\Macc}\frac{\mathrm{d}\Mwind}{\mathrm{d}\log R}
\end{align*}
to compare ejection to accretion rates. In steady state, and assuming a constant $\xi$, it is easy to show that
\begin{align*}
\Mwind=\Macc(R_\mathrm{in})\left(\left(\frac{R_\mathrm{out}}{R_\mathrm{in}}\right)^\xi-1\right)
\end{align*}
for a wind-emitting disk with inner radius $R_\mathrm{int}$ and outer radius $R_\mathrm{out}$ ({  these two radii can in principle be arbitrarily chosen in the wind-emitting region, but we will assume here that they correspond to the radial extension of the wind-emitting region)}. It can be shown that if accretion is the only energy source for the outflow, then energy conservation requires $\xi<1$ \citep[eq. 11.40]{Ferreira.Pelletier95,Lesur21}. This limit is usually exceeded in the first generation of simulations ($\xi\sim 1$ in \citealt{Bethune.Lesur.ea17}, $\xi=1.5$ in \citealt{Bai17}), indicating that these outflows cannot be driven solely by accretion energy, and that another energy source must contribute to the outflow. Since all of these early solutions included some sort of prescribed coronal heating, it was quickly realized that these outflows could be of magneto-thermal origin (i.e. the outflow is a mixture of photo-evaporation and MHD wind).

\subsubsection{\textbf{Magneto-thermal winds}}
The very high mass loading in these simulations triggered the search for more accurate models of the wind thermodynamics, including some of the processes involved in photo-evaporation (\ref{sec:mhd:thermal}). Using this type of approach, \cite{Wang.Bai.ea18} found ejection indices slightly lower than in the first global models, with $\xi=0.5-1$, a result also recovered by \cite{Gressel.Ramsey.ea20}. Hence, the thermal driving turned out to be less important than anticipated in earlier models. \cite{Rodenkirch2020} showed solutions of a hot magneto-thermal wind with the transition to a photoevaporative dominated wind for very small magnetic fields for $\beta > 10^7$. Using numerical self-similar solutions, \cite{Lesur21a} also found solutions with $\xi\sim 0.5$ without any coronal heating, confirming that massive outflows with $\xi\lesssim 1$ are mostly a result of the low magnetizations (high $\beta_p$) required to match the accretion rates expected in YSOs than a thermal effect. Overall, since
$\xi\sim 1$ in all of the MO simulations published to date with $\bp\gg 1$, one expects $\Mwind\gtrsim\Macc$ if $R_\mathrm{out}/R_\mathrm{in}\gtrsim 2$ (i.e. if the wind is launched from a significant fraction of the disk surface). Hence, very massive outflows should be the rule, more than the exception, from a theoretical standpoint.

The fact that these outflows are very massive also has an impact on their kinematics and observational signature. For instance, it is customary to measure the amount of angular momentum extracted by a MO by its lever arm $\lambda$ (see eq.\ref{eq:acc_ang}). Theoretically, $\lambda$ can be related to $\xi$ through $2\xi\simeq 1/(\lambda-1)$ \citep[eq. 11.40]{Lesur21}. Hence, the outflows emitted in magneto-thermal models all have $\lambda\simeq 1.5$.
Because $\lambda$ is a direct measure of the angular momentum "excess", it can also be deduced by measuring the azimuthal and poloidal velocity components of a given outflow \citep{Anderson.Li.ea03} which gives a relatively direct test of these models on observed jets and molecular winds (ref to I. Pascucci's chapter)

Because of these very low $\lambda$ and high $\xi$, these magneto-thermal outflows are relatively far in parameter space from the magneto-centrifugal mechanism of \cite{Blandford.Payne82} from which previous accretion-ejection solutions have been derived \citep{Wardle.Koenigl93,Ferreira97}. Indeed, these early solutions all had $\bp \simeq 1$, with $\lambda\gg 1$ and $\xi \ll 1$ and lead to sonic accretion speeds which makes them incompatible with the known accretion rate and surface density of PPDs. As pointed out by \cite{Bai.Ye.ea16}, the new $\bp\gg 1$ solutions are closer to the magnetic tower scenario of \cite{Lynden-Bell03} since they are mostly driven by the warping of the toroidal magnetic field at the disk ionized surface. However, it is worth noting that MOs exist at all $\bp$ and make a continuum between these two extreme limits \citep{Jacquemin-Ide.Ferreira.ea19,Lesur21a}, hence MHD-driven outflows should be much more general in accreting systems than thought in the past.

\subsubsubsection{\textbf{Outlook}}
Finally, if MOs are key in the angular momentum transport of PPDs, it also implies that our understanding of their secular evolution should be revised. First and foremost, if the mass accretion rate $\Macc$ is driven by the outflow stress (eq.~\ref{eq:acc_ang}), then $\Macc$ is not proportional to the density gradient as in the viscous disk model. As a result, such a disk might not spread radially as efficiently as viscous disks (though see \citealt{yang21}) nor fill density gaps spontaneously. Related to this question, there are hints that planet migration (rate and direction) is probably affected by MOs \citep{McNally20,Kimmig.Dullemond.ea20} though these early results have to be confirmed by models including self-consistent MOs. This will likely have measurable consequences on planet population synthesis models and therefore predicted planet populations (see chapter led by S-J Paardekopper).

Since MOs are structures driven by the large-scale field threading the disk, the strength of this field becomes crucial to predict the evolution of the system. Indeed, all of the models to date show that the mass loss $\zeta$ and lever arm $\lambda$ are decreasing functions of $\bp$. As a result, the mass accretion and ejection rates are partially controlled by the local field strength, which in principle can evolve with time. As a result, the secular evolution of PPDs might be dictated by the evolution of their magnetic flux. One of the consequences is that it becomes possible to create large cavities in a disk as a result of magnetic field concentration in the inner disk regions, leading for instance to an alternative model of transition disks (see \S\ref{sec:obs:winds}).

\begin{figure}[h!]
\centering
\includegraphics[width=0.95\linewidth]{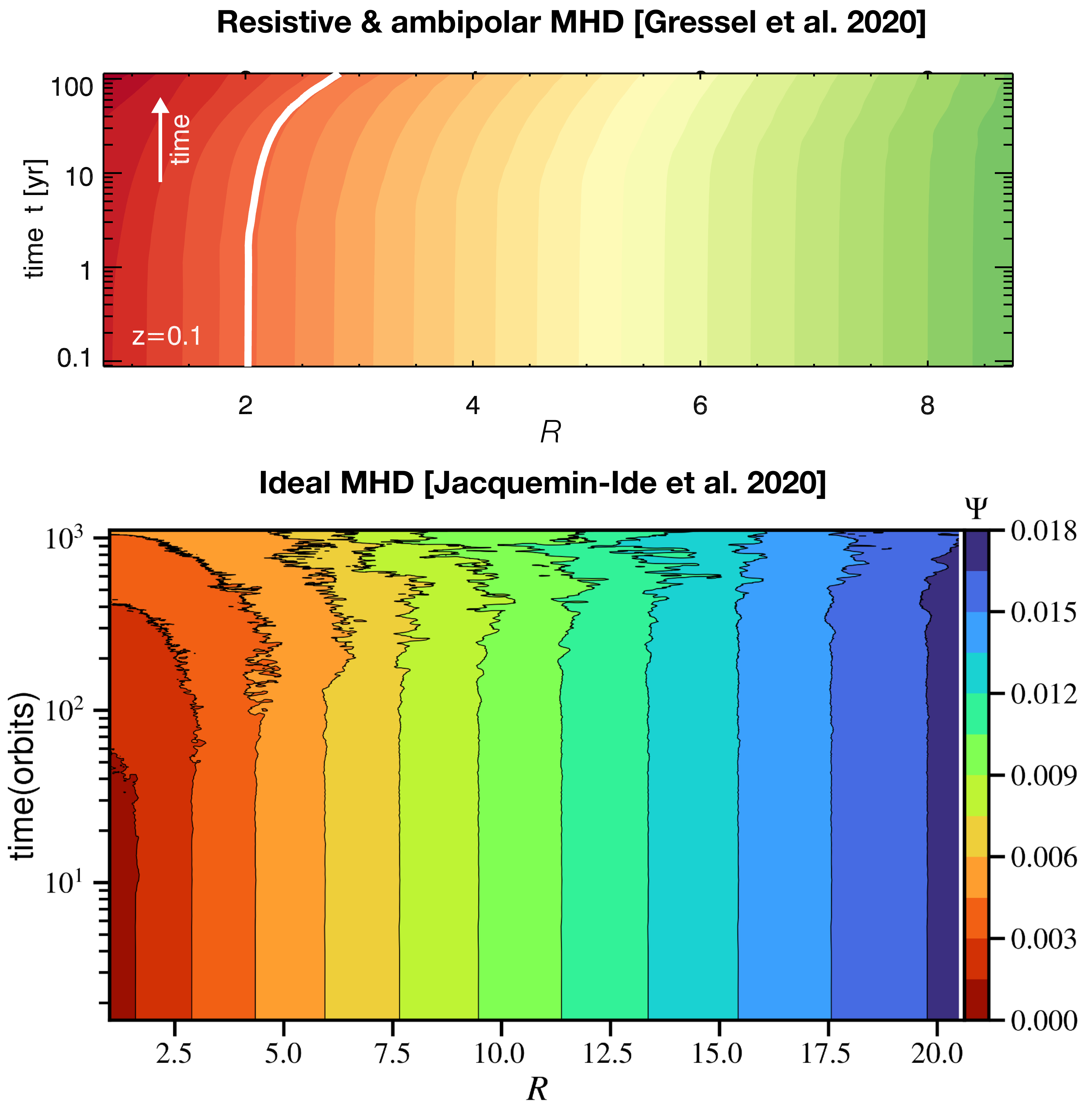}
\caption{\label{fig:field-transport}Temporal evolution of the magnetic flux function with respect to radius in a non-ideal simulation (top, \citealt{Gressel.Ramsey.ea20}) and in an ideal (bottom, from \citealt{Jacquemin-Ide.Lesur.ea20}), both taken in the disk midplane. The isosurface represent the footpoint of magnetic field lines. Field lines diffuse outwards in the non-ideal regime of protoplanetary disks ($R>1\,\mathrm{AU}$) but are advected inwards in the ideal MHD regime.}
\end{figure}

These prospects are tightly linked to the question of the large-scale magnetic field strength and transport in these disks, which is still largely unknown and poorly described. From a theoretical point of view, there is no consensus on how this large-scale field should evolve. Global simulations in the non-ideal MHD regime \citep{Bai.Stone16,Gressel.Ramsey.ea20,Lesur21a} suggest that the large-scale field tend to diffuse outwards for $R\gtrsim 1$ AU, with a velocity increasing with the field strength, but some theoretical models suggest the field can be transported inwards \citep{Leung.Ogilvie19}. In the ideal MHD regime, which should be valid for $R\lesssim 1 AU$ in PPDs (\S\ref{sec:mhd:framework}), the magnetic field is inversely transported inwards \citep{Zhu.Stone18,Jacquemin-Ide.Lesur.ea20}, at a speed that also increases with the field strength. Hence, the behavior of disks in the non-ideal and ideal regimes regarding field transport appears to be opposite (fig.~\ref{fig:field-transport}). If this picture is confirmed, it would suggest the formation of an inner strongly magnetized disk with an outer disk that slowly loses its initial magnetic field. There is little doubt that such a scenario would in turn impact planet formation mechanisms (e.g. \citealt{Suzuki.Ogihara.ea16}), but it still needs to be confirmed.

\subsection{Instability transition zones and interaction with the wind/dust }\label{sec:mhd:transitions}

Since the astonishing observation of HL Tau in 2015 by the ALMA telescope we now know that protoplanetary disks usually display rings and gaps, and sometimes nonaxisymmetric structures such as spirals. Since then, many studies investigated for the nature of these perturbations which are seen both in the gas and the dust component. An important role is played here by the transition zones between different types of hydro or magneto-hydrodynamical instabilities, each of them leading to a different level of turbulence and so different levels of accretion stress. These variations lead to perturbations in the gas and are often accompanied by secondary instabilities like the Rossby Wave Instability \citep[RWI,][]{Lovelace1999}. These perturbations in the gas surface density, lead to even larger perturbations for the solid material, especially for intermediate grain sizes whose radial drift is strongly modulated by the gas-pressure perturbations. In the following, we will review the most prominent transition zones in protoplanetary disks for which we expect long-lived perturbations in the gas and dust.

\begin{figure*}[th]
\centering
\includegraphics[width=\linewidth]{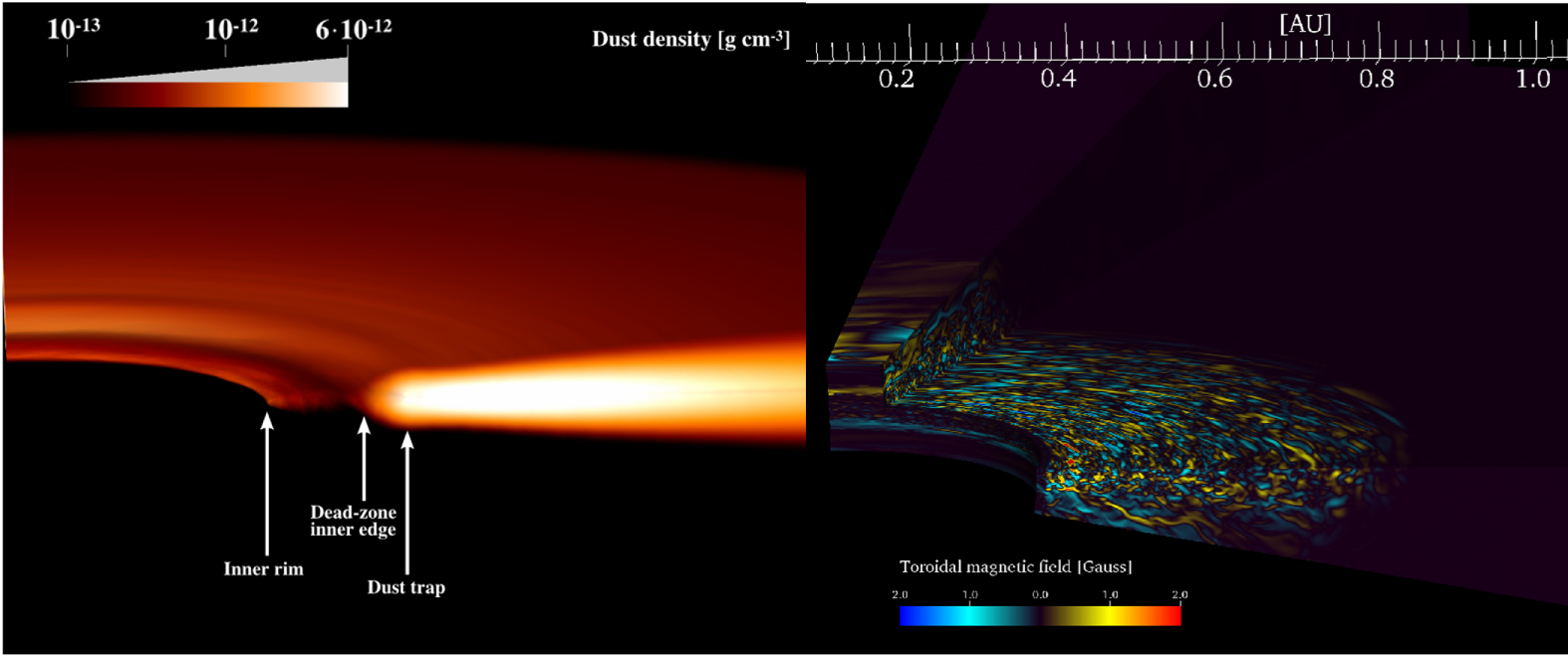}
\caption{\label{fig:iedge}3D global radiation non-ideal MHD simulations of the inner dead-zone edge for a disk around a Herbig star \citep{2017ApJ...835..230F} including irradiation heating and dust sublimation. Left: 3D volume rendering of the dust density.  Right: 3D contour plot of the toroidal magnetic field. Annotated are the location of the dust sublimation (0.47 au), the inner dead-zone edge (0.8 au), and the location of dust trapping (1au). On the right, the domain’s top half is cut away to show the magnetic fields on the midplane. The inner disk shows fully established turbulence by the magneto-rotational instability.}
\end{figure*}

\subsubsection{\textbf{MRI Dead-zone edges}}
The most prominent zone is the transition between the MRI-active and MRI-damped regions in protoplanetary disks. This transition region could lead to variations in the stress-to-pressure ratio $\alpha$ between 10 to 1000 dependent on the magnetic field strength, the ionization degree and the level of the hydrodynamical turbulence \citep{Turner.Fromang.ea14,2015A&A...574A..10L,2017ApJ...835..230F}. The resulting perturbation in the gas surface density often leads to secondary instabilities like the RWI which can generate vortices at these transition zones. Such vortex formations were observed at the inner dead-zone edge \citep{LyraMacLow12,2015A&A...573A.132F,2017ApJ...835..230F} at the outer dead-zone edge \citep{Regaly2012,2015A&A...574A..10L,2015A&A...574A..68F} and even for pure hydrodynamical instability transition zones \citep{flock20}. Those perturbations in the surface density and in the pressure, are ideal places for planet formation \citep{2014ApJ...780...53C} as they lead to the reduction of the radial drift or even directly to the trapping of dust grains \citep{2016A&A...596A..81P,2019ApJ...871...10U,2021MNRAS.504..280J,2022MNRAS.509.5974J}. This concentration of dust grains in a ring structure could be even followed by further dust concentration along the azimuthal direction inside the vortices caused by the RWI \citep{2016A&A...590A..17R,2017ApJ...835..118M}. Fig.~\ref{fig:iedge} shows the dust density and the magnetic field structure around the dead-zone inner edge for a protoplanetary disk around a Herbig type star.

The detailed location of the dead-zone inner edge lies between 0.05 and 0.1 au for a young T Tauri like star \citep{2018ApJ...861..144M,2019A&A...630A.147F} while it can reach 1 au for a Herbig star \citep{2016ApJ...827..144F} due to the greater stellar luminosity. In certain situations when the disk is very optically thick, the dead-zone inner edge could be even further outside, e.g. in the case of large accretion rate and strong accretion heating due to high optical thickness \citep{2018ApJ...861..144M,2019ApJ...881...56S,2021MNRAS.504..280J}.

The exact location of the outer dead-zone edge is more difficult to determine as it depends on the external ionization sources, the amount of dust and gas, and the dust size and porosity. Dependent on these parameters, the dead-zone outer edge usually lies between 10 to 100 au \citep{Turner.Fromang.ea14,2015A&A...574A..10L,2015A&A...574A..68F,2016A&A...596A..81P,delage22}.

\subsubsection{\textbf{Wind and magneto/hydro instabilities}}
Further radial variations in the surface density can be caused by the transition zones of the different hydrodynamical instabilities (\S\ref{S:hydro}). In the following, we want to emphasize the interaction between the wind-dominated upper layers of the protoplanetary disks and the magneto/hydrodynamical dominated midplane regions in the disk.

3D stratified simulations of the MRI showed both, vertical transport of energy (Poynting flux) and vertical transport of gas material induced by buoyant motions \citep{2014ApJ...780...46I,2014ApJ...784..121S} which could support the mass loading at the wind base. In such simulations it was found that the MRI causes a magnetized corona region in the disk with fast radial inflow motions, which can influence the structure of the overall disk wind \citep{2020MNRAS.498..750L}. This magnetized corona region together with a wind layer above was also found in recent 3D stratified ideal MHD simulations \citep{Zhu.Stone18,Jacquemin-Ide.Lesur.ea20}.

If magnetized winds dominate angular momentum transport in non-ideal regions, and therefore gas accretion, it is very likely that the disk spontaneously organizes into rings and gaps regularly spaced because of the non-linear dependence of the wind torque on the surface density \citep[see also chapter on structure formation in the same book]{Bethune.Lesur.ea17,Suriano19,Riols2020,Cui2021b}. In these regions, the VSI (\S\ref{vsi}) could also be at play. Here, the VSI is very efficient in the vertical mixing of gas and dust. In these non-ideal regimes, the VSI could be important for the mixing of gas and dust material into the wind layers \citep{cui20}.

\section{\textbf{PRESENT \& FUTURE TESTS OF THEORETICAL MODELS}}
\label{observations_section}
\begin{figure*}[th]
\centering
\includegraphics[width=\linewidth]{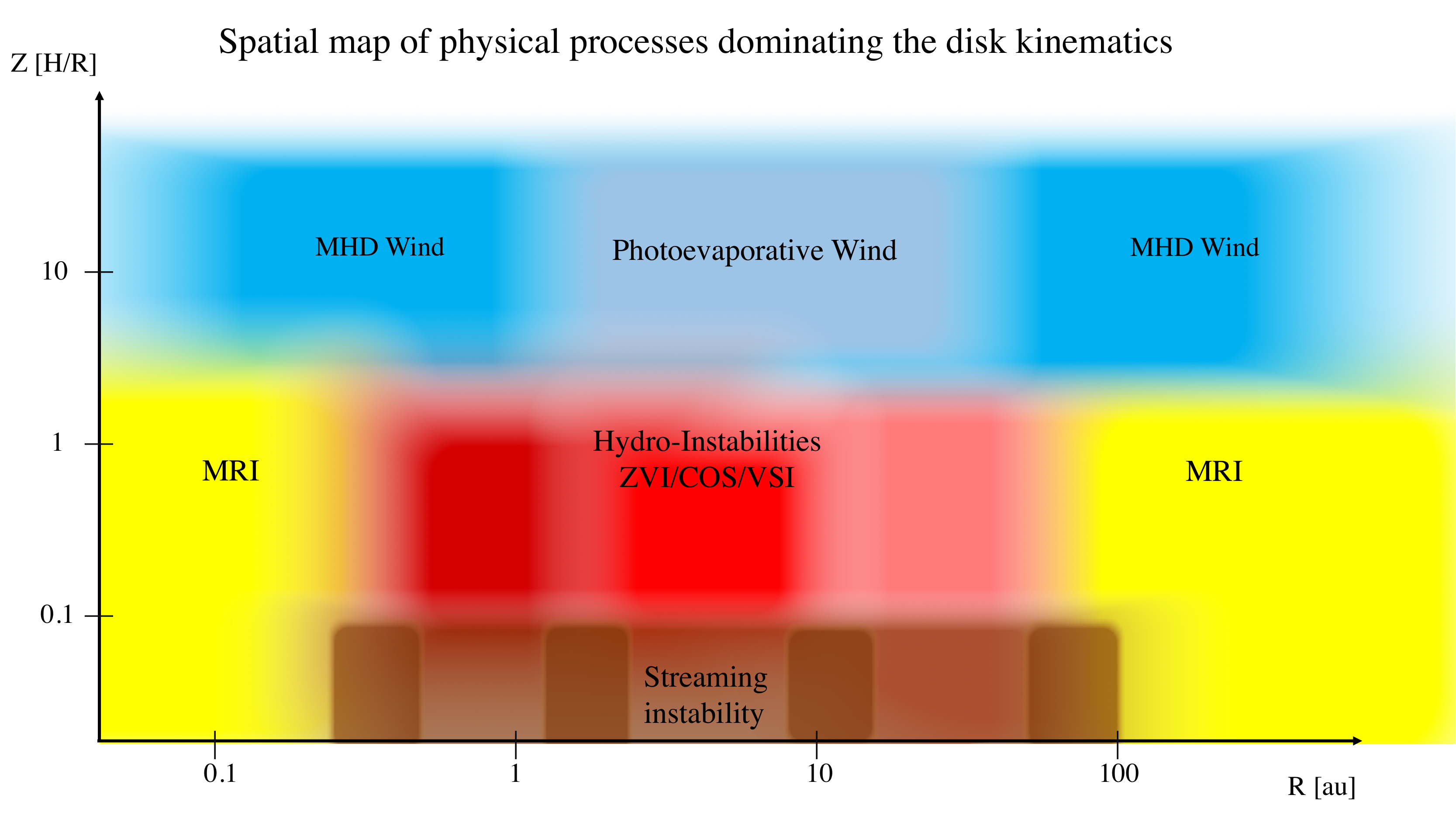}

\caption{\label{fig:occurance}Spatial map showing the dominant physical process for the dynamics and evolution of protoplanetary disks. We assume a shallow power law surface density profile in the non-self gravitating regime with $M_\mathrm{disk}\sim$ a few \% of $M_\mathrm{star}$. {  The ZVI, COS, and the VSI have distinct requirements on the disk structure and cooling times (\S\ref{hydro_application}). Thus their occurrence and distribution within the dead zone can be model-dependent.
}}
\end{figure*}
\subsection{Occurrence diagram}

Fig.~\ref{fig:occurance} presents the dominant physical processes which we currently believe control the dynamics and evolution of protoplanetary disks. Dependent on the local density of the gas and dust, the total dust surface area, the stellar and external irradiation, and the amount of magnetic flux, the spatial extent and the strength of these individual physical processes vary. {  As we have noted in \S\ref{sec:mhd:transitions} these instabilities interact with each other. The MRI can mix and transport gas and dust material into the wind layers. Between these different instability zones we expect pressure perturbation due to the different strength of the turbulence level. In the following we will discuss in more detail the structure of the different gas and dust dynamics shown in Fig.~\ref{fig:occurance}.}

\subsubsection{\textbf{Disk layer $H/R \lessapprox 4$}}
In the inner disk regions, the conditions for MRI activity are met between the disk co-rotation radius and right behind the silicate sublimation zone. We expect a second MRI-active region in the outermost disk regions which are ionized by high energy cosmic rays and or non-thermal radiations. Vertically, the MRI-active regions extend to several $H$, and is found to be more vigorous in the disk corona layers above 2 $H$ (\S \ref{sec:mhd:mri}). Behind the silicate sublimation radius, where temperatures decrease below 900 K, the gas ionization fraction drops, damping the MRI activity as the influence of the MHD non-ideal terms increases (\S\ref{sec:mhd:framework}). In this regime of high optical thickness, hydrodynamical instabilities, more specifically the ZVI, the COV and the VSI could become active (\S\ref{zvi}, \S\ref{cov}, \S\ref{vsi}). The radial and vertical spatial extent of these hydrodynamical instabilities depend strongly on the local opacity which sets the timescale of the radiative transfer and so the heating and cooling in the disk (\S\ref{hydro_application}). We highlight that the opacity and the ionization degree depend both on the local amount and size of the dust material embedded in the gas disk. Dust-drag instabilities occur mostly in regions with a local dust-to-gas mass ratio of order unity. One promising candidate of these dust drag instabilities is the streaming instability, operating close to the disk midplane between the silicate sublimation zone and the outer disk regions undergoing strong vertical mixing (\S\ref{sec:dust:streaming}). The vertical spatial extent of the SI is focused on a region around one-tenth of a pressure scale height above the midplane, and its overall activity assumes a modest to small amount of vertical mixing ($\ass \le 10^{-4}$) by the hydrodynamical instabilities. We expect modulations of the surface density between the different instability regions because of variations of the accretion stress. Depending on the strength of these surface density modulations, we foresee dust concentrations to occur due to the reduction of the radial drift velocity (\ref{eq:drift_vel}). In these regions, we expect an enhanced SI activity or even gravitational instability of the dust layer (\S\ref{sec:dust:streaming}, \S\ref{sec:dust:sgi}) . The surface density modulations can also trigger the growth of the Rossby-Wave-Instability leading to vortex formation and further dust concentration (\S\ref{sec:mhd:transitions}).

\subsubsection{\textbf{Wind layer $H/R \gtrapprox 5$}}
MHD driven wind activity starts at the upper disk layers, in which the coupling between the gas and the magnetic field becomes stronger (FUV layer) and for which the magnetic field strength controls the motion of the gas ($\beta \sim 1$, \S\ref{sec:mhd:mhdw}). The radial extent of the MHD wind could in principle cover the whole disk, depending on the strength and distribution of the magnetic flux. The strength of the photoevaporative driven wind can be stronger or weaker, dependent on the X-Ray, FUV and EUV fluxes (\S\ref{sec:mhd:thermal}). The photoevaporative effects become important outside 1 astronomical unity (for approximately a 1~$M_{\odot}$ star), where the influence of the stellar gravitational potential can be easier be overcome. The magneto- and hydrodynamical instabilities, the photoevaporative and magnetically-driven winds all together shape the evolution of the protoplanetary disk, while their individual strength and significance might vary during the lifetime of the disk.

\subsection{Turbulence}
\label{sec:obs:turbulence}
\begin{figure*}
\centering
\includegraphics[width=0.4\linewidth]{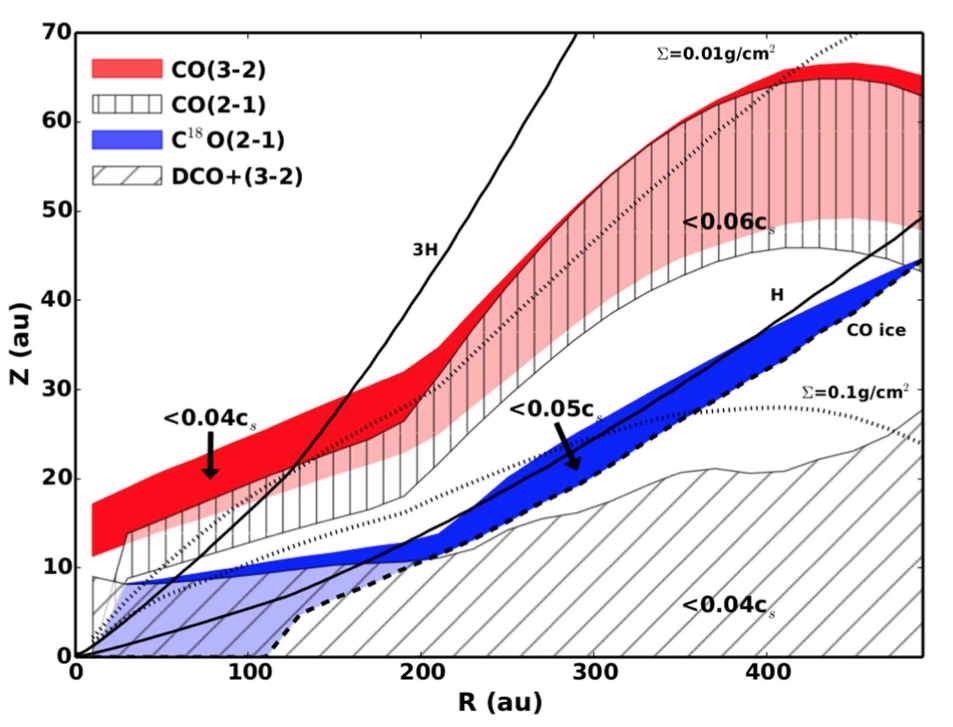}
\includegraphics[width=0.4\linewidth]{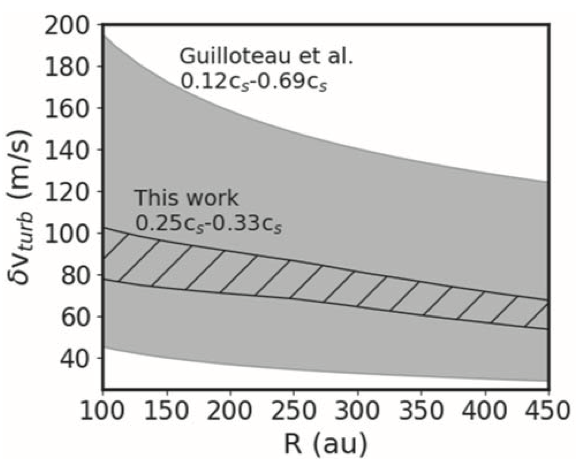}
\label{fig:turbulent_constraints}
\caption{{\it Left:} Upper limits to turbulent velocities in HD~163296 as a function of radius $R$ and mid-plane height $z$. The colors denote the species from which there is a majority of emission; the CO transitions are at large $z$ whereas C$^{18}$O and DCO+ are from lower in the disk. Also included on the plot are lines of constant $\Sigma$, $H$ and 3$H$ ($H$ being the gas scale height) and where CO is ice. In this source, turbulence is at most a few percent of the sound speed. From \cite{flaherty17}. {\it Right:} The turbulent velocity as a function of radius as measured from CO emission in DM Tau (hashed lines), compared to the results from previous work using CS (\citealt{Guilloteau2012}; grey shaded region).  As opposed to HD~163296, DM Tau exhibits strong turbulence, consistent with theoretical predictions \cite{Flaherty2020}.}
\end{figure*}

The notion of gas disks being rendered turbulent has recently been tested with a number of independent observational efforts. The most direct observational constraints on turbulence come from the turbulent broadening of molecular gas emission lines. Most recently, a series of studies quantified the strength of turbulence at large radial distances ($> 30$~AU) in nearby protoplanetary disks using multiple molecular species to probe different heights above the mid-plane (see Pinte et al. chapter for details). These observations generally constrain $\ass < 10^{-4}$--$10^{-3}$ for four out of five sources for which this analysis was carried out \citep{Flaherty2015,Teague2016,flaherty17,Flaherty2018}.  In the remaining source (DM Tau), $\ass \sim 0.01$--0.5 \citep{Guilloteau2012,Flaherty2020}.  A comparison of a disk with a weak turbulence level throughout (HD 163296) and the strongly turbulent DM Tau disk is shown in Fig.~\ref{fig:turbulent_constraints}

While these recent studies were largely constrained to the outer disk ($>30$AU), earlier work by \cite{Carr2004} used band head emission of CO lines originating from upper disk regions and in the inner $0.1$AU of SVS 13. They found a normalized turbulent velocity $\vturb \sim 1$ in these regions, in agreement with theoretical predictions \citep{Gammie1996,Fromang2006c,Simon2011,Simon2015b}\footnote{A constraint on $\ass$ cannot be made here because $\ass$ is effectively the height-integrated turbulence, and these observations probed only the upper disk layers.}.

While the most direct measurement of disk turbulence is line broadening, there are two other approaches to independently quantify the level of turbulence in disks.  The first is quantifying the diffusion of dust grains due to turbulence; since dust grains are (at least somewhat) coupled to gas motions, this approach provides an indirect, yet independent constraint. Assuming a standard dust settling model, \cite{Pinte2016} constrained $\ass$ to be a few $10^{-4}$ in HL Tau. Focusing instead on the radial structure of the rings \cite{Dullemond2018} found a {\it lower} limit of $\ass \approx 10^{-4}$ for a handful of sources, though this precise value is dependent on the poorly constrained grain sizes and width of the gaseous pressure bumps.  \cite{Ohashi2019} conducted radiative transfer modeling that included self-scattering polarization to constrain the grain size and distribution in HD~163296; they found that while $\ass \lesssim 1.5\times10^{-3}$ in one of the gaps (at 50 AU), $\ass$ could be as large as 0.3 in the 90 AU gap.

These constraints are generally consistent with modeling the SED and comparing with those from a sample of Herbig stars, T Tauri stars, and brown dwarfs, as was done by \cite{Mulders2012}. They found $\ass \sim 10^{-4}$ assuming a standard grain size distribution (as well as a constant $\ass$ throughout the disk), though their observations were also consistent with $\ass \sim 10^{-2}$ if the grain size distribution had fewer small grains or a lower gas-to-dust ratio.

Since radial angular momentum transport is likely driven by non-thermal motions (though, these do not have to be turbulent per se; e.g., spiral density waves), the degree of viscous spreading of disks can serve as another indirect constraint on turbulence. The primary quantification of viscous spreading is a measurement of the outer edge of the disk as a function of disk age and there have been several attempts to put constraints on $\ass$ based on such a measurement, both in terms of the dust disk (e.g. \citealt{Rosotti2019}) and the gas disk (e.g., \citealt{Ansdell2018,Trapman2019}).

 \citep{Najita2018} examined a number of disk size measurements from the literature, and found that generally, younger Class I disks were more compact than their Class II counterparts, lending support to the notion of viscous angular momentum transport.  \cite{Ansdell2018} measured the radius that contains 90\% of the CO J=2-1 flux in Lupus and found a large range of $\ass$ values, ranging from $3\times10^{-4}$ to 0.09. \cite{Trapman2019} followed up on this work and concluded that using the 90\% CO flux technique leads to an overestimate of $\ass$ values, finding instead that disks in Lupus are consistent with viscous spreading with $\ass \approx 10^{-4}$--$10^{-3}$.

There are a number of complications that warrant caution in interpreting these results on viscous spreading.  For instance, in the case
of measuring disk sizes from the dust, the radial drift of particles has to be accounted for, particularly for mm-sized grains (such as observed with ALMA in the continuum) as these grains are only marginally coupled to the gas at large radial distances. Furthermore, as pointed out by \cite{Rosotti2019}, it is possible that existing surveys of dust disks lack the sensitivity to detect viscous spreading.  Finally, in the case of the gas disk, changes to abundances for chemical tracers (such as CO) have to be accounted for as well as processes, such as photoevaporation \citep{Winter2020}.

Taken together, all of these studies are consistent with $\ass \sim 10^{-4}$. However, caution is warranted here as well. First, the precise value of $\ass$ is still subject to considerable uncertainty, either due to variations from source to source (e.g., DM Tau \citealt{Guilloteau2012,Flaherty2020}) or limitations of the methods employed (e.g., \citealt{Rosotti2019}). Furthermore, it is inaccurate to strictly equate turbulent velocities and $\ass$, as is done in both the molecular line and dust diffusion measurements; both of these approaches assume $\ass \propto v_{\rm turb}^2$ (see Pinte et al. chapter for instance).  However, for angular momentum to be transported outward, the radial and azimuthal turbulent velocities must be correlated.  For instance, considering only the Reynolds stress, $\ass \propto \delta v_r \delta v_{\phi}$, which is not in general equivalent to $v_{\rm turb}^2$.\footnote{On the other hand, it is appropriate to say that $\adust \propto v_{\rm turb}^2$, as this definition quantifies the dust diffusion (see Section~\ref{sec:intro:dust}). However, we are restricting ourselves to a discussion of the magnitude of turbulent angular momentum transport in this section.} Thus, a disk can be strongly turbulent, while transporting angular momentum outward, inward, or not at all.

These uncertainties aside, $\ass = 10^{-4}$ is too small for angular momentum transport to be completely dominated by turbulent/viscous processes by about two orders of magnitude (eq.\ref{eq:mdot}).  Thus, at face value, this suggests that some other mechanism (e.g., MHD winds, \S\ref{sec:mhd:mhdw}) are also contributing to angular momentum transport, a conclusion that is consistent with theoretical models and numerical simulations that state MHD winds should dominate the accretion process (e.g., \citealt{Gressel.Turner.ea15,Bethune.Lesur.ea17,Bai17}).
Ultimately, however, with ALMA continuing its full operations well into the coming decade, observational constraints on turbulence in disks will only improve with time. Even resolving the large-scale turbulent eddies directly might become possible with ALMA using spatially and spectrally resolved line emission \citep{barraza21}. Such observations of the line emission will help to identify the physical mechanism generating the turbulent motions in protoplanetary disks.

\subsection{Winds}
\label{sec:obs:winds}
Disk winds have been detected through small blue-shifts (a few km/s) in a number of emission lines from molecular, atomic, low-ionization species (see the chapter by Pascucci et al and \citet{Ercolano2017}) and there have been several theoretical efforts aiming at modeling the observations to constrain wind properties and infer their driving mechanisms \citep[e.g.][]{Ercolano2010, Ercolano2016, Weber2020, Nemer2020, Gressel.Ramsey.ea20}. Despite the focussed observational and theoretical effort in this field in the last five years, several important questions remain open, in particular, the relative occurrence of thermal (photoevaporative) and MHD-driven winds is still unknown.

The determination of the mass loss rate and radial profile is crucial to constrain the nature of the disk wind, as thermal winds are not expected to be launched very close to the star, where gravity is strong and the escape temperatures are high. The determination of the emission radius of a (blueshifted) line profile by assumption of Kepler broadening is problematic, given the velocity gradients within the wind \citep{Weber2020}. For this reason, alternatives are currently being explored to directly observe and test disk winds. Two promising avenues are provided by (i) scattered light observations of high inclination objects and (ii) future high-angular resolution radio observations (e.g. with the nGVLA).

Several theoretical efforts have focused on the entrainments of dust grains in thermal and MHD disk winds \citep{Owen2011b, Hutchison2016a, Hutchison2016b, Miyake2016, Franz2020, Hutchison2021, Booth2021}. These models predict significant differences in the grain sizes that could be entrained in thermal and MHD winds as well as in the spatial extent of the wind that would be populated by grains of different sizes. These might be observable signatures that may help identifying the nature of the winds, particularly for objects with high inclinations. \citet{Franz2020} conclude that the detection of winds in scattered light might be challenging with current instrumentation, but it may still be possible for disks where small grains are well mixed with the gas in the disk atmosphere. Indeed a possible detection of a dusty disk wind may have occurred in the scattered light observations of My Lup presented by \citet{Avenhaus2018}.

Spatially-resolved observations of free-free emission at cm wavelengths, which could be obtained by future radio telescopes, such as the \textit{Next Generation Very Large Array} (ngVLA), present a novel and exciting avenue to constrain the nature of disk winds. \citet{Ricci2021} demonstrate that for the case of TW~Hydra, theoretical models of X-ray photoevaporation \citep{Picogna2019} and simple analytical MHD wind models \citep{Weber2020} predict very different radial profiles for the cm emission, which could be resolved by the ngVLA.

Disks with inner cavities, so-called "Transition disks", have also been studied intensively, as they may be objects on the verge of dispersal through photoevaporative winds, particularly in the case where the measured accretion rate on the central object is smaller than $10^{-8} M_{\odot}/yr$. Disks with large cavities that are still vigorously accreting, are difficult to be explained by photoevaporative models \citep{Ercolano2017} and could be the result of planet-disk interactions or formed by magnetic winds \citep{Combet.Ferreira08,Wang.Goodman17}. This last scenario is appealing since large cavity transition disks tend to exhibit accretion rates comparable to standard disks, despite the presence of a large and deep cavity \citep{vanderMarel.Williams.ea18}, indicating transonic accretion velocities in the cavity. In the future, deep ALMA observations will allow the analysis of bright molecular lines that are optically thick and originate in the warm surface layers of disks. This will allow measuring $v(R)$ using the 1st velocity moment of spectral datacubes \citep[e.g.][]{Casassus2021} and explore the kinematical properties of the intracavity gas of these objects.

\subsection{Solar system bodies} \label{sec:obs:ss}
Planetesimals in the Solar System -- remnants of the planet formation stage more than 4.5 billion years ago -- provide an extensive ground for comparing simulations of turbulence and planetesimal formation to the observed outcome in the Solar System. The asteroid belt is dominated in mass by the largest bodies (Ceres, Vesta and Pallas). These may in fact already be considered to be protoplanets that underwent some degree of accretional growth after their formation \citep{JM15}. Vesta differentiated and separated its iron core from a rocky mantle, while Ceres formed later and experienced only heating enough to melt the ice to form phyllosilicates.

Asteroids with a characteristic radius of approximately 100 km constitute a second mass peak in the asteroid belt. \cite{Bottke+etal2005} first suggested that this size could be the peak of the birth size distribution of planetesimals and that smaller asteroids are fragments of this birth population, while the larger sizes are the result of collisional growth \citep{morbidelli09}. The Initial Mass Function (IMF) of planetesimals formed by the streaming instability was evolved numerically by mutual collisions and pebble accretion in \cite{JM15}. This study showed indeed that the asteroids between 100 and 500 km in size may result from a combination of pebble accretion and mutual collisions within a birth size distribution peaking at 100 km. The accretion mode transitions to become dominated by pebble accretion once the bodies reach 1000 km in diameter. The smallest of these bodies are akin to Vesta and Ceres which could thus represent bodies on the threshold to undergoing rapid growth towards planetary bodies.

It has been proposed that small asteroids are completely dominated by fragments of larger asteroids, with no discernible contribution from primordial planetesimals for sizes less than 45 km \cite{Delbo+etal2017}. This observation is in some tension with the IMF of planetesimals which appears to follow a power law towards smaller masses \citep{JM15,SA16,SA17}. However, more recently \cite{LYS19} used very high resolution simulations to demonstrate that the IMF may turn down from a power law at small sizes, so that the IMF approaches more a log-normal shape with tapering both at the small and at the large \citep{SYJ17} end. \cite{Schreiber2018} proposed that turbulence caused by the streaming instability itself suppresses fragmentation on scales below the characteristic scale of the filament. Future simulations at very high resolution or using adaptive mesh refinement techniques will be needed to shed light on the shape of the IMF for bodies down to 10 km.

The size distribution of the Kuiper belt has a power law similar to the asteroid belt for small sizes, while the largest bodies in the cold classical belt have a much steeper size distribution \citep{Fraser+etal2014}. The latter could be the original exponential tapering from the birth process \citep{AS19}.

The binarity of planetesimals is another imprint of the formation process. The cold classicals have a binary fraction of at least 20\% \citep{Noll+etal2008} and another 20\% are contact binaries \citep{ThirouinSheppard2019}. \cite{NYR10} showed in N-body simulations of collapsing pebble clouds how such clouds would contract to form binary planetesimals (see Section~\ref{sec:dust:planets} for more details). This appears to be a unique prediction of planetesimal formation from overdense clumps of pebbles (which could form by the streaming instability or in a pressure bump). The formation of contact binaries is poorly understood but could be a result of gas drag and Kozai-Lidov cycles for initially highly inclined binary orbits \citep{Naoz+etal2010,mckinnon20,LYJ21}. The cold classical Kuiper belt contact binary Arrokoth was visited by the New Horizons mission in 2019 and the perfect shape of the two components indicates collision speeds lower than a few m/s \citep{mckinnon20}.

Another prediction of planetesimal formation by the streaming instability is that planetesimals that avoid heating and catastrophic disruption should survive as piles of primordial pebbles from the solar protoplanetary disk (\citealt{WJ14,WJ17}; see Section~\ref{sec:dust:planets} for more details). The Rosetta mission to the comet 67P/Churyumov-Gerasimenko found a number of indications that this is the case: the interior structure is homogeneous on scales less than a few meters while the density is low consistent with porosity at the pebble scale \cite{LG16}, (2) the size distribution of dust at the surface appears to peak in the mm-cm size range and the outflow of material from the surface has a broader distribution from mm sizes (interpreted to be primordial pebbles) to 10 m chunks of likely consolidated pebble-rich material \citep{blum17}.

Evidence from debris disks is harder to interpret. Classical cascade models start typically with small planetesimals that fragment faster \cite{KrivovWyatt2021}. However, 100-km-sized pebble piles may also be consistent with the observed time-scales of dust production \cite{Krivov+etal2018}.

\section{\textbf{CONCLUSIONS AND PERSPECTIVES}}

In this chapter, we reviewed the multifaceted theories aimed at describing planet-forming disks, from the clumping of tiny dust grains to the launching of large-scale winds.  We covered various instabilities and feedback processes whose implications for planet formation are only beginning to become clear.

The innermost region of the disk near the star is well-ionized and must be strongly turbulent because of the MRI (\S\ref{sec:mhd:mri}), but this is not necessarily the case for the outer ($R\gtrsim 1\au$) regions (\S\ref{sec:mhd:framework}), which are spatially-resolved for instance with ALMA.  For these outer regions, predicting the dynamics and the strengths of the MRI and other instabilities requires knowing how the ionization state varies with position.
Even if the MRI is suppressed, several thermally-driven instabilities have been established as potential sources of hydrodynamic turbulence and structure formation in PPDs (\S\ref{S:hydro}). Their basic properties, including the underlying physical mechanisms and onset requirements, are understood. The main outstanding issues are whether and how these instabilities operate under realistic PPD conditions, and what kinds of turbulence or other flows they trigger (in terms of $\ass$ and $\adust$ for instance). To this end, radiation hydrodynamic simulations should be developed, treating heating and cooling processes consistently with the disk structure -- key factors that determine whether the instabilities will operate and how they will evolve. This endeavor should be carried out paying close attention to observational constraints, which now include direct probes of turbulence in the disk bulk (\S\ref{sec:obs:turbulence}).

Turbulence and other flows in the gas in turn impact dust coagulation, since planetesimal formation now appears to involve an interplay between linear instability, nonlinear saturation, and triggers of strong clumping of solids (\S\ref{section_dust}). Much progress has been made on the conditions for triggering strong clumping, which depend on dust size, solid-to-gas column density ratio, and radial pressure gradient (\S\ref{sec:dust:streaming}).  There has been progress also on the resulting planetesimal mass functions with self-gravity included in the numerical simulations (\S\ref{sec:dust:planets}).
While earlier studies focused on mono-disperse dust size distributions, investigation of the streaming instability with a spread of dust sizes and of the interaction of planetesimal formation with external turbulence have just begun, yet already offer some surprises such as the quenching of the linear streaming instability in some cases (\S\ref{SSS:silin}). In the near future, global simulations should be one of the major fronts in advancing this topic.

Global-scale magnetic fields are also expected in PPDs as relicts from the systems' formation (\S\ref{sec:mhd:framework}).  These can dominate the large-scale gas dynamics by launching thermo-MHD disk winds (\ref{sec:mhd:thermal}-\S\ref{sec:mhd:mhdw}). In contrast to earlier wind models, these can operate with weak global magnetic fields while accounting for the accretion rates observed in class 2 objects, even where the disks are otherwise non-turbulent. A major uncertainty here is the field strength, which is not predicted by any complete theory, and for which we have few empirical constraints. Modeling the long-term evolution of the large-scale field and better measuring it, either through astrophysical observations or by analyzing meteorites from the early solar system, should be focuses of future research.

The combined outcome of these diverse, dynamic processes is likely to be a disk that is no longer smooth and monotonically-varying with distance from the star.  Radial variations in the accretion stress, due to transitions in the ionization state, the main heating and cooling processes, and the magnetic field transport will inevitably lead to structure in the gas surface density profile and so the pressure (\S\ref{sec:mhd:transitions}). Local pressure variations naturally have outsized impacts on the dust surface density profile through the spatial gradients they induce in the dust-gas drift speed (\S\ref{sec:intro:dust}). From the apparent ubiquity and complexity of these interactions, it is clear that models of disk evolution, planet formation, and planet-disk interaction can advance only if we move away from viscous, alpha-type disk models, which attempt to gloss over fundamental physical and chemical ingredients. The next generation of models will properly treat  turbulence, dust transport, and wind-driven accretion, either explicitly or through sub-grid or effective models.  Such sub-grid models may be based on simulations at higher resolution or including more processes over a limited time or space domain. Modeling efforts furthermore should be expanded to more closely consider disks at various evolutionary stages around stars of various masses, to advance our understanding of how disks' diversity shapes the planet population across our Galaxy.

\section*{\textbf{ACKNOWLEDGMENTS}}
GL acknowledges support from the European Research Council (ERC) under the European Union Horizon 2020 research and innovation program (Grant agreement No. 815559 (MHDiscs)). BE acknoweldges support from DFG Research
Unit  "Transition discs" (FOR 2634/1, ER 685/8-2) and from the Excellence Cluster ORIGINS
which is funded by the Deutsche Forschungsgemeinschaft
(DFG, German Research Foundation) under Germany's
Excellence Strategy - EXC-2094 - 390783311. MF acknowledges funding from the European Research Council (ERC) under the European Union’s Horizon 2020 research and innovation program (grant agreement No. 757957). MKL is supported by the Ministry of Science and Technology of Taiwan (grants 107-2112-M-001-043-MY3, 110-2112-M-001-034-, 110-2124-M-002-012-, 111-2124-M-002-013-) and an Academia Sinica Career Development Award (AS-CDA-110-M06). CCY acknowledges the support from NASA via the Emerging Worlds program (grant number 80NSSC20K0347) and via the Astrophysics Theory Program (grant number 80NSSC21K0141). JB and PM acknowledge grant NSF AST-1510708 and computational resources from multiple XSEDE awards. JJG  was supported by NASA grant 17-ATP17-0094. AJ acknowledges funding from the European Research Foundation (ERC Consolidator Grant 724687-PLANETESYS), the Knut and Alice Wallenberg Foundation (Wallenberg Scholar Grant 2019.0442), the Swedish Research Council (Project Grant 2018-04867), the Danish National Research Foundation (DNRF Chair Grant DNRF159) and
the Göran Gustafsson Foundation. WL, JBS, OMU, CCY, and ANY acknowledge the support from NASA via the Theoretical and Computational Astrophysics Networks program (grant number 80NSSC21K0497). WL acknowledges grant NSF AST-2007422, and computational resources from XSEDE through allocation TG-AST140014. JBS acknowledges support from NASA under Emerging Worlds through grant 80NSSC20K0702. NJT contributed at the Jet Propulsion Laboratory, California Institute of Technology, under contract with NASA and with the support of NASA grant 17-XRP17\_ 2-0081. JS acknowledges support from the Royal Society Te Ap\=arangi, New Zealand through Rutherford Discovery Fellowship RDF-U001804. OMU acknowledges NASA’s New Horizons Kuiper Belt Extended Mission and NASA ISFM.
\bibliography{fullbib_final}

\begin{thebibliography}{401}
\parskip=0pt \itemsep=0pt \small \baselineskip=11pt
\providecommand{\natexlab}[1]{#1}

\bibitem[\protect\astroncite{\emph{{Abod} et~al.}}{2019}]{AS19}
{Abod} C.~P. et~al., 2019 \emph{\apj}, \emph{883}, 2, 192.

\bibitem[\protect\astroncite{\emph{{Adachi} et~al.}}{1976}]{Adachi1976}
{Adachi} I. et~al., 1976 \emph{Progress of Theoretical Physics}, \emph{56},
  1756.

\bibitem[\protect\astroncite{\emph{{Akimkin} et~al.}}{2020}]{Akimkin2020}
{Akimkin} V.~V. et~al., 2020 \emph{\apj}, \emph{889}, 1, 64.

\bibitem[\protect\astroncite{\emph{{Alexander} et~al.}}{2014}]{Alexander2014}
{Alexander} R. et~al., 2014 \emph{Protostars and Planets VI} (H.~{Beuther},
  R.~S. {Klessen}, C.~P. {Dullemond}, and T.~{Henning}), p. 475.

\bibitem[\protect\astroncite{\emph{{Alexander}
  et~al.}}{2006{\natexlab{a}}}]{Alexander2006a}
{Alexander} R.~D. et~al., 2006{\natexlab{a}} \emph{\mnras}, \emph{369}, 1, 216.

\bibitem[\protect\astroncite{\emph{{Alexander}
  et~al.}}{2006{\natexlab{b}}}]{Alexander2006b}
{Alexander} R.~D. et~al., 2006{\natexlab{b}} \emph{\mnras}, \emph{369}, 1, 229.

\bibitem[\protect\astroncite{\emph{Anderson et~al.}}{2003}]{Anderson.Li.ea03}
Anderson J.~M. et~al., 2003 \emph{\apjl}, \emph{590}, L107.

\bibitem[\protect\astroncite{\emph{{Andrews} et~al.}}{2009}]{Andrews2009}
{Andrews} S.~M. et~al., 2009 \emph{\apj}, \emph{700}, 2, 1502.

\bibitem[\protect\astroncite{\emph{{Ansdell} et~al.}}{2018}]{Ansdell2018}
{Ansdell} M. et~al., 2018 \emph{\apj}, \emph{859}, 1, 21.

\bibitem[\protect\astroncite{\emph{{Arakawa} and {Krijt}}}{2021}]{Arakawa2021}
{Arakawa} S. and {Krijt} S., 2021 \emph{\apj}, \emph{910}, 2, 130.

\bibitem[\protect\astroncite{\emph{{Arlt} and {Urpin}}}{2004}]{arlt04}
{Arlt} R. and {Urpin} V., 2004 \emph{\aap}, \emph{426}, 755.

\bibitem[\protect\astroncite{\emph{{Armitage}}}{2011}]{Armitage11}
{Armitage} P.~J., 2011 \emph{\araa}, \emph{49}, 1, 195.

\bibitem[\protect\astroncite{\emph{{Auffinger} and
  {Laibe}}}{2018}]{Auffinger2018}
{Auffinger} J. and {Laibe} G., 2018 \emph{\mnras}, \emph{473}, 1, 796.

\bibitem[\protect\astroncite{\emph{{Avenhaus} et~al.}}{2018}]{Avenhaus2018}
{Avenhaus} H. et~al., 2018 \emph{\apj}, \emph{863}, 1, 44.

\bibitem[\protect\astroncite{\emph{{Bai}}}{2014}]{Bai2014_Hall}
{Bai} X.-N., 2014 \emph{\apj}, \emph{791}, 2, 137.

\bibitem[\protect\astroncite{\emph{{Bai}}}{2015}]{2015ApJ...798...84B}
{Bai} X.-N., 2015 \emph{\apj}, \emph{798}, 2, 84.

\bibitem[\protect\astroncite{\emph{Bai}}{2017}]{Bai17}
Bai X.-N., 2017 \emph{\apj}, \emph{845}, 1, 75.

\bibitem[\protect\astroncite{\emph{{Bai} and {Goodman}}}{2009}]{Bai2009}
{Bai} X.-N. and {Goodman} J., 2009 \emph{\apj}, \emph{701}, 1, 737.

\bibitem[\protect\astroncite{\emph{{Bai} and {Stone}}}{2010}]{BS10c}
{Bai} X.-N. and {Stone} J.~M., 2010 \emph{\apj}, \emph{722}, 2, 1437.

\bibitem[\protect\astroncite{\emph{Bai and Stone}}{2010}]{Bai2010}
Bai X.-N. and Stone J.~M., 2010 \emph{\apjs}, \emph{190}, 2, 297.

\bibitem[\protect\astroncite{\emph{{Bai} and {Stone}}}{2010}]{BS10b}
{Bai} X.-N. and {Stone} J.~M., 2010 \emph{\apjl}, \emph{722}, 2, L220.

\bibitem[\protect\astroncite{\emph{Bai and Stone}}{2013}]{Bai.Stone13}
Bai X.-N. and Stone J.~M., 2013 \emph{\apj}, \emph{769}, 1, 76.

\bibitem[\protect\astroncite{\emph{{Bai} and
  {Stone}}}{2014}]{2014ApJ...796...31B}
{Bai} X.-N. and {Stone} J.~M., 2014 \emph{\apj}, \emph{796}, 1, 31.

\bibitem[\protect\astroncite{\emph{{Bai} and {Stone}}}{2017}]{Bai.Stone16}
{Bai} X.-N. and {Stone} J.~M., 2017 \emph{\apj}, \emph{836}, 1, 46.

\bibitem[\protect\astroncite{\emph{Bai et~al.}}{2016}]{Bai.Ye.ea16}
Bai X.-N. et~al., 2016 \emph{\apj}, \emph{818}, 2, 152.

\bibitem[\protect\astroncite{\emph{{Baines} et~al.}}{1965}]{Baines1965}
{Baines} M.~J. et~al., 1965 \emph{\mnras}, \emph{130}, 63.

\bibitem[\protect\astroncite{\emph{{Balbus} and
  {Hawley}}}{1991}]{1991ApJ...376..214B}
{Balbus} S.~A. and {Hawley} J.~F., 1991 \emph{\apj}, \emph{376}, 214.

\bibitem[\protect\astroncite{\emph{{Balbus} and
  {Terquem}}}{2001}]{Balbus+Terquem2001}
{Balbus} S.~A. and {Terquem} C., 2001 \emph{\apj}, \emph{552}, 1, 235.

\bibitem[\protect\astroncite{\emph{{Barge} et~al.}}{2016}]{Barge2016}
{Barge} P. et~al., 2016 \emph{\aap}, \emph{592}, A136.

\bibitem[\protect\astroncite{\emph{{Barker} and {Latter}}}{2015}]{barker15}
{Barker} A.~J. and {Latter} H.~N., 2015 \emph{\mnras}, \emph{450}, 21.

\bibitem[\protect\astroncite{\emph{{Barranco}}}{2009}]{barranco09}
{Barranco} J.~A., 2009 \emph{\apj}, \emph{691}, 907.

\bibitem[\protect\astroncite{\emph{{Barranco} and {Marcus}}}{2005}]{barranco05}
{Barranco} J.~A. and {Marcus} P.~S., 2005 \emph{\apj}, \emph{623}, 1157.

\bibitem[\protect\astroncite{\emph{{Barranco} et~al.}}{2018}]{barranco2018}
{Barranco} J.~A. et~al., 2018 \emph{\apj}, \emph{869}, 2, 127.

\bibitem[\protect\astroncite{\emph{{Barraza-Alfaro} et~al.}}{2021}]{barraza21}
{Barraza-Alfaro} M. et~al., 2021 \emph{\aap}, \emph{653}, A113.

\bibitem[\protect\astroncite{\emph{{Ben{\'\i}tez-Llambay}
  et~al.}}{2019}]{BKP19}
{Ben{\'\i}tez-Llambay} P. et~al., 2019 \emph{\apjs}, \emph{241}, 2, 25.

\bibitem[\protect\astroncite{\emph{B{\'e}thune
  et~al.}}{2017}]{Bethune.Lesur.ea17}
B{\'e}thune W. et~al., 2017 \emph{\aap}, \emph{600}, A75.

\bibitem[\protect\astroncite{\emph{{Birnstiel} and
  {Andrews}}}{2014}]{Birnstiel2014}
{Birnstiel} T. and {Andrews} S.~M., 2014 \emph{\apj}, \emph{780}, 2, 153.

\bibitem[\protect\astroncite{\emph{{Birnstiel} et~al.}}{2009}]{Birnstiel2009}
{Birnstiel} T. et~al., 2009 \emph{\aap}, \emph{503}, 1, L5.

\bibitem[\protect\astroncite{\emph{Birnstiel et~al.}}{2012}]{Birnstiel2012}
Birnstiel T. et~al., 2012 \emph{\aap}, \emph{539}, A148.

\bibitem[\protect\astroncite{\emph{{Bischoff} et~al.}}{2020}]{Bischoff2020}
{Bischoff} D. et~al., 2020 \emph{\mnras}, \emph{497}, 3, 2517.

\bibitem[\protect\astroncite{\emph{Blandford and
  Payne}}{1982}]{Blandford.Payne82}
Blandford R.~D. and Payne D.~G., 1982 \emph{\mnras}, \emph{199}, 883.

\bibitem[\protect\astroncite{\emph{{Blum}}}{2018}]{Blum2018}
{Blum} J., 2018 \emph{\ssr}, \emph{214}, 52.

\bibitem[\protect\astroncite{\emph{Blum and Wurm}}{2008}]{Blum2008}
Blum J. and Wurm G., 2008 \emph{Annual Review of Astronomy and Astrophysics},
  \emph{46}, 1, 21.

\bibitem[\protect\astroncite{\emph{{Blum}
  et~al.}}{2017{\natexlab{a}}}]{Blum2017}
{Blum} J. et~al., 2017{\natexlab{a}} \emph{\mnras}, \emph{469}, S755.

\bibitem[\protect\astroncite{\emph{{Blum} et~al.}}{2017{\natexlab{b}}}]{blum17}
{Blum} J. et~al., 2017{\natexlab{b}} \emph{\mnras}, \emph{469}, S755.

\bibitem[\protect\astroncite{\emph{{Booth} and {Clarke}}}{2021}]{Booth2021}
{Booth} R.~A. and {Clarke} C.~J., 2021 \emph{\mnras}, \emph{502}, 2, 1569.

\bibitem[\protect\astroncite{\emph{{Booth} et~al.}}{2018}]{Booth2018}
{Booth} R.~A. et~al., 2018 \emph{\mnras}, \emph{475}, 1, 167.

\bibitem[\protect\astroncite{\emph{{Bottke} et~al.}}{2005}]{Bottke+etal2005}
{Bottke} W.~F. et~al., 2005 \emph{\icarus}, \emph{175}, 1, 111.

\bibitem[\protect\astroncite{\emph{Brauer et~al.}}{2008}]{Brauer2008}
Brauer F. et~al., 2008 \emph{\aap}, \emph{480}, 3, 859.

\bibitem[\protect\astroncite{\emph{{Burke} and
  {Hollenbach}}}{1983}]{Burke_Hollenbach_1983}
{Burke} J.~R. and {Hollenbach} D.~J., 1983 \emph{\apj}, \emph{265}, 223.

\bibitem[\protect\astroncite{\emph{Carr et~al.}}{2004}]{Carr2004}
Carr J.~S. et~al., 2004 \emph{\apj}, \emph{603}, 1, 213.

\bibitem[\protect\astroncite{\emph{{Carrera} et~al.}}{2015}]{CJD15}
{Carrera} D. et~al., 2015 \emph{\aap}, \emph{579}, A43.

\bibitem[\protect\astroncite{\emph{{Carrera} et~al.}}{2021}]{CS21}
{Carrera} D. et~al., 2021 \emph{\aj}, \emph{161}, 2, 96.

\bibitem[\protect\astroncite{\emph{{Casassus} et~al.}}{2019}]{Casassus2019}
{Casassus} S. et~al., 2019 \emph{\mnras}, \emph{483}, 3, 3278.

\bibitem[\protect\astroncite{\emph{{Casassus} et~al.}}{2021}]{Casassus2021}
{Casassus} S. et~al., 2021 \emph{arXiv e-prints}, arXiv:2104.08379.

\bibitem[\protect\astroncite{\emph{{Chatterjee} and
  {Tan}}}{2014}]{2014ApJ...780...53C}
{Chatterjee} S. and {Tan} J.~C., 2014 \emph{\apj}, \emph{780}, 1, 53.

\bibitem[\protect\astroncite{\emph{{Chen} and {Lin}}}{2020}]{chen20}
{Chen} K. and {Lin} M.-K., 2020 \emph{\apj}, \emph{891}, 2, 132.

\bibitem[\protect\astroncite{\emph{{Chiang}}}{2008}]{chiang08}
{Chiang} E., 2008 \emph{\apj}, \emph{675}, 1549-1558.

\bibitem[\protect\astroncite{\emph{{Chiang} and {Youdin}}}{2010}]{cy10}
{Chiang} E. and {Youdin} A.~N., 2010 \emph{Annual Review of Earth and Planetary
  Sciences}, \emph{38}, 493.

\bibitem[\protect\astroncite{\emph{{Chiang} and {Goldreich}}}{1997}]{chiang97}
{Chiang} E.~I. and {Goldreich} P., 1997 \emph{\apj}, \emph{490}, 368.

\bibitem[\protect\astroncite{\emph{Clair et~al.}}{1970}]{Clair1970}
Clair B. P.~L. et~al., 1970 \emph{Journal of Atmospheric Sciences}, \emph{27},
  2, 308.

\bibitem[\protect\astroncite{\emph{Combet and
  Ferreira}}{2008}]{Combet.Ferreira08}
Combet C. and Ferreira J., 2008 \emph{\aap}, \emph{479}, 481.

\bibitem[\protect\astroncite{\emph{{Cui} and {Bai}}}{2020}]{cui20}
{Cui} C. and {Bai} X.-N., 2020 \emph{\apj}, \emph{891}, 1, 30.

\bibitem[\protect\astroncite{\emph{{Cui} and {Bai}}}{2021}]{Cui2021b}
{Cui} C. and {Bai} X.-N., 2021 \emph{\mnras}.

\bibitem[\protect\astroncite{\emph{Cui and Lin}}{2021}]{cui21}
Cui C. and Lin M.-K., 2021 \emph{Monthly Notices of the Royal Astronomical
  Society}, \emph{505}, 2, 2983.

\bibitem[\protect\astroncite{\emph{{Cuzzi} et~al.}}{1993}]{Cuzzi1993}
{Cuzzi} J.~N. et~al., 1993 \emph{\icarus}, \emph{106}, 102.

\bibitem[\protect\astroncite{\emph{{Cuzzi} et~al.}}{2001}]{CH01}
{Cuzzi} J.~N. et~al., 2001 \emph{\apj}, \emph{546}, 1, 496.

\bibitem[\protect\astroncite{\emph{{Cuzzi}
  et~al.}}{2014}]{Cuzzi_etal_2014ApJS..210...21C}
{Cuzzi} J.~N. et~al., 2014 \emph{\apjs}, \emph{210}, 2, 21.

\bibitem[\protect\astroncite{\emph{{Delage} et~al.}}{2022}]{delage22}
{Delage} T.~N. et~al., 2022 \emph{\aap}, \emph{658}, A97.

\bibitem[\protect\astroncite{\emph{{Delbo'} et~al.}}{2017}]{Delbo+etal2017}
{Delbo'} M. et~al., 2017 \emph{Science}, \emph{357}, 6355, 1026.

\bibitem[\protect\astroncite{\emph{{Desch} and
  {Cuzzi}}}{2000}]{Desch+Cuzzi2000}
{Desch} S.~J. and {Cuzzi} J.~N., 2000 \emph{\icarus}, \emph{143}, 1, 87.

\bibitem[\protect\astroncite{\emph{{Desch} and
  {Turner}}}{2015}]{Desch+Turner2015}
{Desch} S.~J. and {Turner} N.~J., 2015 \emph{\apj}, \emph{811}, 2, 156.

\bibitem[\protect\astroncite{\emph{{Desch} et~al.}}{2018}]{Desch2018}
{Desch} S.~J. et~al., 2018 \emph{\apjs}, \emph{238}, 1, 11.

\bibitem[\protect\astroncite{\emph{{Dewdney} et~al.}}{2009}]{Dewdney2009}
{Dewdney} P.~E. et~al., 2009 \emph{IEEE Proceedings}, \emph{97}, 8, 1482.

\bibitem[\protect\astroncite{\emph{{Dominik} et~al.}}{2007}]{Dominik2007}
{Dominik} C. et~al., 2007 \emph{Protostars and Planets V} (B.~{Reipurth},
  D.~{Jewitt}, and K.~{Keil}), p. 783.

\bibitem[\protect\astroncite{\emph{{Dubrulle} et~al.}}{1995}]{Dubrulle1995}
{Dubrulle} B. et~al., 1995 \emph{\icarus}, \emph{114}, 237.

\bibitem[\protect\astroncite{\emph{{Dullemond} et~al.}}{2018}]{Dullemond2018}
{Dullemond} C.~P. et~al., 2018 \emph{\apjl}, \emph{869}, 2, L46.

\bibitem[\protect\astroncite{\emph{Epstein}}{1923}]{Epstein1923}
Epstein P.~S., 1923 \emph{PhRv}, \emph{22}, 1.

\bibitem[\protect\astroncite{\emph{{Ercolano} and {Owen}}}{2010}]{Ercolano2010}
{Ercolano} B. and {Owen} J.~E., 2010 \emph{\mnras}, \emph{406}, 3, 1553.

\bibitem[\protect\astroncite{\emph{{Ercolano} and {Owen}}}{2016}]{Ercolano2016}
{Ercolano} B. and {Owen} J.~E., 2016 \emph{\mnras}, \emph{460}, 4, 3472.

\bibitem[\protect\astroncite{\emph{{Ercolano} and
  {Pascucci}}}{2017}]{Ercolano2017}
{Ercolano} B. and {Pascucci} I., 2017 \emph{Royal Society Open Science},
  \emph{4}, 4, 170114.

\bibitem[\protect\astroncite{\emph{{Estrada} et~al.}}{2016}]{Estrada_etal_2016}
{Estrada} P.~R. et~al., 2016 \emph{\apj}, \emph{818}, 2, 200.

\bibitem[\protect\astroncite{\emph{{Faure} et~al.}}{2015}]{2015A&A...573A.132F}
{Faure} J. et~al., 2015 \emph{\aap}, \emph{573}, A132.

\bibitem[\protect\astroncite{\emph{Ferreira}}{1997}]{Ferreira97}
Ferreira J., 1997 \emph{\aap}, \emph{319}, 340.

\bibitem[\protect\astroncite{\emph{Ferreira and
  Pelletier}}{1995}]{Ferreira.Pelletier95}
Ferreira J. and Pelletier G., 1995 \emph{\aap}, \emph{295}, 807.

\bibitem[\protect\astroncite{\emph{{Flaherty} et~al.}}{2020}]{Flaherty2020}
{Flaherty} K. et~al., 2020 \emph{\apj}, \emph{895}, 2, 109.

\bibitem[\protect\astroncite{\emph{{Flaherty} et~al.}}{2015}]{Flaherty2015}
{Flaherty} K.~M. et~al., 2015 \emph{\apj}, \emph{813}, 2, 99.

\bibitem[\protect\astroncite{\emph{{Flaherty} et~al.}}{2017}]{flaherty17}
{Flaherty} K.~M. et~al., 2017 \emph{\apj}, \emph{843}, 150.

\bibitem[\protect\astroncite{\emph{{Flaherty} et~al.}}{2018}]{Flaherty2018}
{Flaherty} K.~M. et~al., 2018 \emph{\apj}, \emph{856}, 2, 117.

\bibitem[\protect\astroncite{\emph{{Flock} and {Mignone}}}{2021}]{FM21}
{Flock} M. and {Mignone} A., 2021 \emph{\aap}, \emph{650}, A119.

\bibitem[\protect\astroncite{\emph{{Flock} et~al.}}{2013}]{2013A&A...560A..43F}
{Flock} M. et~al., 2013 \emph{\aap}, \emph{560}, A43.

\bibitem[\protect\astroncite{\emph{{Flock} et~al.}}{2015}]{2015A&A...574A..68F}
{Flock} M. et~al., 2015 \emph{\aap}, \emph{574}, A68.

\bibitem[\protect\astroncite{\emph{{Flock} et~al.}}{2016}]{2016ApJ...827..144F}
{Flock} M. et~al., 2016 \emph{\apj}, \emph{827}, 2, 144.

\bibitem[\protect\astroncite{\emph{{Flock}
  et~al.}}{2017{\natexlab{a}}}]{2017ApJ...835..230F}
{Flock} M. et~al., 2017{\natexlab{a}} \emph{\apj}, \emph{835}, 2, 230.

\bibitem[\protect\astroncite{\emph{{Flock}
  et~al.}}{2017{\natexlab{b}}}]{flock17}
{Flock} M. et~al., 2017{\natexlab{b}} \emph{\apj}, \emph{850}, 2, 131.

\bibitem[\protect\astroncite{\emph{{Flock} et~al.}}{2019}]{2019A&A...630A.147F}
{Flock} M. et~al., 2019 \emph{\aap}, \emph{630}, A147.

\bibitem[\protect\astroncite{\emph{{Flock} et~al.}}{2020}]{flock20}
{Flock} M. et~al., 2020 \emph{\apj}, \emph{897}, 2, 155.

\bibitem[\protect\astroncite{\emph{{Flores-Rivera} et~al.}}{2020}]{flores20}
{Flores-Rivera} L. et~al., 2020 \emph{\aap}, \emph{644}, A50.

\bibitem[\protect\astroncite{\emph{Frank et~al.}}{2014}]{Frank.Ray.ea14}
Frank A. et~al., 2014 \emph{Protostars and Planets VI}, \emph{-1}, 451.

\bibitem[\protect\astroncite{\emph{{Franz} et~al.}}{2020}]{Franz2020}
{Franz} R. et~al., 2020 \emph{\aap}, \emph{635}, A53.

\bibitem[\protect\astroncite{\emph{{Fraser} et~al.}}{2014}]{Fraser+etal2014}
{Fraser} W.~C. et~al., 2014 \emph{\apj}, \emph{782}, 2, 100.

\bibitem[\protect\astroncite{\emph{{Fricke}}}{1968}]{fricke68}
{Fricke} K., 1968 \emph{\zap}, \emph{68}, 317.

\bibitem[\protect\astroncite{\emph{{Fromang} and {Lesur}}}{2019}]{Fromang2019}
{Fromang} S. and {Lesur} G., 2019 \emph{EAS Publications Series}, vol.~82 of
  \emph{EAS Publications Series}, pp. 391--413.

\bibitem[\protect\astroncite{\emph{Fromang and Nelson}}{2006}]{Fromang2006c}
Fromang S. and Nelson R.~P., 2006 \emph{\aap}, \emph{457}, 343.

\bibitem[\protect\astroncite{\emph{{Fukuhara} et~al.}}{2021}]{fukuhara21}
{Fukuhara} Y. et~al., 2021 \emph{\apj}, \emph{914}, 2, 132.

\bibitem[\protect\astroncite{\emph{Gammie}}{1996}]{Gammie1996}
Gammie C.~F., 1996 \emph{\apj}, \emph{457}, 355.

\bibitem[\protect\astroncite{\emph{Garaud et~al.}}{2013}]{Garaud2013}
Garaud P. et~al., 2013 \emph{\apj}, \emph{764}, 2, 146.

\bibitem[\protect\astroncite{\emph{{Gerbig} et~al.}}{2020}]{GM20}
{Gerbig} K. et~al., 2020 \emph{\apj}, \emph{895}, 2, 91.

\bibitem[\protect\astroncite{\emph{{Glassgold}
  et~al.}}{2004}]{Glassgold_etal_2004}
{Glassgold} A.~E. et~al., 2004 \emph{\apj}, \emph{615}, 2, 972.

\bibitem[\protect\astroncite{\emph{{Glassgold} et~al.}}{2017}]{Glassgold+2017}
{Glassgold} A.~E. et~al., 2017 \emph{\mnras}, \emph{472}, 2, 2447.

\bibitem[\protect\astroncite{\emph{{Goldreich} and
  {Schubert}}}{1967}]{goldreich67}
{Goldreich} P. and {Schubert} G., 1967 \emph{\apj}, \emph{150}, 571.

\bibitem[\protect\astroncite{\emph{{Goldreich} and {Ward}}}{1973}]{gw73}
{Goldreich} P. and {Ward} W.~R., 1973 \emph{\apj}, \emph{183}, 1051.

\bibitem[\protect\astroncite{\emph{{Gole} and
  {Simon}}}{2018}]{2018ApJ...869...84G}
{Gole} D.~A. and {Simon} J.~B., 2018 \emph{\apj}, \emph{869}, 1, 84.

\bibitem[\protect\astroncite{\emph{{Gole} et~al.}}{2020}]{GS20}
{Gole} D.~A. et~al., 2020 \emph{\apj}, \emph{904}, 2, 132.

\bibitem[\protect\astroncite{\emph{{Gonzalez} et~al.}}{2017}]{Gonzalez2017}
{Gonzalez} J.~F. et~al., 2017 \emph{\mnras}, \emph{467}, 2, 1984.

\bibitem[\protect\astroncite{\emph{{Goodman} and {Pindor}}}{2000}]{Goodman2000}
{Goodman} J. and {Pindor} B., 2000 \emph{\icarus}, \emph{148}, 537.

\bibitem[\protect\astroncite{\emph{{Gorti} et~al.}}{2009}]{Gorti2009}
{Gorti} U. et~al., 2009 \emph{\apj}, \emph{705}, 1237.

\bibitem[\protect\astroncite{\emph{Gressel et~al.}}{2015}]{Gressel.Turner.ea15}
Gressel O. et~al., 2015 \emph{\apj}, \emph{801}, 2, 84.

\bibitem[\protect\astroncite{\emph{Gressel et~al.}}{2020}]{Gressel.Ramsey.ea20}
Gressel O. et~al., 2020 \emph{\apj}, \emph{896}, 2, 126.

\bibitem[\protect\astroncite{\emph{{G{\"u}del}}}{2015}]{Guedel2015}
{G{\"u}del} M., 2015 \emph{European Physical Journal Web of Conferences}, vol.
  102 of \emph{European Physical Journal Web of Conferences}, p. 00015.

\bibitem[\protect\astroncite{\emph{Guilloteau et~al.}}{2012}]{Guilloteau2012}
Guilloteau S. et~al., 2012 \emph{\aap}, \emph{548}, 70.

\bibitem[\protect\astroncite{\emph{{Hartlep} and {Cuzzi}}}{2020}]{HC20}
{Hartlep} T. and {Cuzzi} J.~N., 2020 \emph{\apj}, \emph{892}, 2, 120.

\bibitem[\protect\astroncite{\emph{{Hartlep} et~al.}}{2017}]{HCW17}
{Hartlep} T. et~al., 2017 \emph{\pre}, \emph{95}, 3, 033115.

\bibitem[\protect\astroncite{\emph{{Hartmann} et~al.}}{1998}]{Hartmann1998}
{Hartmann} L. et~al., 1998 \emph{\apj}, \emph{495}, 1, 385.

\bibitem[\protect\astroncite{\emph{{Hartmann} et~al.}}{2016}]{Hartmann2016}
{Hartmann} L. et~al., 2016 \emph{\araa}, \emph{54}, 135.

\bibitem[\protect\astroncite{\emph{{Haworth} et~al.}}{2020}]{haworth20}
{Haworth} T.~J. et~al., 2020 \emph{\mnras}, \emph{494}, 3, 4130.

\bibitem[\protect\astroncite{\emph{{Hirai} et~al.}}{2018}]{2018ApJ...853..174H}
{Hirai} K. et~al., 2018 \emph{\apj}, \emph{853}, 2, 174.

\bibitem[\protect\astroncite{\emph{{Hirose}
  et~al.}}{2014}]{2014ApJ...787....1H}
{Hirose} S. et~al., 2014 \emph{\apj}, \emph{787}, 1, 1.

\bibitem[\protect\astroncite{\emph{{Hollenbach} and
  {McKee}}}{1979}]{Hollenbach_McKee_1979}
{Hollenbach} D. and {McKee} C.~F., 1979 \emph{\apjs}, \emph{41}, 555.

\bibitem[\protect\astroncite{\emph{{Homma} et~al.}}{2019}]{Homma2019}
{Homma} K.~A. et~al., 2019 \emph{AAS/Division for Extreme Solar Systems
  Abstracts}, vol.~51 of \emph{AAS/Division for Extreme Solar Systems
  Abstracts}, p. 317.03.

\bibitem[\protect\astroncite{\emph{{Hubeny}}}{1990}]{hubeny90}
{Hubeny} I., 1990 \emph{\apj}, \emph{351}, 632.

\bibitem[\protect\astroncite{\emph{{Hutchison} and
  {Clarke}}}{2021}]{Hutchison2021}
{Hutchison} M.~A. and {Clarke} C.~J., 2021 \emph{\mnras}, \emph{501}, 1, 1127.

\bibitem[\protect\astroncite{\emph{{Hutchison}
  et~al.}}{2016{\natexlab{a}}}]{Hutchison2016a}
{Hutchison} M.~A. et~al., 2016{\natexlab{a}} \emph{\mnras}, \emph{461}, 1, 742.

\bibitem[\protect\astroncite{\emph{{Hutchison}
  et~al.}}{2016{\natexlab{b}}}]{Hutchison2016b}
{Hutchison} M.~A. et~al., 2016{\natexlab{b}} \emph{\mnras}, \emph{463}, 3,
  2725.

\bibitem[\protect\astroncite{\emph{{Inutsuka} and
  {Sano}}}{2005}]{Inutsuka+Sano2005}
{Inutsuka} S.-i. and {Sano} T., 2005 \emph{\apjl}, \emph{628}, 2, L155.

\bibitem[\protect\astroncite{\emph{{Io} and
  {Suzuki}}}{2014}]{2014ApJ...780...46I}
{Io} Y. and {Suzuki} T.~K., 2014 \emph{\apj}, \emph{780}, 1, 46.

\bibitem[\protect\astroncite{\emph{{Ishitsu} et~al.}}{2009}]{ishitsu09}
{Ishitsu} N. et~al., 2009 \emph{arXiv e-prints}, arXiv:0905.4404.

\bibitem[\protect\astroncite{\emph{{Ivlev} et~al.}}{2016}]{Ivlev+2016}
{Ivlev} A.~V. et~al., 2016 \emph{\apj}, \emph{833}, 1, 92.

\bibitem[\protect\astroncite{\emph{{Jacquemin-Ide}
  et~al.}}{2019}]{Jacquemin-Ide.Ferreira.ea19}
{Jacquemin-Ide} J. et~al., 2019 \emph{\mnras}, \emph{490}, 3, 3112.

\bibitem[\protect\astroncite{\emph{{Jacquemin-Ide}
  et~al.}}{2021}]{Jacquemin-Ide.Lesur.ea20}
{Jacquemin-Ide} J. et~al., 2021 \emph{\aap}, \emph{647}, A192.

\bibitem[\protect\astroncite{\emph{Jacquet et~al.}}{2011}]{Jacquet2011}
Jacquet E. et~al., 2011 \emph{\mnras}, \emph{415}, 4, 3591.

\bibitem[\protect\astroncite{\emph{{Jankovic}
  et~al.}}{2021}]{2021MNRAS.504..280J}
{Jankovic} M.~R. et~al., 2021 \emph{\mnras}, \emph{504}, 1, 280.

\bibitem[\protect\astroncite{\emph{{Jankovic}
  et~al.}}{2022}]{2022MNRAS.509.5974J}
{Jankovic} M.~R. et~al., 2022 \emph{\mnras}, \emph{509}, 4, 5974.

\bibitem[\protect\astroncite{\emph{{Jaupart} and {Laibe}}}{2020}]{Jaupart2020}
{Jaupart} E. and {Laibe} G., 2020 \emph{\mnras}, \emph{492}, 4, 4591.

\bibitem[\protect\astroncite{\emph{{Johansen} and {Youdin}}}{2007}]{JY07}
{Johansen} A. and {Youdin} A., 2007 \emph{\apj}, \emph{662}, 1, 627.

\bibitem[\protect\astroncite{\emph{{Johansen} et~al.}}{2006}]{johansen06}
{Johansen} A. et~al., 2006 \emph{\apj}, \emph{643}, 2, 1219.

\bibitem[\protect\astroncite{\emph{{Johansen} et~al.}}{2007}]{JO07}
{Johansen} A. et~al., 2007 \emph{\nat}, \emph{448}, 7157, 1022.

\bibitem[\protect\astroncite{\emph{{Johansen} et~al.}}{2009}]{JYM09}
{Johansen} A. et~al., 2009 \emph{\apjl}, \emph{704}, 2, L75.

\bibitem[\protect\astroncite{\emph{{Johansen} et~al.}}{2011}]{JKH11}
{Johansen} A. et~al., 2011 \emph{\aap}, \emph{529}, A62.

\bibitem[\protect\astroncite{\emph{{Johansen} et~al.}}{2015}]{JM15}
{Johansen} A. et~al., 2015 \emph{Science Advances}, \emph{1}, 1500109.

\bibitem[\protect\astroncite{\emph{{Kataoka}
  et~al.}}{2013{\natexlab{a}}}]{Kataoka2013b}
{Kataoka} A. et~al., 2013{\natexlab{a}} \emph{\aap}, \emph{557}, L4.

\bibitem[\protect\astroncite{\emph{{Kataoka}
  et~al.}}{2013{\natexlab{b}}}]{Kataoka2013a}
{Kataoka} A. et~al., 2013{\natexlab{b}} \emph{\aap}, \emph{554}, A4.

\bibitem[\protect\astroncite{\emph{Kerswell}}{2002}]{kerswell04}
Kerswell R.~R., 2002 \emph{Annual Review of Fluid Mechanics}, \emph{34}, 1, 83.

\bibitem[\protect\astroncite{\emph{Kimmig
  et~al.}}{2020}]{Kimmig.Dullemond.ea20}
Kimmig C.~N. et~al., 2020 \emph{\aap}, \emph{633}, A4.

\bibitem[\protect\astroncite{\emph{{Klahr} and {Hubbard}}}{2014}]{klahr14}
{Klahr} H. and {Hubbard} A., 2014 \emph{\apj}, \emph{788}, 21.

\bibitem[\protect\astroncite{\emph{{Klahr} and {Schreiber}}}{2020}]{KS20}
{Klahr} H. and {Schreiber} A., 2020 \emph{\apj}, \emph{901}, 1, 54.

\bibitem[\protect\astroncite{\emph{{Klahr} and {Schreiber}}}{2021}]{KS21}
{Klahr} H. and {Schreiber} A., 2021 \emph{\apj}, \emph{911}, 1, 9.

\bibitem[\protect\astroncite{\emph{{Klahr} and {Bodenheimer}}}{2003}]{klahr03}
{Klahr} H.~H. and {Bodenheimer} P., 2003 \emph{\apj}, \emph{582}, 869.

\bibitem[\protect\astroncite{\emph{{Knobloch} and
  {Spruit}}}{1982}]{Knobloch_Spruit_1982}
{Knobloch} E. and {Spruit} H.~C., 1982 \emph{\aap}, \emph{113}, 2, 261.

\bibitem[\protect\astroncite{\emph{{K{\"o}nigl} et~al.}}{2010}]{Konigl2010}
{K{\"o}nigl} A. et~al., 2010 \emph{\mnras}, \emph{401}, 1, 479.

\bibitem[\protect\astroncite{\emph{{Krapp} et~al.}}{2019}]{Krapp2019}
{Krapp} L. et~al., 2019 \emph{\apjl}, \emph{878}, 2, L30.

\bibitem[\protect\astroncite{\emph{{Krapp} et~al.}}{2020}]{Krapp2020}
{Krapp} L. et~al., 2020 \emph{\mnras}.

\bibitem[\protect\astroncite{\emph{{Kratter} and {Lodato}}}{2016}]{kratter16}
{Kratter} K. and {Lodato} G., 2016 \emph{\araa}, \emph{54}, 271.

\bibitem[\protect\astroncite{\emph{{Krivov} and
  {Wyatt}}}{2021}]{KrivovWyatt2021}
{Krivov} A.~V. and {Wyatt} M.~C., 2021 \emph{\mnras}, \emph{500}, 1, 718.

\bibitem[\protect\astroncite{\emph{{Krivov} et~al.}}{2018}]{Krivov+etal2018}
{Krivov} A.~V. et~al., 2018 \emph{\mnras}, \emph{474}, 2, 2564.

\bibitem[\protect\astroncite{\emph{{Kruijer} et~al.}}{2017}]{Kruijer2017}
{Kruijer} T.~S. et~al., 2017 \emph{Proceedings of the National Academy of
  Science}, \emph{114}, 26, 6712.

\bibitem[\protect\astroncite{\emph{{Kruijer} et~al.}}{2020}]{Kruijer2020}
{Kruijer} T.~S. et~al., 2020 \emph{Nature Astronomy}, \emph{4}, 32.

\bibitem[\protect\astroncite{\emph{{Kunz} and
  {Lesur}}}{2013}]{2013MNRAS.434.2295K}
{Kunz} M.~W. and {Lesur} G., 2013 \emph{\mnras}, \emph{434}, 3, 2295.

\bibitem[\protect\astroncite{\emph{{Laibe}}}{2014}]{Laibe2014}
{Laibe} G., 2014 \emph{\mnras}, \emph{437}, 4, 3037.

\bibitem[\protect\astroncite{\emph{Lambrechts et~al.}}{2016}]{Lambrechts2016}
Lambrechts M. et~al., 2016 \emph{\aap}, \emph{591}, A133.

\bibitem[\protect\astroncite{\emph{{Latter}}}{2016}]{latter16}
{Latter} H.~N., 2016 \emph{\mnras}, \emph{455}, 2608.

\bibitem[\protect\astroncite{\emph{{Latter} and {Papaloizou}}}{2018}]{latter18}
{Latter} H.~N. and {Papaloizou} J., 2018 \emph{\mnras}, \emph{474}, 3, 3110.

\bibitem[\protect\astroncite{\emph{{Latter} and {Rosca}}}{2017}]{Latter2017}
{Latter} H.~N. and {Rosca} R., 2017 \emph{\mnras}, \emph{464}, 2, 1923.

\bibitem[\protect\astroncite{\emph{{Lee} et~al.}}{2010}]{lee10}
{Lee} A.~T. et~al., 2010 \emph{\apj}, \emph{718}, 1367.

\bibitem[\protect\astroncite{\emph{Lesur}}{2021a}]{Lesur21}
Lesur G., 2021a \emph{Journal of Plasma Physics}, \emph{87}, 1, 205870101.

\bibitem[\protect\astroncite{\emph{{Lesur}}}{2021b}]{Lesur21a}
{Lesur} G., 2021b \emph{\aap}, \emph{650}, A35.

\bibitem[\protect\astroncite{\emph{{Lesur} and {Papaloizou}}}{2009}]{lesur09b}
{Lesur} G. and {Papaloizou} J.~C.~B., 2009 \emph{\aap}, \emph{498}, 1.

\bibitem[\protect\astroncite{\emph{{Lesur} and {Papaloizou}}}{2010}]{lesur10}
{Lesur} G. and {Papaloizou} J.~C.~B., 2010 \emph{\aap}, \emph{513}, A60.

\bibitem[\protect\astroncite{\emph{Lesur et~al.}}{2014}]{Lesur.Kunz.ea14}
Lesur G. et~al., 2014 \emph{\aap}, \emph{566}, 56.

\bibitem[\protect\astroncite{\emph{{Lesur} and
  {Latter}}}{2016}]{lesurlatter2016}
{Lesur} G.~R.~J. and {Latter} H., 2016 \emph{\mnras}, \emph{462}, 4549.

\bibitem[\protect\astroncite{\emph{Leung and Ogilvie}}{2019}]{Leung.Ogilvie19}
Leung P. K.~C. and Ogilvie G.~I., 2019 \emph{\mnras}, \emph{487}, 5155.

\bibitem[\protect\astroncite{\emph{{Leung} and
  {Ogilvie}}}{2020}]{2020MNRAS.498..750L}
{Leung} P. K.~C. and {Ogilvie} G.~I., 2020 \emph{\mnras}, \emph{498}, 1, 750.

\bibitem[\protect\astroncite{\emph{{Li} et~al.}}{2000}]{li00}
{Li} H. et~al., 2000 \emph{\apj}, \emph{533}, 1023.

\bibitem[\protect\astroncite{\emph{{Li} et~al.}}{2001}]{li01}
{Li} H. et~al., 2001 \emph{\apj}, \emph{551}, 874.

\bibitem[\protect\astroncite{\emph{{Li} and {Youdin}}}{2021}]{LY21}
{Li} R. and {Youdin} A.~N., 2021 \emph{\apj}, \emph{919}, 2, 107.

\bibitem[\protect\astroncite{\emph{{Li} et~al.}}{2018}]{LYS18}
{Li} R. et~al., 2018 \emph{\apj}, \emph{862}, 1, 14.

\bibitem[\protect\astroncite{\emph{{Li} et~al.}}{2019}]{LYS19}
{Li} R. et~al., 2019 \emph{\apj}, \emph{885}, 1, 69.

\bibitem[\protect\astroncite{\emph{{Lin}}}{2019}]{lin19}
{Lin} M.-K., 2019 \emph{\mnras}, \emph{485}, 4, 5221.

\bibitem[\protect\astroncite{\emph{{Lin}}}{2021}]{Lin2021}
{Lin} M.-K., 2021 \emph{\apj}, \emph{907}, 2, 64.

\bibitem[\protect\astroncite{\emph{{Lin} and {Kratter}}}{2016}]{lin16}
{Lin} M.-K. and {Kratter} K.~M., 2016 \emph{\apj}, \emph{824}, 2, 91.

\bibitem[\protect\astroncite{\emph{{Lin} and {Youdin}}}{2015}]{lin15}
{Lin} M.-K. and {Youdin} A.~N., 2015 \emph{\apj}, \emph{811}, 17.

\bibitem[\protect\astroncite{\emph{{Lin} and {Youdin}}}{2017}]{Lin2017}
{Lin} M.-K. and {Youdin} A.~N., 2017 \emph{\apj}, \emph{849}, 2, 129.

\bibitem[\protect\astroncite{\emph{{Liu} et~al.}}{2020}]{LL20}
{Liu} B. et~al., 2020 \emph{\aap}, \emph{638}, A88.

\bibitem[\protect\astroncite{\emph{{Lobo Gomes} et~al.}}{2015}]{gomes15}
{Lobo Gomes} A. et~al., 2015 \emph{\apj}, \emph{810}, 2, 94.

\bibitem[\protect\astroncite{\emph{{Lommen} et~al.}}{2009}]{Lommen2009}
{Lommen} D. et~al., 2009 \emph{\aap}, \emph{495}, 3, 869.

\bibitem[\protect\astroncite{\emph{{Lorek} et~al.}}{2016}]{LG16}
{Lorek} S. et~al., 2016 \emph{\aap}, \emph{587}, A128.

\bibitem[\protect\astroncite{\emph{{Love} et~al.}}{1994}]{Love1994}
{Love} S.~G. et~al., 1994 \emph{\icarus}, \emph{111}, 1, 227.

\bibitem[\protect\astroncite{\emph{{Lovelace} et~al.}}{1999}]{Lovelace1999}
{Lovelace} R.~V.~E. et~al., 1999 \emph{\apj}, \emph{513}, 2, 805.

\bibitem[\protect\astroncite{\emph{{Lynden-Bell}}}{2003}]{Lynden-Bell03}
{Lynden-Bell} D., 2003 \emph{\mnras}, \emph{341}, 4, 1360.

\bibitem[\protect\astroncite{\emph{{Lyra}}}{2014}]{lyra14}
{Lyra} W., 2014 \emph{\apj}, \emph{789}, 77.

\bibitem[\protect\astroncite{\emph{{Lyra} and {Klahr}}}{2011}]{lyraklahr2011}
{Lyra} W. and {Klahr} H., 2011 \emph{\aap}, \emph{527}, A138+.

\bibitem[\protect\astroncite{\emph{{Lyra} and {Mac Low}}}{2012}]{LyraMacLow12}
{Lyra} W. and {Mac Low} M.-M., 2012 \emph{\apj}, \emph{756}, 1, 62.

\bibitem[\protect\astroncite{\emph{{Lyra} and {Umurhan}}}{2019}]{lyra19}
{Lyra} W. and {Umurhan} O.~M., 2019 \emph{\pasp}, \emph{131}, 1001, 072001.

\bibitem[\protect\astroncite{\emph{{Lyra} et~al.}}{2015}]{2015A&A...574A..10L}
{Lyra} W. et~al., 2015 \emph{\aap}, \emph{574}, A10.

\bibitem[\protect\astroncite{\emph{{Lyra} et~al.}}{2018}]{lyra18}
{Lyra} W. et~al., 2018 \emph{Research Notes of the American Astronomical
  Society}, \emph{2}, 4, 195.

\bibitem[\protect\astroncite{\emph{{Lyra} et~al.}}{2021}]{LYJ21}
{Lyra} W. et~al., 2021 \emph{\icarus}, \emph{356}, 113831.

\bibitem[\protect\astroncite{\emph{{Malygin}
  et~al.}}{2014}]{2014A&A...568A..91M}
{Malygin} M.~G. et~al., 2014 \emph{\aap}, \emph{568}, A91.

\bibitem[\protect\astroncite{\emph{{Malygin} et~al.}}{2017}]{malygin17}
{Malygin} M.~G. et~al., 2017 \emph{\aap}, \emph{605}, A30.

\bibitem[\protect\astroncite{\emph{{Mamatsashvili}
  et~al.}}{2020}]{2020ApJ...904...47M}
{Mamatsashvili} G. et~al., 2020 \emph{\apj}, \emph{904}, 1, 47.

\bibitem[\protect\astroncite{\emph{{Manara} et~al.}}{2016}]{Manara2016}
{Manara} C.~F. et~al., 2016 \emph{\aap}, \emph{591}, L3.

\bibitem[\protect\astroncite{\emph{{Manger} and {Klahr}}}{2018}]{manger18}
{Manger} N. and {Klahr} H., 2018 \emph{\mnras}, \emph{480}, 2, 2125.

\bibitem[\protect\astroncite{\emph{{Manger} et~al.}}{2020}]{manger20}
{Manger} N. et~al., 2020 \emph{\mnras}, \emph{499}, 2, 1841.

\bibitem[\protect\astroncite{\emph{{Marcus} et~al.}}{2013}]{MPJH13}
{Marcus} P.~S. et~al., 2013 \emph{PhRvL}, \emph{111}, 8, 084501.

\bibitem[\protect\astroncite{\emph{{Marcus} et~al.}}{2015}]{MPJBHL15}
{Marcus} P.~S. et~al., 2015 \emph{\apj}, \emph{808}, 87.

\bibitem[\protect\astroncite{\emph{{Marcus} et~al.}}{2016}]{MPJB16}
{Marcus} P.~S. et~al., 2016 \emph{\apj}, \emph{833}, 148.

\bibitem[\protect\astroncite{\emph{McKinnon et~al.}}{2020}]{mckinnon20}
McKinnon W.~B. et~al., 2020 \emph{Science}, \emph{367}, 6481.

\bibitem[\protect\astroncite{\emph{{McNally} and {Pessah}}}{2015}]{mcnally15}
{McNally} C.~P. and {Pessah} M.~E., 2015 \emph{\apj}, \emph{811}, 2, 121.

\bibitem[\protect\astroncite{\emph{{McNally}
  et~al.}}{2014}]{2014ApJ...791...62M}
{McNally} C.~P. et~al., 2014 \emph{\apj}, \emph{791}, 1, 62.

\bibitem[\protect\astroncite{\emph{{McNally} et~al.}}{2020}]{McNally20}
{McNally} C.~P. et~al., 2020 \emph{\mnras}, \emph{493}, 3, 4382.

\bibitem[\protect\astroncite{\emph{{Meheut}
  et~al.}}{2015}]{2015A&A...579A.117M}
{Meheut} H. et~al., 2015 \emph{\aap}, \emph{579}, A117.

\bibitem[\protect\astroncite{\emph{{Michikoshi} et~al.}}{2007}]{michikoshi07}
{Michikoshi} S. et~al., 2007 \emph{\apj}, \emph{657}, 521.

\bibitem[\protect\astroncite{\emph{{Michikoshi} et~al.}}{2012}]{michikoshi12}
{Michikoshi} S. et~al., 2012 \emph{\apj}, \emph{746}, 35.

\bibitem[\protect\astroncite{\emph{{Miranda}
  et~al.}}{2017}]{2017ApJ...835..118M}
{Miranda} R. et~al., 2017 \emph{\apj}, \emph{835}, 2, 118.

\bibitem[\protect\astroncite{\emph{{Miyake} et~al.}}{2016}]{Miyake2016}
{Miyake} T. et~al., 2016 \emph{\apj}, \emph{821}, 1, 3.

\bibitem[\protect\astroncite{\emph{{Mohanty}
  et~al.}}{2018}]{2018ApJ...861..144M}
{Mohanty} S. et~al., 2018 \emph{\apj}, \emph{861}, 2, 144.

\bibitem[\protect\astroncite{\emph{{Morbidelli} et~al.}}{2009}]{morbidelli09}
{Morbidelli} A. et~al., 2009 \emph{\icarus}, \emph{204}, 558.

\bibitem[\protect\astroncite{\emph{{Mori} et~al.}}{2019}]{2019ApJ...872...98M}
{Mori} S. et~al., 2019 \emph{\apj}, \emph{872}, 1, 98.

\bibitem[\protect\astroncite{\emph{{Mulders} and
  {Dominik}}}{2012}]{Mulders2012}
{Mulders} G.~D. and {Dominik} C., 2012 \emph{\aap}, \emph{539}, A9.

\bibitem[\protect\astroncite{\emph{{Murphy} and
  {Pessah}}}{2015}]{2015ApJ...802..139M}
{Murphy} G.~C. and {Pessah} M.~E., 2015 \emph{\apj}, \emph{802}, 2, 139.

\bibitem[\protect\astroncite{\emph{{Musiolik} and {Wurm}}}{2019}]{Musiolik2019}
{Musiolik} G. and {Wurm} G., 2019 \emph{\apj}, \emph{873}, 1, 58.

\bibitem[\protect\astroncite{\emph{{Najita} and {Bergin}}}{2018}]{Najita2018}
{Najita} J.~R. and {Bergin} E.~A., 2018 \emph{\apj}, \emph{864}, 2, 168.

\bibitem[\protect\astroncite{\emph{{Nakagawa} et~al.}}{1986}]{Nakagawa1986}
{Nakagawa} Y. et~al., 1986 \emph{\icarus}, \emph{67}, 375.

\bibitem[\protect\astroncite{\emph{{Nakatani}
  et~al.}}{2018{\natexlab{a}}}]{Nakatani+2018a}
{Nakatani} R. et~al., 2018{\natexlab{a}} \emph{\apj}, \emph{857}, 1, 57.

\bibitem[\protect\astroncite{\emph{{Nakatani}
  et~al.}}{2018{\natexlab{b}}}]{Nakatani+2018b}
{Nakatani} R. et~al., 2018{\natexlab{b}} \emph{\apj}, \emph{865}, 1, 75.

\bibitem[\protect\astroncite{\emph{{Naoz} et~al.}}{2010}]{Naoz+etal2010}
{Naoz} S. et~al., 2010 \emph{\apj}, \emph{719}, 2, 1775.

\bibitem[\protect\astroncite{\emph{{Natta} et~al.}}{2007}]{NT07}
{Natta} A. et~al., 2007 \emph{Protostars and Planets V} (B.~{Reipurth},
  D.~{Jewitt}, and K.~{Keil}), p. 767.

\bibitem[\protect\astroncite{\emph{{Nelson} et~al.}}{2013}]{nelson13}
{Nelson} R.~P. et~al., 2013 \emph{\mnras}, \emph{435}, 2610.

\bibitem[\protect\astroncite{\emph{{Nemer} et~al.}}{2020}]{Nemer2020}
{Nemer} A. et~al., 2020 \emph{\apjl}, \emph{904}, 2, L27.

\bibitem[\protect\astroncite{\emph{{Nesvorn{\'y}} et~al.}}{2010}]{NYR10}
{Nesvorn{\'y}} D. et~al., 2010 \emph{\aj}, \emph{140}, 3, 785.

\bibitem[\protect\astroncite{\emph{{Nesvorn{\'y}} et~al.}}{2019}]{NL19}
{Nesvorn{\'y}} D. et~al., 2019 \emph{Nature Astronomy}, \emph{3}, 808.

\bibitem[\protect\astroncite{\emph{{Nesvorn{\'y}} et~al.}}{2021}]{NL21}
{Nesvorn{\'y}} D. et~al., 2021 \emph{The Planetary Science Journal}, \emph{2},
  1, 27.

\bibitem[\protect\astroncite{\emph{{Noll} et~al.}}{2008}]{Noll+etal2008}
{Noll} K.~S. et~al., 2008 \emph{\icarus}, \emph{194}, 2, 758.

\bibitem[\protect\astroncite{\emph{{Ohashi} and {Kataoka}}}{2019}]{Ohashi2019}
{Ohashi} S. and {Kataoka} A., 2019 \emph{\apj}, \emph{886}, 2, 103.

\bibitem[\protect\astroncite{\emph{{Okuzumi}}}{2009}]{Okuzumi2009}
{Okuzumi} S., 2009 \emph{\apj}, \emph{698}, 2, 1122.

\bibitem[\protect\astroncite{\emph{Okuzumi et~al.}}{2012}]{Okuzumi2012}
Okuzumi S. et~al., 2012 \emph{\apj}, \emph{752}, 2, 106.

\bibitem[\protect\astroncite{\emph{{Okuzumi} et~al.}}{2019}]{Okuzumi+2019}
{Okuzumi} S. et~al., 2019 \emph{\apj}, \emph{878}, 2, 133.

\bibitem[\protect\astroncite{\emph{{Orlanski} and
  {Bryan}}}{1969}]{orlanski1969}
{Orlanski} I. and {Bryan} K., 1969 \emph{\jgr}, \emph{74}, 6975.

\bibitem[\protect\astroncite{\emph{{Owen} et~al.}}{2010}]{Owen2010}
{Owen} J.~E. et~al., 2010 \emph{\mnras}, \emph{401}, 3, 1415.

\bibitem[\protect\astroncite{\emph{{Owen} et~al.}}{2011}]{Owen2011b}
{Owen} J.~E. et~al., 2011 \emph{\mnras}, \emph{411}, 2, 1104.

\bibitem[\protect\astroncite{\emph{{Paardekooper}
  et~al.}}{2020}]{Paardekooper2020}
{Paardekooper} S.-J. et~al., 2020 \emph{\mnras}, \emph{499}, 3, 4223.

\bibitem[\protect\astroncite{\emph{{Paardekooper}
  et~al.}}{2021}]{Paardekooper2021}
{Paardekooper} S.-J. et~al., 2021 \emph{\mnras}, \emph{502}, 2, 1579.

\bibitem[\protect\astroncite{\emph{{Pan}}}{2020}]{Pan2020a}
{Pan} L., 2020 \emph{\apj}, \emph{898}, 1, 8.

\bibitem[\protect\astroncite{\emph{{Pan} and {Yu}}}{2020}]{Pan2020}
{Pan} L. and {Yu} C., 2020 \emph{\apj}, \emph{898}, 1, 7.

\bibitem[\protect\astroncite{\emph{{Pan} et~al.}}{2011}]{PP11}
{Pan} L. et~al., 2011 \emph{\apj}, \emph{740}, 1, 6.

\bibitem[\protect\astroncite{\emph{{Parkin}}}{2014}]{2014MNRAS.441.2078P}
{Parkin} E.~R., 2014 \emph{\mnras}, \emph{441}, 3, 2078.

\bibitem[\protect\astroncite{\emph{{Pelegr{\'{\i}}} and
  {Sangr{\~a}}}}{1998}]{pelegri1998}
{Pelegr{\'{\i}}} J.~L. and {Sangr{\~a}} P., 1998 \emph{\jgr}, \emph{103}, 30.

\bibitem[\protect\astroncite{\emph{{Perez-Becker} and
  {Chiang}}}{2011}]{Perez-Becker+Chiang2011}
{Perez-Becker} D. and {Chiang} E., 2011 \emph{\apj}, \emph{735}, 1, 8.

\bibitem[\protect\astroncite{\emph{{Petersen}
  et~al.}}{2007{\natexlab{a}}}]{peterson07a}
{Petersen} M.~R. et~al., 2007{\natexlab{a}} \emph{\apj}, \emph{658}, 1236.

\bibitem[\protect\astroncite{\emph{{Petersen}
  et~al.}}{2007{\natexlab{b}}}]{peterson07b}
{Petersen} M.~R. et~al., 2007{\natexlab{b}} \emph{\apj}, \emph{658}, 1252.

\bibitem[\protect\astroncite{\emph{{Pfeil} and {Klahr}}}{2019}]{pfeil19}
{Pfeil} T. and {Klahr} H., 2019 \emph{\apj}, \emph{871}, 2, 150.

\bibitem[\protect\astroncite{\emph{{Pfeil} and {Klahr}}}{2021}]{pfeil20}
{Pfeil} T. and {Klahr} H., 2021 \emph{\apj}, \emph{915}, 2, 130.

\bibitem[\protect\astroncite{\emph{{Phillips}}}{1972}]{phillips1972}
{Phillips} O.~M., 1972 \emph{Deep Sea Research and Oceanographic Abstracts},
  \emph{19}, 79.

\bibitem[\protect\astroncite{\emph{{Picogna} et~al.}}{2018}]{picogna18}
{Picogna} G. et~al., 2018 \emph{\aap}, \emph{616}, A116.

\bibitem[\protect\astroncite{\emph{{Picogna} et~al.}}{2019}]{Picogna2019}
{Picogna} G. et~al., 2019 \emph{\mnras}, \emph{487}, 1, 691.

\bibitem[\protect\astroncite{\emph{{Pignatale} et~al.}}{2018}]{Pignatale2018}
{Pignatale} F.~C. et~al., 2018 \emph{\apjl}, \emph{867}, 2, L23.

\bibitem[\protect\astroncite{\emph{{Pineda} et~al.}}{2020}]{pineda20}
{Pineda} J.~E. et~al., 2020 \emph{Nature Astronomy}, \emph{4}, 1158.

\bibitem[\protect\astroncite{\emph{{Pinilla} and {Youdin}}}{2017}]{pinilla17}
{Pinilla} P. and {Youdin} A., 2017 \emph{Astrophysics and Space Science
  Library}, vol. 445 of \emph{Astrophysics and Space Science Library}
  (M.~{Pessah} and O.~{Gressel}), p.~91.

\bibitem[\protect\astroncite{\emph{{Pinilla}
  et~al.}}{2016}]{2016A&A...596A..81P}
{Pinilla} P. et~al., 2016 \emph{\aap}, \emph{596}, A81.

\bibitem[\protect\astroncite{\emph{{Pinte} et~al.}}{2016}]{Pinte2016}
{Pinte} C. et~al., 2016 \emph{\apj}, \emph{816}, 1, 25.

\bibitem[\protect\astroncite{\emph{{Potapov} et~al.}}{2020}]{Potapov2020}
{Potapov} A. et~al., 2020 \emph{\prl}, \emph{124}, 22, 221103.

\bibitem[\protect\astroncite{\emph{{Rab} et~al.}}{2017}]{Rab+2017}
{Rab} C. et~al., 2017 \emph{\aap}, \emph{603}, A96.

\bibitem[\protect\astroncite{\emph{{Raettig} et~al.}}{2013}]{raettig13}
{Raettig} N. et~al., 2013 \emph{\apj}, \emph{765}, 2, 115.

\bibitem[\protect\astroncite{\emph{Raettig et~al.}}{2015}]{raettig15}
Raettig N. et~al., 2015 \emph{\apj}, \emph{804}, 1, 35.

\bibitem[\protect\astroncite{\emph{{Raettig} et~al.}}{2021}]{raettig21}
{Raettig} N. et~al., 2021 \emph{arXiv e-prints}, arXiv:2103.04476.

\bibitem[\protect\astroncite{\emph{{Reg{\'a}ly} et~al.}}{2012}]{Regaly2012}
{Reg{\'a}ly} Z. et~al., 2012 \emph{\mnras}, \emph{419}, 2, 1701.

\bibitem[\protect\astroncite{\emph{{Ricci} et~al.}}{2021}]{Ricci2021}
{Ricci} L. et~al., 2021 \emph{arXiv e-prints}, arXiv:2104.03400.

\bibitem[\protect\astroncite{\emph{{Richard} et~al.}}{2016}]{richard16}
{Richard} S. et~al., 2016 \emph{\mnras}, \emph{456}, 4, 3571.

\bibitem[\protect\astroncite{\emph{{Riols} and {Lesur}}}{2018}]{RL18}
{Riols} A. and {Lesur} G., 2018 \emph{\aap}, \emph{617}, A117.

\bibitem[\protect\astroncite{\emph{{Riols} et~al.}}{2020}]{Riols2020}
{Riols} A. et~al., 2020 \emph{\aap}, \emph{639}, A95.

\bibitem[\protect\astroncite{\emph{{Riols} et~al.}}{2021}]{Riols2021}
{Riols} A. et~al., 2021 \emph{\mnras}, \emph{506}, 1, 1407.

\bibitem[\protect\astroncite{\emph{{Robinson} et~al.}}{2020}]{RF20}
{Robinson} J.~E. et~al., 2020 \emph{\aap}, \emph{643}, A55.

\bibitem[\protect\astroncite{\emph{{Rodenkirch} et~al.}}{2020}]{Rodenkirch2020}
{Rodenkirch} P.~J. et~al., 2020 \emph{\aap}, \emph{633}, A21.

\bibitem[\protect\astroncite{\emph{{Rosotti} et~al.}}{2019}]{Rosotti2019}
{Rosotti} G.~P. et~al., 2019 \emph{\mnras}, \emph{486}, 4, 4829.

\bibitem[\protect\astroncite{\emph{{Ross} and
  {Latter}}}{2018}]{2018MNRAS.477.3329R}
{Ross} J. and {Latter} H.~N., 2018 \emph{\mnras}, \emph{477}, 3, 3329.

\bibitem[\protect\astroncite{\emph{{Rucska} and {Wadsley}}}{2021}]{RW21}
{Rucska} J.~J. and {Wadsley} J.~W., 2021 \emph{\mnras}, \emph{500}, 1, 520.

\bibitem[\protect\astroncite{\emph{{R{\"u}diger}
  et~al.}}{2018}]{2018PhR...741....1R}
{R{\"u}diger} G. et~al., 2018 \emph{\physrep}, \emph{741}, 1.

\bibitem[\protect\astroncite{\emph{{Ruge} et~al.}}{2016}]{2016A&A...590A..17R}
{Ruge} J.~P. et~al., 2016 \emph{\aap}, \emph{590}, A17.

\bibitem[\protect\astroncite{\emph{{Safronov}}}{1969}]{Safronov1969}
{Safronov} V.~S., 1969 \emph{{Evoliutsiia doplanetnogo oblaka.}}

\bibitem[\protect\astroncite{\emph{{Salmeron} et~al.}}{2011}]{Salmeron2011}
{Salmeron} R. et~al., 2011 \emph{\mnras}, \emph{412}, 2, 1162.

\bibitem[\protect\astroncite{\emph{{Salvesen} et~al.}}{2016}]{Salvesen16}
{Salvesen} G. et~al., 2016 \emph{\mnras}, \emph{457}, 1, 857.

\bibitem[\protect\astroncite{\emph{{Scardoni} et~al.}}{2021}]{SBC21}
{Scardoni} C.~E. et~al., 2021 \emph{\mnras}, \emph{504}, 1, 1495.

\bibitem[\protect\astroncite{\emph{{Scepi} et~al.}}{2018}]{2018A&A...620A..49S}
{Scepi} N. et~al., 2018 \emph{\aap}, \emph{620}, A49.

\bibitem[\protect\astroncite{\emph{{Sch{\"a}fer} et~al.}}{2017}]{SYJ17}
{Sch{\"a}fer} U. et~al., 2017 \emph{\aap}, \emph{597}, A69.

\bibitem[\protect\astroncite{\emph{{Sch{\"a}fer} et~al.}}{2020}]{schafer20}
{Sch{\"a}fer} U. et~al., 2020 \emph{\aap}, \emph{635}, A190.

\bibitem[\protect\astroncite{\emph{{Schaffer} et~al.}}{2018}]{SYJ18}
{Schaffer} N. et~al., 2018 \emph{\aap}, \emph{618}, A75.

\bibitem[\protect\astroncite{\emph{{Schaffer} et~al.}}{2021}]{SJL21}
{Schaffer} N. et~al., 2021 \emph{arXiv e-prints}, arXiv:2106.15302.

\bibitem[\protect\astroncite{\emph{{Schobert}
  et~al.}}{2019}]{2019ApJ...881...56S}
{Schobert} B.~N. et~al., 2019 \emph{\apj}, \emph{881}, 1, 56.

\bibitem[\protect\astroncite{\emph{{Schr{\"a}pler}
  et~al.}}{2018}]{Schrapler2018}
{Schr{\"a}pler} R. et~al., 2018 \emph{\apj}, \emph{853}, 1, 74.

\bibitem[\protect\astroncite{\emph{{Schreiber} and
  {Klahr}}}{2018}]{Schreiber2018}
{Schreiber} A. and {Klahr} H., 2018 \emph{\apj}, \emph{861}, 1, 47.

\bibitem[\protect\astroncite{\emph{{Schultz} and
  {Schumacher}}}{1999}]{schultz99}
{Schultz} D.~M. and {Schumacher} P.~N., 1999 \emph{Monthly Weather Review},
  \emph{127}, 12, 2709.

\bibitem[\protect\astroncite{\emph{{Seifert} et~al.}}{2021}]{Seifert+2021}
{Seifert} R.~A. et~al., 2021 \emph{arXiv e-prints}, arXiv:2103.10971.

\bibitem[\protect\astroncite{\emph{{Sekiya}}}{1983}]{sek83}
{Sekiya} M., 1983 \emph{Progress of Theoretical Physics}, \emph{69}, 1116.

\bibitem[\protect\astroncite{\emph{{Sekiya} and {Onishi}}}{2018}]{SO18}
{Sekiya} M. and {Onishi} I.~K., 2018 \emph{\apj}, \emph{860}, 2, 140.

\bibitem[\protect\astroncite{\emph{{Semenov} et~al.}}{2003}]{Semenov_etal_2003}
{Semenov} D. et~al., 2003 \emph{\aap}, \emph{410}, 611.

\bibitem[\protect\astroncite{\emph{{Seok} and {Li}}}{2017}]{Seok2017}
{Seok} J.~Y. and {Li} A., 2017 \emph{\apj}, \emph{835}, 2, 291.

\bibitem[\protect\astroncite{\emph{Shakura and
  Sunyaev}}{1973}]{Shakura.Sunyaev73}
Shakura N.~I. and Sunyaev R.~A., 1973 \emph{\aap}, \emph{24}, 337.

\bibitem[\protect\astroncite{\emph{{Shariff} and {Cuzzi}}}{2011}]{Shariff2011}
{Shariff} K. and {Cuzzi} J.~N., 2011 \emph{\apj}, \emph{738}, 1, 73.

\bibitem[\protect\astroncite{\emph{{Shi} and {Chiang}}}{2013}]{shi13}
{Shi} J.-M. and {Chiang} E., 2013 \emph{\apj}, \emph{764}, 20.

\bibitem[\protect\astroncite{\emph{{Shtemler} and {Mond}}}{2019}]{shtemler19}
{Shtemler} Y. and {Mond} M., 2019 \emph{\mnras}, \emph{488}, 3, 4207.

\bibitem[\protect\astroncite{\emph{{Shtemler} and {Mond}}}{2020}]{shtemler20}
{Shtemler} Y. and {Mond} M., 2020 \emph{\mnras}, \emph{499}, 3, 3222.

\bibitem[\protect\astroncite{\emph{{Simon} and
  {Armitage}}}{2014}]{2014ApJ...784...15S}
{Simon} J.~B. and {Armitage} P.~J., 2014 \emph{\apj}, \emph{784}, 1, 15.

\bibitem[\protect\astroncite{\emph{Simon et~al.}}{2011}]{Simon2011}
Simon J.~B. et~al., 2011 \emph{\apj}, \emph{743}, 1, 17.

\bibitem[\protect\astroncite{\emph{{Simon} et~al.}}{2013}]{Simon.bai.ea13b}
{Simon} J.~B. et~al., 2013 \emph{\apj}, \emph{764}, 1, 66.

\bibitem[\protect\astroncite{\emph{Simon et~al.}}{2013}]{Simon.Bai.ea13a}
Simon J.~B. et~al., 2013 \emph{\apj}, \emph{775}, 1, 73.

\bibitem[\protect\astroncite{\emph{Simon et~al.}}{2015}]{Simon2015b}
Simon J.~B. et~al., 2015 \emph{\apj}, \emph{808}, 2, 180.

\bibitem[\protect\astroncite{\emph{{Simon} et~al.}}{2016}]{SA16}
{Simon} J.~B. et~al., 2016 \emph{\apj}, \emph{822}, 1, 55.

\bibitem[\protect\astroncite{\emph{{Simon} et~al.}}{2017}]{SA17}
{Simon} J.~B. et~al., 2017 \emph{\apjl}, \emph{847}, 2, L12.

\bibitem[\protect\astroncite{\emph{{Simon} et~al.}}{2018}]{2018ApJ...865...10S}
{Simon} J.~B. et~al., 2018 \emph{\apj}, \emph{865}, 1, 10.

\bibitem[\protect\astroncite{\emph{{Spiegel}}}{1957}]{Spiegel_1957}
{Spiegel} E.~A., 1957 \emph{\apj}, \emph{126}, 202.

\bibitem[\protect\astroncite{\emph{{Squire} and
  {Hopkins}}}{2018{\natexlab{a}}}]{Squire2018a}
{Squire} J. and {Hopkins} P.~F., 2018{\natexlab{a}} \emph{\mnras}, \emph{477},
  5011.

\bibitem[\protect\astroncite{\emph{{Squire} and
  {Hopkins}}}{2018{\natexlab{b}}}]{Squire2018}
{Squire} J. and {Hopkins} P.~F., 2018{\natexlab{b}} \emph{\apj}, \emph{856},
  L15.

\bibitem[\protect\astroncite{\emph{{Squire} and {Hopkins}}}{2020}]{Squire2020a}
{Squire} J. and {Hopkins} P.~F., 2020 \emph{\mnras}, \emph{498}, 1, 1239.

\bibitem[\protect\astroncite{\emph{{Steinpilz} et~al.}}{2019}]{Steinpilz2019}
{Steinpilz} T. et~al., 2019 \emph{\apj}, \emph{874}, 1, 60.

\bibitem[\protect\astroncite{\emph{{Steinpilz}
  et~al.}}{2020{\natexlab{a}}}]{Steinpilz2020b}
{Steinpilz} T. et~al., 2020{\natexlab{a}} \emph{Nature Physics}, \emph{16}, 2,
  225.

\bibitem[\protect\astroncite{\emph{{Steinpilz}
  et~al.}}{2020{\natexlab{b}}}]{Steinpilz2020a}
{Steinpilz} T. et~al., 2020{\natexlab{b}} \emph{New Journal of Physics},
  \emph{22}, 9, 093025.

\bibitem[\protect\astroncite{\emph{{Stoll} and {Kley}}}{2014}]{stoll14}
{Stoll} M.~H.~R. and {Kley} W., 2014 \emph{\aap}, \emph{572}, A77.

\bibitem[\protect\astroncite{\emph{{Stoll} and {Kley}}}{2016}]{stoll16}
{Stoll} M. H.~R. and {Kley} W., 2016 \emph{\aap}, \emph{594}, A57.

\bibitem[\protect\astroncite{\emph{{Stoll}
  et~al.}}{2017{\natexlab{a}}}]{stoll17}
{Stoll} M. H.~R. et~al., 2017{\natexlab{a}} \emph{\aap}, \emph{599}, L6.

\bibitem[\protect\astroncite{\emph{{Stoll}
  et~al.}}{2017{\natexlab{b}}}]{stoll17b}
{Stoll} M. H.~R. et~al., 2017{\natexlab{b}} \emph{\aap}, \emph{604}, A28.

\bibitem[\protect\astroncite{\emph{{Suriano} et~al.}}{2019}]{Suriano19}
{Suriano} S.~S. et~al., 2019 \emph{\mnras}, \emph{484}, 1, 107.

\bibitem[\protect\astroncite{\emph{{Suzuki} and
  {Inutsuka}}}{2014}]{2014ApJ...784..121S}
{Suzuki} T.~K. and {Inutsuka} S.-i., 2014 \emph{\apj}, \emph{784}, 2, 121.

\bibitem[\protect\astroncite{\emph{Suzuki et~al.}}{2016}]{Suzuki.Ogihara.ea16}
Suzuki T.~K. et~al., 2016 \emph{\aap}, \emph{596}, A74.

\bibitem[\protect\astroncite{\emph{{Takahashi} and
  {Inutsuka}}}{2014}]{takahashi14}
{Takahashi} S.~Z. and {Inutsuka} S.-i., 2014 \emph{\apj}, \emph{794}, 55.

\bibitem[\protect\astroncite{\emph{{Takahashi} and
  {Inutsuka}}}{2016}]{Takahashi2016}
{Takahashi} S.~Z. and {Inutsuka} S.-i., 2016 \emph{\aj}, \emph{152}, 184.

\bibitem[\protect\astroncite{\emph{{Tanga} et~al.}}{2004}]{tanga04}
{Tanga} P. et~al., 2004 \emph{\aap}, \emph{427}, 1105.

\bibitem[\protect\astroncite{\emph{{Tassoul}}}{1978}]{tassoul78}
{Tassoul} J., 1978 \emph{{Theory of rotating stars}}.

\bibitem[\protect\astroncite{\emph{{Teague} et~al.}}{2016}]{Teague2016}
{Teague} R. et~al., 2016 \emph{\aap}, \emph{592}, A49.

\bibitem[\protect\astroncite{\emph{{Teed} and {Latter}}}{2021}]{TeedLatter21}
{Teed} R.~J. and {Latter} H.~N., 2021 \emph{\mnras}.

\bibitem[\protect\astroncite{\emph{{Teiser} et~al.}}{2021}]{Teiser2021}
{Teiser} J. et~al., 2021 \emph{\apjl}, \emph{908}, 2, L22.

\bibitem[\protect\astroncite{\emph{{Thi} et~al.}}{2019}]{Thi+2019}
{Thi} W.~F. et~al., 2019 \emph{\aap}, \emph{632}, A44.

\bibitem[\protect\astroncite{\emph{{Thirouin} and
  {Sheppard}}}{2019}]{ThirouinSheppard2019}
{Thirouin} A. and {Sheppard} S.~S., 2019 \emph{\aj}, \emph{157}, 6, 228.

\bibitem[\protect\astroncite{\emph{{Tominaga} et~al.}}{2019}]{tominaga19}
{Tominaga} R.~T. et~al., 2019 \emph{\apj}, \emph{881}, 1, 53.

\bibitem[\protect\astroncite{\emph{{Tominaga} et~al.}}{2020}]{tominaga20}
{Tominaga} R.~T. et~al., 2020 \emph{\apj}, \emph{900}, 2, 182.

\bibitem[\protect\astroncite{\emph{{Trapman} et~al.}}{2020}]{Trapman2019}
{Trapman} L. et~al., 2020 \emph{\aap}, \emph{640}, A5.

\bibitem[\protect\astroncite{\emph{Turner et~al.}}{2014}]{Turner.Fromang.ea14}
Turner N.~J. et~al., 2014 \emph{Protostar and {{Planets VI}}} (H.~Beuther,
  R.~S. Klessen, C.~P. Dullemond, and T.~K. Henning), pp. 411--432, {University
  of Arizona Press}.

\bibitem[\protect\astroncite{\emph{{Ueda} et~al.}}{2019}]{2019ApJ...871...10U}
{Ueda} T. et~al., 2019 \emph{\apj}, \emph{871}, 1, 10.

\bibitem[\protect\astroncite{\emph{{Umebayashi} and
  {Nakano}}}{2009}]{Umebayashi+Nakano2009}
{Umebayashi} T. and {Nakano} T., 2009 \emph{\apj}, \emph{690}, 1, 69.

\bibitem[\protect\astroncite{\emph{{Umurhan} et~al.}}{2013}]{umurhan13}
{Umurhan} O.~M. et~al., 2013 \emph{European Physical Journal Web of
  Conferences}, vol.~46 of \emph{European Physical Journal Web of Conferences},
  p. 3003.

\bibitem[\protect\astroncite{\emph{{Umurhan}
  et~al.}}{2016{\natexlab{a}}}]{umurhanshariff2016}
{Umurhan} O.~M. et~al., 2016{\natexlab{a}} \emph{\apj}, \emph{830}, 95.

\bibitem[\protect\astroncite{\emph{{Umurhan}
  et~al.}}{2016{\natexlab{b}}}]{umurhan16c}
{Umurhan} O.~M. et~al., 2016{\natexlab{b}} \emph{\aap}, \emph{586}, A33.

\bibitem[\protect\astroncite{\emph{{Umurhan} et~al.}}{2020}]{Umurhan2020}
{Umurhan} O.~M. et~al., 2020 \emph{\apj}, \emph{895}, 1, 4.

\bibitem[\protect\astroncite{\emph{{Urpin}}}{2003}]{urpin03}
{Urpin} V., 2003 \emph{\aap}, \emph{404}, 397.

\bibitem[\protect\astroncite{\emph{{Urpin} and {Brandenburg}}}{1998}]{urpin98}
{Urpin} V. and {Brandenburg} A., 1998 \emph{\mnras}, \emph{294}, 399.

\bibitem[\protect\astroncite{\emph{{van der Marel}
  et~al.}}{2018}]{vanderMarel.Williams.ea18}
{van der Marel} N. et~al., 2018 \emph{\apj}, \emph{854}, 2, 177.

\bibitem[\protect\astroncite{\emph{{Venuti} et~al.}}{2014}]{Venuti14}
{Venuti} L. et~al., 2014 \emph{\aap}, \emph{570}, A82.

\bibitem[\protect\astroncite{\emph{{Vinkovi{\'c}} and
  {{\v{C}}emelji{\'c}}}}{2021}]{VC21}
{Vinkovi{\'c}} D. and {{\v{C}}emelji{\'c}} M., 2021 \emph{\mnras}, \emph{500},
  1, 506.

\bibitem[\protect\astroncite{\emph{{Visser} et~al.}}{2021}]{VDD21}
{Visser} R.~G. et~al., 2021 \emph{\aap}, \emph{647}, A126.

\bibitem[\protect\astroncite{\emph{{Volponi}}}{2014}]{volponi14}
{Volponi} F., 2014 \emph{\mnras}, \emph{441}, 1, 813.

\bibitem[\protect\astroncite{\emph{{Volponi}}}{2016}]{volponi16}
{Volponi} F., 2016 \emph{\mnras}, \emph{460}, 1, 560.

\bibitem[\protect\astroncite{\emph{{Wahlberg Jansson} and
  {Johansen}}}{2014}]{WJ14}
{Wahlberg Jansson} K. and {Johansen} A., 2014 \emph{\aap}, \emph{570}, A47.

\bibitem[\protect\astroncite{\emph{{Wahlberg Jansson} and
  {Johansen}}}{2017}]{WJ17}
{Wahlberg Jansson} K. and {Johansen} A., 2017 \emph{\mnras}, \emph{469}, S149.

\bibitem[\protect\astroncite{\emph{{Wahlberg Jansson} et~al.}}{2017}]{WJBB17}
{Wahlberg Jansson} K. et~al., 2017 \emph{\apj}, \emph{835}, 1, 109.

\bibitem[\protect\astroncite{\emph{{Wang} and
  {Balmforth}}}{2020}]{wangbalmforth2020}
{Wang} C. and {Balmforth} N.~J., 2020 \emph{Journal of Fluid Mechanics},
  \emph{883}, A12.

\bibitem[\protect\astroncite{\emph{{Wang} and
  {Balmforth}}}{2021}]{Wang_Balmforth_2021}
{Wang} C. and {Balmforth} N.~J., 2021 \emph{Journal of Fluid Mechanics},
  \emph{917}, A48.

\bibitem[\protect\astroncite{\emph{Wang and Goodman}}{2017}]{Wang.Goodman17}
Wang L. and Goodman J.~J., 2017 \emph{\apj}, \emph{835}, 1, 59.

\bibitem[\protect\astroncite{\emph{Wang et~al.}}{2019}]{Wang.Bai.ea18}
Wang L. et~al., 2019 \emph{\apj}, \emph{874}, 1, 90.

\bibitem[\protect\astroncite{\emph{{Wang}}}{2016}]{wangmarcus2016}
{Wang} M., 2016 \emph{Baroclinic Critical Layers and Zombie Vortex Instability
  in Stratified Rotational Shear Flow}, Ph.D. thesis, University of California,
  Berkeley.

\bibitem[\protect\astroncite{\emph{{Ward}}}{1976}]{war76}
{Ward} W.~R., 1976 \emph{{Frontiers of Astrophysics, {\rm ed. E.H. Avrett}}},
  pp. 1--40.

\bibitem[\protect\astroncite{\emph{{Ward}}}{2000}]{war00}
{Ward} W.~R., 2000 \emph{Origin of the Earth and Moon}, pp. 75--84.

\bibitem[\protect\astroncite{\emph{{Wardle}}}{2007}]{Wardle2007}
{Wardle} M., 2007 \emph{\apss}, \emph{311}, 1-3, 35.

\bibitem[\protect\astroncite{\emph{Wardle and
  K{\"o}nigl}}{1993}]{Wardle.Koenigl93}
Wardle M. and K{\"o}nigl A., 1993 \emph{\apj}, \emph{410}, 218.

\bibitem[\protect\astroncite{\emph{{Wardle} and {Ng}}}{1999}]{Wardle+Ng1999}
{Wardle} M. and {Ng} C., 1999 \emph{\mnras}, \emph{303}, 2, 239.

\bibitem[\protect\astroncite{\emph{{Weber} et~al.}}{2020}]{Weber2020}
{Weber} M.~L. et~al., 2020 \emph{\mnras}, \emph{496}, 1, 223.

\bibitem[\protect\astroncite{\emph{{Whipple}}}{1966}]{Whipple1966}
{Whipple} F.~L., 1966 \emph{Science}, \emph{153}, 3731, 54.

\bibitem[\protect\astroncite{\emph{{Whipple}}}{1972}]{Whipple1972}
{Whipple} F.~L., 1972 \emph{From Plasma to Planet} (A.~{Elvius}), p. 211.

\bibitem[\protect\astroncite{\emph{{Whipple}}}{1973}]{Whipple1973}
{Whipple} F.~L., 1973 \emph{{Radial Pressure in the Solar Nebula as Affecting
  the Motions of Planetesimals}}, vol. 319, p. 355.

\bibitem[\protect\astroncite{\emph{{Windmark} et~al.}}{2012}]{Windmark2012}
{Windmark} F. et~al., 2012 \emph{\aap}, \emph{544}, L16.

\bibitem[\protect\astroncite{\emph{{Winter} et~al.}}{2020}]{Winter2020}
{Winter} A.~J. et~al., 2020 \emph{\mnras}, \emph{497}, 1, L40.

\bibitem[\protect\astroncite{\emph{{Woitke} et~al.}}{2016}]{Woitke_etal_2016}
{Woitke} P. et~al., 2016 \emph{\aap}, \emph{586}, A103.

\bibitem[\protect\astroncite{\emph{{W{\"o}lfer} et~al.}}{2019}]{Woelfer2019}
{W{\"o}lfer} L. et~al., 2019 \emph{\mnras}, \emph{490}, 4, 5596.

\bibitem[\protect\astroncite{\emph{{Wurster}}}{2021}]{Wurster2021}
{Wurster} J., 2021 \emph{\mnras}, \emph{501}, 4, 5873.

\bibitem[\protect\astroncite{\emph{{Xu} and {Bai}}}{2021}]{XB21}
{Xu} Z. and {Bai} X.-N., 2021 \emph{arXiv e-prints}, arXiv:2108.10486.

\bibitem[\protect\astroncite{\emph{{Yang} and {Johansen}}}{2014}]{YJ14}
{Yang} C.-C. and {Johansen} A., 2014 \emph{\apj}, \emph{792}, 2, 86.

\bibitem[\protect\astroncite{\emph{{Yang} and {Johansen}}}{2016}]{YJ16}
{Yang} C.-C. and {Johansen} A., 2016 \emph{\apjs}, \emph{224}, 2, 39.

\bibitem[\protect\astroncite{\emph{{Yang} and {Zhu}}}{2021}]{YZ21}
{Yang} C.-C. and {Zhu} Z., 2021 \emph{\mnras}, \emph{508}, 4, 5538.

\bibitem[\protect\astroncite{\emph{{Yang} et~al.}}{2017}]{YJC17}
{Yang} C.~C. et~al., 2017 \emph{\aap}, \emph{606}, A80.

\bibitem[\protect\astroncite{\emph{{Yang} et~al.}}{2018}]{YMJ18}
{Yang} C.-C. et~al., 2018 \emph{\apj}, \emph{868}, 1, 27.

\bibitem[\protect\astroncite{\emph{{Yang} and {Bai}}}{2021}]{yang21}
{Yang} H. and {Bai} X.-N., 2021 \emph{\apj}, \emph{922}, 2, 201.

\bibitem[\protect\astroncite{\emph{Yellin-Bergovoy et~al.}}{2021}]{yellin21}
Yellin-Bergovoy R. et~al., 2021 \emph{Geophysical \& Astrophysical Fluid
  Dynamics}, \emph{0}, 0, 1.

\bibitem[\protect\astroncite{\emph{Youdin and Johansen}}{2007}]{Youdin2007}
Youdin A. and Johansen A., 2007 \emph{\apj}, \emph{662}, 1, 613.

\bibitem[\protect\astroncite{\emph{{Youdin}}}{2005}]{y05a}
{Youdin} A.~N., 2005 \emph{{astro-ph/0508659}}.

\bibitem[\protect\astroncite{\emph{{Youdin}}}{2011}]{Youdin2011}
{Youdin} A.~N., 2011 \emph{\apj}, \emph{731}, 2, 99.

\bibitem[\protect\astroncite{\emph{Youdin and Goodman}}{2005}]{Youdin2005}
Youdin A.~N. and Goodman J., 2005 \emph{\apj}, \emph{620}, 1, 459.

\bibitem[\protect\astroncite{\emph{{Youdin} and {Lithwick}}}{2007}]{YL07}
{Youdin} A.~N. and {Lithwick} Y., 2007 \emph{\icarus}, \emph{192}, 2, 588.

\bibitem[\protect\astroncite{\emph{{Youdin} and {Shu}}}{2002}]{ys02}
{Youdin} A.~N. and {Shu} F.~H., 2002 \emph{\apj}, \emph{580}, 494.

\bibitem[\protect\astroncite{\emph{Zhu and Stone}}{2018}]{Zhu.Stone18}
Zhu Z. and Stone J.~M., 2018 \emph{\apj}, \emph{857}, 1, 34.

\bibitem[\protect\astroncite{\emph{{Zhu} and {Yang}}}{2021}]{Zhu2021}
{Zhu} Z. and {Yang} C.-C., 2021 \emph{\mnras}, \emph{501}, 1, 467.

\bibitem[\protect\astroncite{\emph{{Zhu} et~al.}}{2015}]{ZSB15}
{Zhu} Z. et~al., 2015 \emph{\apj}, \emph{801}, 2, 81.

\bibitem[\protect\astroncite{\emph{{Zhuravlev}}}{2019}]{Zhuravlev2019}
{Zhuravlev} V.~V., 2019 \emph{\mnras}, \emph{489}, 3, 3850.

\bibitem[\protect\astroncite{\emph{{Zhuravlev}}}{2020}]{Zhuravlev2020}
{Zhuravlev} V.~V., 2020 \emph{\mnras}, \emph{494}, 1, 1395.

\bibitem[\protect\astroncite{\emph{Zsom et~al.}}{2010}]{Zsom2010}
Zsom A. et~al., 2010 \emph{\aap}, \emph{513}, A57.

\end{thebibliography}
\bibliographystyle{pp7}

\end{document}